\documentclass[rmp,twocolumn,amsmath,eqsecnum]{revtex4}

\usepackage{bm}
\usepackage{graphicx}
\usepackage[bookmarks,colorlinks=true,citecolor=blue]{hyperref}
\pdfoutput=1

\newcommand{\eps}{\varepsilon}

\newcommand{\dtot}[2]{\frac{d #1}{d #2}}
\newcommand{\dpar}[2]{\frac{\partial #1}{\partial #2}}

\newcommand{\rmd}{d}

\newcommand{\ket}[1]{\lvert #1 \rangle}
\newcommand{\bra}[1]{\langle #1 \rvert}
\newcommand{\bracket}[1]{\langle #1 \rangle}

\DeclareMathOperator{\TR}{Tr}

\renewcommand{\cal}{\mathcal}
\newcommand{\dparh}[2]{\frac{\partial #1}{\hbar\partial #2}}
\newcommand{\intqq}{\int_\text{BZ} \frac{dq}{2\pi}\,}

\newcommand{\teps}{\tilde{\eps}}

\DeclareMathOperator{\Tr}{Tr}

\begin{document}

\title{Berry Phase Effects on Electronic Properties}

\author{Di Xiao}
\affiliation{Materials Science \& Technology Division, Oak Ridge
  National Laboratory, Oak Ridge, TN 37831, USA}

\author{Ming-Che Chang}
\affiliation{Department of Physics, National Taiwan Normal University,
Taipei 11677, Taiwan}

\author{Qian Niu}
\affiliation{Department of Physics, The University of Texas at Austin,
  Austin, TX 78712, USA}

\date{\today}

\begin{abstract}
Ever since its discovery, the Berry phase has permeated through all
branches of physics.  Over the last three decades, it was gradually
realized that the Berry phase of the electronic wave function can have
a profound effect on material properties and is responsible for a
spectrum of phenomena, such as ferroelectricity, orbital magnetism,
various (quantum/anomalous/spin) Hall effects, and quantum charge
pumping.  This progress is summarized in a pedagogical manner in this
review.  We start with a brief summary of necessary background,
followed by a detailed discussion of the Berry phase effect in a
variety of solid state applications.  A common thread of the review is
the semiclassical formulation of electron dynamics, which is a
versatile tool in the study of electron dynamics in the presence of
electromagnetic fields and more general perturbations.  Finally, we
demonstrate a re-quantization method that converts a semiclassical
theory to an effective quantum theory.  It is clear that the Berry
phase should be added as a basic ingredient to our understanding of
basic material properties.
\end{abstract}

\maketitle

\tableofcontents


\section{Introduction}

\subsection{Topical overview}

In 1984, Michael Berry wrote a paper that has generated immense
interests throughout the different fields of physics including quantum
chemistry \cite{berry1984}.  This is about adiabatic evolution of an eigenenergy state
when the external parameters of a quantum system change slowly and
make up a loop in the parameter space.  In the absence of degeneracy,
the eigenstate will surely come back to itself when finishing the
loop, but there will be a phase difference equal to the time integral
of the energy (divided by $\hbar$) plus an extra which is later
commonly called the Berry phase.

The Berry phase has three key properties that make the concept
important.  First, it is gauge invariant. The eigen-wavefunction is
defined by a homogeneous linear equation (the eigenvalue equation), so
one has the gauge freedom of multiplying it with an overall phase
factor which can be parameter dependent.  The Berry phase is unchanged
(up to unessential integer multiple of $2\pi$) by such a phase factor,
provided the eigen-wavefunction is kept to be single valued over the
loop. This property makes the Berry phase physical, and the early
experimental studies were focused on measuring it directly through
interference phenomena.

Second, the Berry phase is geometrical. It can be written as a
line-integral over the loop in the parameter space, and does not
depend on the exact rate of change along the loop. This property makes
it possible to express the Berry phase in terms of local geometrical
quantities in the parameter space.  Indeed, Berry himself showed that
one can write the Berry phase as an integral of a field, which we now
call as the Berry curvature, over a surface suspending the loop. A
large class of applications of the Berry phase concept occur when the
parameters themselves are actually dynamical variables of slow degrees
of freedom. The Berry curvature plays an essential role in the
effective dynamics of these slow variables. The vast majority of
applications considered in this review are of this nature.

Third, the Berry phase has close analogies to gauge field theories and
differential geometry \cite{simon1983}.  This makes the Berry phase a beautiful,
intuitive and powerful unifying concept, especially valuable in
today's ever specializing physical science. In primitive terms, the
Berry phase is like the Aharonov-Bohm phase of a charge particle
traversing a loop including a magnetic flux, while the Berry curvature
is like the magnetic field. The integral of the Berry curvature over
closed surfaces, such as a sphere or torus, is known to be topological
and quantized as integers (Chern numbers).  This is analogous to the
Dirac monopoles of magnetic charges that must be quantized in order to
have a consistent theory of quantum mechanical theory for charged
particles in magnetic fields. Interestingly, while the magnetic
monopoles are yet to be detected in the real world, the topological
Chern numbers have already found correspondence with the quantized
Hall conductance plateaus in the spectacular quantum Hall phenomenon
\cite{thouless1982}.

This review is about applications of the Berry phase concept in solid
state physics.  In this field, we are typically interested in
macroscopic phenomena which are slow in time and smooth in space in
comparison with the atomic scales.  Not surprisingly, the vast
majority of applications of the Berry phase concept are found
here. This field of physics is also extremely diverse, and we have
many layers of theoretical frameworks with different degrees of
transparency and validity  \cite{shapere1989,gphase2003}. Therefore,
a unifying organizing principle such as the Berry phase concept is
particularly valuable.

We will focus our attention on electronic properties, which play a
dominant role in various aspects of material properties.  The
electrons are the glue of materials and they are also the agents
responding swiftly to external fields and giving rise to strong and
useful signals. A basic paradigm of theoretical framework is based on
the assumption that electrons are in Bloch waves traveling more or
less independently in periodic potentials of the lattice, except that
the Pauli exclusion principle has to be satisfied and
electron-electron interactions are taken care of in some
self-consistent manner. Much of our intuition on electron transport
are derived from the semiclassical picture that the electrons behave
almost as free particles in response to external fields provided one
uses the band energy in place of the free-particle dispersion. Partly
for this reason, first-principles studies of electronic properties
tend to only document the energy band structures and various density
profiles.

There have been overwhelming evidences that such a simple picture
cannot give complete account of effects to first order in the
fields. The anomalous velocity is a prime example, whose existence
were predicted in the early 50s \cite{karplus1954} and whose experimental relevance were
established only recently \cite{lee2004,lee2004a}. The usual derivation of the anomalous
velocity is based on a linear response analysis: the velocity operator
has off-diagonal elements and electric field mixes the bands, so that
the expectation value of the velocity acquires an additional term
proportional to the field other than the usual group velocity in the
original band \cite{adams1959,blount1962c}. The anomalous velocity were first recognized in mid and
late 90s as an effect of the Berry curvature intrinsic to each of the
bands, whose existence has really nothing to do with the external
field \cite{chang1995,chang1996,sundaram1999}.  This understanding enabled us to make a direct connection with
the topological Chern number formulation of the quantum Hall effect \cite{thouless1982,kohmoto1985},
giving incentive as well as confidence in our pursuit of the
eventually successful intrinsic explanation of the anomalous Hall
effect \cite{taguchi2001,jungwirth2002,fang2003,yao2004,zeng2006}. .

Interestingly enough, the traditional view cannot even explain some
basic effects to zeroth order of the fields. The two basic
electromagnetic properties of solids as a medium are the electric
polarization and magnetization, which can exist in the absence of
electric and magnetic fields in ferroelectric and ferromagnetic
materials. Their classic definition were based on the picture of bound
charges and currents, but these are clearly inadequate for the
electronic contribution and it was known that the polarization and
orbital magnetization cannot be determined from the charge and current
densities in the bulk of a crystal at all. A breakthrough on electric
polarization were made in early 90s by linking it with the phenomenon
of adiabatic charge transport and expressing it in terms of the Berry
phase \footnote{Also called Zak's
phase, it is independent of the Berry curvature which only
characterize Berry phases over loops continuously shrinkable to
zero \cite{zak1989}.} across the entire Brillouin zone \cite{resta1992,king-smith1993,vanderbilt1993}.  Based on the Berry phase formula, one can now routinely
calculate polarization related properties using first principles
methods, with a typical precision of the density functional
theory. The breakthrough on orbital magnetization came only recently,
showing that it not only consists of the orbital moments of the quasi
particles but also contains a Berry curvature contribution of
topological origin \cite{xiao2005,thonhauser2005,shi2007}.

In this article, we will follow the traditional semiclassical formalism
of quasiparticle dynamics, only to make it more rigorous by including
the Berry curvatures in the various facets of the phase space
including the adiabatic time parameter. All of the above mentioned
effects are transparently revealed with complete precision of the full
quantum theory.  Related effects on anomalous thermoelectric, valley
Hall and magneto-transport are easily predicted, and other effects due
to crystal deformation and order parameter inhomogeneity can also be
explored without difficulty. Moreover, by including a side-jump term
(which is itself a kind of Berry phase effect) into the usual
Boltzmann equation and associated transport during collisions as well
as anomalous transport between collisions, the semiclassical theory
can also reproduce all the intricacies of linear response theory in
weakly disordered systems \cite{sinitsyn2008}.  On a microscopic level, although the
electron wavepacket dynamics is yet to be directly observed in solids,
the formalism has been replicated for light transport in photonic
crystals, where the associated Berry phase effects are vividly
exhibited in experiments \cite{bliokh2008}. Finally, it is possible to generalize the
semiclassical dynamics in a single band into one with degenerate or
nearly degenerate bands \cite{culcer2005,shindou2005}, and one can study transport phenomena where
interband coherence effects such as in spin transport, only to realize
that the Berry curvatures and quasiparticle magnetic moments become
non-abealian (i.e., matrices).

The semiclassical formalism is certainly amendable to include
quantization effects.  For integrable dynamics, such as Bloch
oscillations and cyclotron orbits, one can use the Bohr-Sommerfeld or
EKB quantization rule. The Berry phase enters naturally as a shift to
the classical action, affecting the energies of the quantized levels,
e.g., the Wannier-Stark ladders and the Landau levels.  A high point
of this kind of applications is the explanation of the intricate
fractal-like Hofstadter spectrum \cite{chang1995,chang1996}. A recent breakthrough has also
enabled us to find the density of quantum states in the phase space
for the general case (including non-integrable systems) \cite{xiao2005}, revealing
Berry-curvature corrections which should have broad impacts on
calculations of equilibrium as well as transport properties. Finally,
one can execute a generalized Peierls substitution relating the
physical variables to the canonical variables, turning the
semiclassical dynamics into a full quantum theory valid to first order
in the fields \cite{chang2008}.  Spin-orbit coupling and various mysterious Yafet terms
are all found a simple explanation from this generalized Peierls
substitution.

Therefore, it is clear that Berry phase effects in solid state physics
are not something just nice to be found here and there, the concept is
essential for a coherent understanding of all the basic phenomena. It
is the purpose of this review to summarize a theoretical framework
which continues the traditional semiclassical point of view but with a
much broader range of validity.  It is necessary and sufficient to
include the Berry phases and gradient energy corrections in addition
to the energy dispersions in order to account all phenomena up to
first order in the fields.

\subsection{Organization of the review}

This review can be divided into three main parts.  In
sec.~\ref{sec:pump} we consider the simplest example of Berry phase in
crystals: the adiabatic transport in a band insulator.  In particular,
we show that induced adiabatic current due to a time-dependent
perturbation can be written as a Berry phase of the electronic wave
functions.  Based on this understanding, the modern theory of electric
polarization is reviewed.  In sec.~\ref{sec:hall} the electron
dynamics in the presence of an electric field is discussed as a
specific example of the time-dependent problem, and the basic formula
developed in Sec.~\ref{sec:pump} can be directly applied.  In this
case, the Berry phase manifest as transverse velocity of the
electrons, which gives rise to a Hall current.  We then apply this
formula to study the quantum, anomalous, and valley Hall effect.

To study the electron dynamics under spatial-dependent perturbations,
we turn to the semiclassical formalism of Bloch electron dynamics,
which has proven to be a powerful tool to investigate the influence of
slowly varying perturbations on the electron dynamics.
Sec.~\ref{sec:wave}, we discuss the construction of the electron wave
packet and show that the wave packet carries an orbital magnetic
moment.  Two applications of the wave packet approach, the orbital
magnetization, and anomalous thermoelectric transport in ferromagnet
are discussed.  In Sec.~\ref{sec:em} the wave packet dynamics in the
presence of electromagnetic fields is studied.  We show that the Berry
phase not only affects the equations of motion, but also modifies the
electron density of states in the phase space, which can be changed by
applying a magnetic field.  The formula of orbital magnetization is
rederived using the modified density of states.  We also presented a
comprehensive study of the magneto-transport in the presence of the
Berry phase.  The electron dynamics under more general perturbations
is discussed in Sec.~\ref{sec:gen}.  We again present two
applications: electron dynamics in deformed crystals and polarization
induced by inhomogeneity.

In the remaining part of the review, we focus on the re-quantization
of the semiclassical formulation. In Sec.~\ref{sec:qz}, the
Bohr-Sommerfeld quantization is reviewed in details. With its help,
one can incorporate the Berry phase effect into the Wannier-Stark
ladders and the Landau levels very easily.  In Sec.~\ref{sec:mbb}, we
show that the same semiclassical approach can be applied to systems
subject to a very strong magnetic field.  One only has to separate the
field into a quantization part and a perturbation. The former should
be treated quantum mechanically to obtain the magnetic Bloch band
spectrum while the latter is treated perturbatively.  Using this
formalism, we discuss the cyclotron motion, the splitting into
magnetic subbands, and the quantum Hall effect.  In
Sec.~\ref{sec:nab}, we review recent advance on the non-Abelian Berry
phase in degenerate bands.  We show that the Berry connection now
plays an explicit role in spin dynamics and is deeply related to the
spin-orbit interaction.  We then cite the relativistic Dirac electrons
and the Kane model in semiconductors as two examples of application.
Finally, we briefly discuss the re-quantization of the semiclassical theory
and the hierarchy of effective quantum theories.

We do not attempt to cover all of the Berry phase effects in this review. 
Interested readers can consult the following books or review articles 
for many more left un-mentioned: \citet{gphase1989,nenciu1991,resta1994,teufel2003,chang2008,resta2000,gphase2003,thouless-TQN}.  
In this review, we focus on a pedagogical and self-contained approach, with the main
machinery being the semiclassical formalism of Bloch electron
dynamics~\cite{chang1995,chang1996,sundaram1999}.  We shall start with
the simplest case, the gradually expand the formalism as more
complicated physical situations are considered.  Whenever a new
ingredient is added, a few applications is provided to demonstrate the
basic ideas.  The vast number of application we discussed is a
reflection of the universality of the Berry phase effect.



\subsection{\label{sec:berry}Basic Concepts of The Berry phase}

In this subsection we introduce the basic concepts of the Berry phase.
Following Berry's original paper~\cite{berry1984}, we first discuss
how the Berry phase arises as a generic feature of the adiabatic
evolution of a quantum state.  We then introduce the local description
of the Berry phase in terms of the Berry curvature.  A two-level model
is used to demonstrate these concepts as well as some important
properties, such as the quantization of the Berry phase.  Our aim is
to provide a minimal but self-contained introduction.  For a detailed
account of the Berry phase, including its mathematical foundation and
applications in a wide range of fields in physics, we refer the readers
to the books by~\citet{gphase1989,gphase2003} and references therein.


\subsubsection{Cyclic adiabatic evolution}

Let us consider a physical system described by a Hamiltonian that
depends on time through a set of parameters, denoted by $\bm R = (R_1,
R_2, \dots)$, i.e.,
\begin{equation}
H = H(\bm R) \;, \qquad \bm R = \bm R(t) \;.
\end{equation}
We are interested in the adiabatic evolution of the system as $\bm
R(t)$ moves slowly along a path $\cal C$ in the parameter space.  For
this purpose, it will be useful to introduce an instantaneous
orthonormal basis from the eigenstates of $H(\bm R)$ at each value of
the parameter $\bm R$, i.e.,
\begin{equation} \label{berry:basis}
H(\bm R) \ket{n(\bm R)} =  \eps_n(\bm R)\ket{n(\bm R)} \;.
\end{equation}
However, Eq.~\eqref{berry:basis} alone does not completely determine
the basis function $\ket{n(\bm R)}$; it still allows an arbitrary $\bm
R$-dependent phase factor of $\ket{n(\bm R)}$.  One can make a phase
choice, also known as a gauge, to remove this arbitrariness.  Here we
require that the phase of the basis function is smooth and
single-valued along the path $\cal C$ in the parameter
space.~\footnote{Strictly speaking, such a phase choice is guaranteed
  only in finite neighborhoods of the parameter space.  In the general
  case, one can proceed by dividing the path into several such
  neighborhoods overlapping with each other, then use the fact that in
  the overlapping region the wave functions are related by a gauge
  transformation of the form~\eqref{berry:gauge}.}

According to the quantum adiabatic theorem~\cite{kato1950,messiah-QM},
a system initially in one of its eigenstates $\ket{n(\bm R(0))}$ will
stay as an instantaneous eigenstate of the Hamiltonian $H(\bm R(t))$
throughout the process.  Therefore the only degree of freedom we have
is the phase of the quantum state.  Write the state at time $t$ as
\begin{equation} \label{berry:ansatz}
\ket{\psi_n(t)} = e^{i\gamma_n(t)} e^{-\frac{i}{\hbar}
\int_0^t dt' \eps_n(\bm R(t'))} \ket{n(\bm R(t))} \;,
\end{equation}
where the second exponential is known as the dynamical phase factor.
Inserting Eq.~\eqref{berry:ansatz} into the time-dependent
Schr\"odinger equation
\begin{equation}
i\hbar \dpar{}{t}\ket{\psi_n(t)} = H(\bm R(t)) \ket{\psi_n(t)}
\end{equation}
and multiplying it from the left by $\bra{n(\bm R(t))}$, one finds that
$\gamma_n$ can be expressed as a path integral in the parameter space
\begin{equation}  \label{berry:gamma}
\gamma_n = \int_{\cal C} d\bm R \cdot \bm{\cal A}_n(\bm R) \;,
\end{equation}
where $\bm{\cal A_n(\bm R)}$ is a vector-valued function
\begin{equation}
\bm{\cal A}_n(\bm R) = i\bracket{n(\bm R)|\dpar{}{\bm R}|n(\bm R)} \;.
\end{equation}
This vector $\bm{\cal A}_n(\bm R)$ is called the Berry connection or
the Berry vector potential.  Equation~\eqref{berry:gamma} shows that
in addition to the dynamical phase, the quantum state will acquire an
additional phase $\gamma_n$ during the adiabatic evolution.

Obviously, $\bm{\cal A}_n(\bm R)$ is gauge-dependent.  If we make a
gauge transformation
\begin{equation}  \label{berry:gauge}
\ket{n(\bm R)} \to e^{i\zeta(\bm R)} \ket{n(\bm R)}
\end{equation}
with $\zeta(\bm R)$ being an arbitrary smooth function, $\bm{\cal
  A}_n(\bm R)$ transforms according to
\begin{equation}
\bm{\cal A}_n(\bm R) \to \bm{\cal A}_n(\bm R)
- \dpar{}{\bm R} \zeta(\bm R) \;.
\end{equation}
Consequently, the phase $\gamma_n$ given by Eq.~\eqref{berry:gamma}
will be changed by $\zeta(\bm R(0)) - \zeta(\bm R(T))$ after the
transformation, where $\bm R(0)$ and $\bm R(T)$ are the initial and
final points of the path $\cal C$.  This observation has led
\citet{fock1928} to conclude that one can always choose a suitable
$\zeta(\bm R)$ such that $\gamma_n$ accumulated along the path $\cal
C$ is canceled out, leaving Eq.~\eqref{berry:ansatz} with only the
dynamical phase.  Because of this, the phase $\gamma_n$ has long been
deemed unimportant and it was usually neglected in the theoretical
treatment of time-dependent problems.

This conclusion remained unchallenged until \citet{berry1984}
reconsidered the cyclic evolution of the system along a \emph{closed}
path $\cal C$ with $\bm R(T) = \bm R(0)$.  The phase choice we made
earlier on the basis function $\ket{n(\bm R)}$ requires $e^{i\zeta(\bm
  R)}$ in the gauge transformation, Eq.~\eqref{berry:gauge}, to be
single-valued, which implies
\begin{equation}
\zeta(\bm R(0)) - \zeta(\bm R(T)) = 2\pi \times \text{integer} \;.
\end{equation}
This shows that $\gamma_n$ can be only changed by an integer multiple
of $2\pi$ under the gauge transformation~\eqref{berry:gauge} and it
cannot be removed.  Therefore for a closed path, $\gamma_n$ becomes a
gauge-invariant physical quantity, now known as the Berry phase or
geometric phase in general; it is given by
\begin{equation} \label{berry:phase}
\gamma_n = \oint_{\cal C} d\bm R \cdot \bm{\cal A}_n(\bm R) \;.
\end{equation}

From the above definition, we can see that the Berry phase only
depends on the geometric aspect of the closed path, and is independent
of how $\bm R(t)$ varies in time.  The explicit time-dependence is
thus not essential in the description of the Berry phase and will be
dropped in the following discussion.


\subsubsection{Berry curvature}

It is useful to define, in analogy to electrodynamics, a gauge field
tensor derived from the Berry vector potential:
\begin{equation} \label{berry:def} \begin{split}
\Omega^n_{\mu\nu}(\bm R) &= \dpar{}{R^\mu} \cal A^n_\nu(\bm R)
- \dpar{}{R_\nu} \cal A^n_\mu (\bm R) \\
&= i\Bigl[\bracket{\dpar{n(\bm R)}{R^\mu}|\dpar{n(\bm R)}{R^\nu}}
- (\nu \leftrightarrow \mu)\Bigr] \;.
\end{split} \end{equation}
This field is called the Berry curvature.  Then according to Stokes's
theorem the Berry phase can be written as a surface integral
\begin{equation} \label{berry:phase1}
\gamma_n = \int_{\cal S} dR^\mu\wedge dR^\nu \,
\frac{1}{2} \Omega^n_{\mu\nu}(\bm R) \;,
\end{equation}
where $\cal S$ is an arbitrary surface enclosed by the path $\cal C$.
It can be verified from Eq.~\eqref{berry:def} that, unlike the Berry
vector potential, the Berry curvature is gauge invariant and
thus observable.

If the parameter space is three-dimensional, Eqs.~\eqref{berry:def}
and \eqref{berry:phase1} can be recasted into a vector form
\begin{align*}
\bm\Omega_n(\bm R) &= \bm\nabla_{\bm R} \times \bm{\cal A}_n(\bm R)\;,
\tag{\ref{berry:def}'} \label{berry:def1} \\
\gamma_n &= \int_{\cal S} d\bm S \cdot \bm\Omega_n(\bm R) \;.
\tag{\ref{berry:phase1}'}
\end{align*}
The Berry curvature tensor $\Omega^n_{\mu\nu}$ and vector
$\bm\Omega_n$ is related by $\Omega^n_{\mu\nu} = \epsilon_{\mu\nu\xi}
(\bm\Omega_n)_\xi$ with $\epsilon_{\mu\nu\xi}$ being the Levi-Civita
antisymmetry tensor.  The vector form gives us an intuitive picture of
the Berry curvature: it can be viewed as the magnetic field in the
parameter space.

Besides the differential formula given in Eq.~\eqref{berry:def}, the
Berry curvature can be also written as a summation over the
eigenstates:
\begin{equation} \label{berry:sum}
\Omega^n_{\mu\nu}(\bm R) = i\sum_{n'\neq n}
\frac{\bracket{n|\dpar{H}{R^\mu}|n'}\bracket{n'|\dpar{H}{R^\nu}|n}
- (\nu \leftrightarrow \mu)}{(\eps_n - \eps_{n'})^2} \;.
\end{equation}
It can be obtained from Eq.~\eqref{berry:def} by using the relation
$\bracket{n|\partial H/\partial\bm R|n'} = \bracket{\partial
  n/\partial \bm R|n'}(\eps_n - \eps_{n'})$ for $n'\neq n$.  The
summation formula has the advantage that no differentiation on the
wave function is involved, therefore it can be evaluated under any
gauge choice.  This property is particularly useful for numerical
calculations, in which the condition of a smooth phase choice of
the eigenstates is not guaranteed in standard diagonalization
algorithms.  It has been used to evaluate the Berry curvature in
crystals with the eigenfunctions supplied from first-principles
calculations~\cite{fang2003,yao2004}.

Equation~\eqref{berry:sum} offers further insight on the origin of the
Berry curvature.  The adiabatic approximation we adopted earlier is
essentially a projection operation, i.e., the dynamics of the system
is restricted to the $n$th energy level.  In view of
Eq.~\eqref{berry:sum}, the Berry curvature can be regarded as the
result of the ``residual'' interaction of those projected-out energy
levels. In fact, if all energy levels are included, it follows from
Eq.~\eqref{berry:sum} that the total Berry curvature vanishes for each
value of $\bm R$,
\begin{equation}
\sum_n \Omega^n_{\mu\nu}(\bm R) = 0 \;.
\end{equation}
This is the local conservation law for the Berry
curvature.~\footnote{The conservation law is obtained under the
  condition that the full Hamiltonian is known.  However, in practice
  one usually deals with effective Hamiltonians which are obtained
  through some projection process of the full Hamiltonian.  Therefore
  there will always be some ``residual'' Berry curvature accompanying
  the effective Hamiltonian.  See \citet{chang2008} and discussions in
  Sec.~\ref{sec:nab}.}  Equation~\eqref{berry:sum} also shows that
$\Omega^n_{\mu\nu}(\bm R)$ becomes singular if two energy levels
$\eps_n(\bm R)$ and $\eps_{n'}(\bm R)$ are brought together at certain
value of $\bm R$.  This degeneracy point corresponds to a monopole in
the parameter space; an explicit example is given below.  If the
degenerate points form a string in the parameter space, it is known as
the Dirac string.

So far we have discussed the situation where a single energy level can
be separated out in the adiabatic evolution.  However, if the energy
levels are degenerate, then the dynamics must be projected to a
subspace spanned by those degenerate eigenstates.  \citet{wilczek1984}
showed that in this situation non-Abelian Berry curvature naturally
arises.  \citet{culcer2005,shindou2005} have discussed the non-Abelian
Berry curvature in the context of degenerate Bloch bands.  In the
following we shall focus on the Abelian formulation and defer the
discussion of the non-Abelian Berry curvature to Sec.~\ref{sec:nab}.

Compared to the Berry phase which is always associated with a closed
path, the Berry curvature is truly a \emph{local} quantity.  It
provides a local description of the geometric properties of the
parameter space.  More importantly, just like a magnetic field can
affect the electron dynamics, the Berry curvature also directly
participates in the dynamics of the adiabatic
parameters~\cite{kuratsuji1985}.  In this sense, the Berry curvature
is a more fundamental quantity than the Berry phase.


\subsubsection{\label{sec:spin}Example: The two-level system}

Let us consider a concrete example: a two-level system.  The purpose
to study this system is two-fold.  Firstly, as a simple model, it
demonstrates the basic concepts as well as several important properties
of the Berry phase.  Secondly, it will be repeatedly used later in
this article, although in different physical context.  It is therefore
useful to go through the basis of this model.

The generic Hamiltonian of a two-level system takes the following form
\begin{equation} \label{berry:ham}
H = \bm h(\bm R) \cdot \bm\sigma \;,
\end{equation}
where $\bm\sigma$ is the Pauli matrices.  Despite its simple form, the
above Hamiltonian describes a number of physical systems in condensed
matter physics for which the Berry phase effect has been discussed.
Examples include spin-orbit coupled systems~\cite{culcer2003,liu2008},
linearly conjugated diatomic polymers~\cite{su1979,rice1982},
one-dimensional ferroelectrics~\cite{onoda2004,vanderbilt1993},
graphene~\cite{semenoff1984,haldane1988}, and Bogoliubov
quasiparticles~\cite{zhang2006}.

Parameterize $\bm h$ by its azimuthal angle $\theta$ and polar angle
$\phi$, $\bm h = h(\sin\theta\cos\phi, \sin\theta\sin\phi,
\cos\theta)$.  The two eigenstates with energies $\pm h$ is
\begin{equation} \label{berry:u}
\ket{u_-} = \binom{\sin\frac{\theta}{2}e^{-i\phi}}
{-\cos\frac{\theta}{2}} \;, \qquad
\ket{u_+} = \binom{\cos\frac{\theta}{2}e^{-i\phi}}
{\sin\frac{\theta}{2}} \;.
\end{equation}
We are, of course, free to add an arbitrary phase to these wave
functions.  Let us consider the lower energy level.  The Berry
connection is given by
\begin{subequations}
\begin{align}
\cal A_\theta &= \bracket{u|i\partial_\theta u} = 0 \;,  \\
\cal A_\phi &= \bracket{u|i\partial_\phi u} = \sin^2\frac{\theta}{2} \;,
\end{align}
\end{subequations}
and the Berry curvature is
\begin{equation} \label{berry:curv}
\Omega_{\theta\phi} = \partial_\theta \cal A_\phi
- \partial_\phi \cal A_\theta = \frac{1}{2}\sin\theta \;.
\end{equation}
However, the phase of $\ket{u_-}$ is not defined at the south pole
($\theta=\pi$).  We can choose another gauge by multiplying
$\ket{u_-}$ by $e^{i\phi}$ so that the wave function is smooth and
single valued everywhere except at the north pole.  Under this gauge
we find $\cal A_\theta = 0$ and $\cal A_\phi =
-\cos^2\frac{\theta}{2}$, and the same expression for the Berry
curvature as in Eq.~\eqref{berry:curv}.  This is not surprising because
the Berry curvature is a gauge-independent quantity and the Berry
connection is not.~\footnote{
Another way to calculate the Berry curvature is to use
Eq.~\eqref{berry:sum} directly.}

If $\bm h(\bm R)$ depends on a set of parameters $\bm R$, then
\begin{equation} \label{berry:R1R2}
\Omega_{R_1R_2} = \frac{1}{2}\dpar{(\phi, \cos\theta)}{(R_1, R_2)} \;.
\end{equation}

Several important properties of the Berry curvature can be revealed by
considering the specific case of $\bm h = (x, y, z)$.  Using
Eq.~\eqref{berry:R1R2}, we find the Berry curvature in its vector form
\begin{equation}  \label{berry:curv1}
\bm\Omega = \frac{1}{2} \frac{\bm h}{h^3} \;.
\end{equation}
One recognizes that Eq.~\eqref{berry:curv1} is the field generated by a
monopole at the origin $\bm h = 0$~\cite{dirac1931,sakurai-QM,wu1975},
where the two energy levels become degenerate.  Therefore the
degeneracy points act as sources and drains of the Berry curvature
flux.  Integrate the Berry curvature over a sphere containing the
monopole, which is the Berry phase on the sphere; we find
\begin{equation}
\frac{1}{2\pi} \int_{S^2} d\theta d\phi\, \Omega_{\theta\phi} = 1 \;.
\end{equation}
In general, the Berry curvature integrated over a \emph{closed}
manifold is quantized in the units of $2\pi$ and equals to the net
number of monopoles inside.  This number is called the Chern number
and is responsible for a number of quantization effects discussed below.


\subsection{\label{sec:zak}Berry phase in Bloch bands}

In the above we have introduced the basic concepts of the Berry phase
for a generic system described by a parameter-dependent Hamiltonian.
We now consider its realization in crystalline solids.  As we shall
see, the band structure of crystals provides a natural platform to
investigate the occurrence of the Berry phase effect.

Within the independent electron approximation, the band structure of a
crystal is determined by the following Hamiltonian for a single
electron:
\begin{equation} \label{zak:ham1}
H = \frac{\hat{p}^2}{2m} + V(\bm r) \;,
\end{equation}
where $V(\bm r + \bm a) = V(\bm r)$ is the periodic potential with
$\bm a$ being the Bravais lattice vector.  According to Bloch's
theorem, the eigenstates of a periodic Hamiltonian satisfy the
following boundary condition
\begin{equation} \label{zak:bound1}
\psi_{n\bm q}(\bm r + \bm a) = e^{i\bm q\cdot \bm a}
\psi_{n\bm q}(\bm r) \;,
\end{equation}
where $n$ is the band index and $\hbar\bm q$ is the crystal momentum,
which resides in the Brillouin zone.  Thus the system is described by
a $\bm q$-independent Hamiltonian with a $\bm
q$-dependent boundary condition, Eq.~\eqref{zak:bound1}.  To comply
with the general formalism of the Berry phase, we make the following
unitary transformation to obtain a $\bm q$-dependent Hamiltonian:
\begin{equation} \label{zak:ham}
H(\bm q) = e^{-i\bm q\cdot r} H e^{i\bm q\cdot \bm r}
= \frac{(\hat{\bm p} + \hbar\bm q)^2}{2m} + V(\bm r) \;.
\end{equation}
The transformed eigenstate $u_{n\bm q}(\bm r) = e^{-i\bm q\cdot \bm
  r}\psi_{n\bm q}(\bm r)$ is just the cell-periodic part of the Bloch
function.  It satisfies the strict periodic boundary condition
\begin{equation} \label{zak:bound}
u_{n\bm q}(\bm r +\bm a) = u_{n\bm q}(\bm r) \;.
\end{equation}
This boundary condition ensures that all the eigenstates live in the
same Hilbert space.  We can thus identify the Brillouin zone as the
parameter space of the transformed Hamiltonian $H(\bm q)$, and
$\ket{u_n(\bm q)}$ as the basis function.

Since the $\bm q$-dependence of the basis function is inherent to the
Bloch problem, various Berry phase effects are expected in crystals.
For example, if $\bm q$ is forced to vary in the momentum space, then
the Bloch state will pick up a Berry phase:
\begin{equation}
\gamma_n = \oint_{\cal C} d\bm q \cdot
\bracket{u_n(\bm q)|i\bm\nabla_{\bm q}|u_n(\bm q)} \;.
\end{equation}
We emphasize that the path $\cal C$ must be closed to make $\gamma_n$
a gauge-invariant quantity with physical significance.

Generally speaking, there are two ways to generate a closed path in
the momentum space.  One can apply a magnetic field, which induces a
cyclotron motion along a closed orbit in the $\bm q$-space.  This way
the Berry phase can manifest in various magneto-oscillatory
effects~\cite{mikitik1999,mikitik2004,mikitik2007}, which have been
observed in metallic compound LaRhIn$_5$~\cite{goodrich2002}, and most
recently, graphene
systems~\cite{zhang2005,novoselov2005,novoselov2006}.  Such a closed
orbit is possible only in two or three-dimensional systems (see Sec.~\ref{sec:qza}).  Following
our discussion in Sec.~\ref{sec:berry}, we can define the Berry
curvature of the energy bands, given by
\begin{equation}
\bm\Omega_n(\bm q) = \bm\nabla_{\bm q} \times \bracket{u_n(\bm q)
|i\bm\nabla_{\bm q}|u_n(\bm q)} \;.
\end{equation}
The Berry curvature $\bm\Omega_n(\bm q)$ is an intrinsic property of
the band structure because it only depends on the wave function.  It
is nonzero in a wide range of materials, in particular, crystals with
broken time-reversal or inversion symmetry.  In fact, once we have
introduced the concept of the Berry curvature, a closed loop is not
necessary because the Berry curvature itself is a local
gauge-invariant quantity.  It is now well recognized that information
of the Berry curvature is essential in a proper description of the
dynamics of Bloch electrons, which has various effects on transport
and thermodynamic properties of crystals.

One can also apply an electric field to cause a linear variation of
$\bm q$.  In this case, a closed path is realized when $\bm q$ sweeps
the entire Brillouin zone.  To see this, we note that the Brillouin
zone has the topology of a torus: the two points $\bm q$ and $\bm q +
\bm G$ can be identified as the same point, where $\bm G$ is the
reciprocal lattice vector.  This can be seen by noting that
$\ket{\psi_n(\bm q)}$ and $\ket{\psi_n(\bm q+ \bm G)}$ satisfy the
same boundary condition in Eq.~\eqref{zak:bound1}, therefore they can
at most differ by a phase factor.  The torus topology is realized by
making the phase choice $\ket{\psi_n(\bm q)} = \ket{\psi_n(\bm q+\bm
  G)}$.  Consequently, $\ket{u_n(\bm q)}$ and $\ket{u_n(\bm q + \bm
  G)}$ satisfy the following phase relation
\begin{equation} \label{zak:bloch}
\ket{u_n(\bm q)} = e^{i\bm G\cdot\bm r}\ket{u_n(\bm q+\bm G)} \;.
\end{equation}
This gauge choice is called the periodic gauge~\cite{resta2000}.

In this case, the Berry phase across the Brillouin zone is called
Zak's phase~\cite{zak1989}
\begin{equation}  \label{zak:phase}
\gamma_n = \int_\text{BZ} d\bm q\cdot
\bracket{u_n(\bm q)|i\bm\nabla_{\bm q}|u_n(\bm q)} \;.
\end{equation}
It plays an important role in the formation of Wannier-Stark
ladders~\cite{wannier1962} (see Sec.~\ref{sec:qzb}).  We note that this phase is entirely due
to the torus topology of the Brillouin zone, and it is the only way to
realize a closed path in a one-dimensional parameter space.  By
analyzing the symmetry properties of Wannier functions~\cite{kohn1959}
of a one-dimensional crystal, \citet{zak1989} showed that $\gamma_n$
is either 0 or $\pi$ in the presence of inversion symmetry; a simple
argument is given in Sec.~\ref{sec:polar}.  If the crystal lacks
inversion symmetry, $\gamma_n$ can assume any value.  Zak's phase is
also related to macroscopic polarization of an
insulator~\cite{resta1994,king-smith1993,sipe1999} (see
Sec.~\ref{sec:polar}).


\section{\label{sec:pump}Adiabatic transport and electric polarization}


One of the earlier examples of the Berry phase effect in crystals is
the adiabatic transport in a one-dimensional band insulator, first
considered by \citet{thouless1983}.  He found that if the potential
varies slowly in time and returns to itself after some time, the
particle transport during the time cycle can be expressed as a Berry
phase and it is an integer.  This idea was later generalized to
many-body systems with interactions and disorder, provided that the
Fermi energy always lies in a bulk energy gap during the
cycle~\cite{niu1984}.  \citet{avron1988} studied the adiabatic
transport in multiply connected systems.  The scheme of adiabatic
transport under one or several controlling parameters has proven very
powerful.  It opened the door to the field of parametric charge
pumping~\cite{niu1990,zhou1999,talyanskii1997,switkes1999,brouwer1998}.
It also provides a firm foundation to the modern theory of
polarization developed in the early
90's~\cite{king-smith1993,resta1994,ortiz1994}.

\subsection{\label{sec:current}Adiabatic current}

Let us consider a one-dimensional band insulator under a slowly
varying time-dependent perturbation.  We assume the perturbation is
periodic in time, i.e., the Hamiltonian satisfies
\begin{equation}
H(t+T) = H(t) \;.
\end{equation}
Since the time-dependent Hamiltonian still has the translational
symmetry of the crystal, its instantaneous eigenstates has the Bloch
form $e^{iqx}\ket{u_n(q, t)}$.  It is convenient to work with the
$q$-space representation of the Hamiltonian $H(q, t)$ [see
  Eq.~\eqref{zak:ham}] with eigenstates $\ket{u_n(q, t)}$.  We note
that under this parametrization, the wave vector $q$ and time $t$ are
put on an equal footing as both are independent cooridinates of the
parameter space.

We are interested in the adiabatic current induced by the variation of
external potentials.  Apart from an unimportant overall phase
factor and up to first order in the rate of the change of the
Hamiltonian, the wave function is given by
\begin{equation} \label{pump:un}
\ket{u_n} - i\hbar\sum_{n'\neq n}
\frac{\ket{u_{n'}}\bracket{u_{n'}|\dpar{u_n}{t}}}{\eps_n - \eps_{n'}} \;.
\end{equation}
The velocity operator in the $q$-representaion has the form $v(q, t) =
\partial H(q, t)/\partial (\hbar q)$.  Hence, the average velocity in
a state of given $q$ is found to first order as
\begin{equation}
v_n(q) = \dparh{\eps_n(q)}{q} - i \sum_{n'\neq n} \Bigl\{
\frac{\bracket{u_n|\dpar{H}{q}|u_{n'}}\bracket{u_{n'}|\dpar{u_n}{t}}}
{\eps_n - \eps_{n'}} - \text{c.c.} \Bigr\}\;,
\end{equation}
where c.c.\ denotes the complex conjugate.  Using the fact that
$\bracket{u_n|\partial H/\partial q|u_{n'}} = (\eps_n - \eps_{n'})
\bracket{\partial u_n/\partial q|u_{n'}}$ and the identity $\sum_{n'}
\ket{u_{n'}} \bra{u_{n'}} = 1$, we find
\begin{equation}
v_n(q) = \dparh{\eps_n(q)}{q} 
- i \Bigl[\bracket{\dpar{u_n}{q}|\dpar{u_n}{t}} 
- \bracket{\dpar{u_n}{t}|\dpar{u_n}{q}}\Bigr] \;.
\end{equation}
The second term is exactly the Berry curvature $\Omega^n_{qt}$ defined
in the parameter space $(q, t)$ [see Eq.~\eqref{berry:def}].
Therefore the above equation can be recasted into a compact form
\begin{equation} \label{pump:v}
v_n(q) = \dparh{\eps_n(q)}{q} - \Omega^n_{qt} \;.
\end{equation}

Upon integration over the Brillouin zone, the zeroth order term
given by the derivative of the band energy vanishes, and only the
first order term survives.  The induced adiabatic current is given by
\begin{equation} \label{pump:j}
j = -\sum_n \intqq \Omega^n_{qt} \;,
\end{equation}
where the sum is over filled bands.  We have thus derived the
remarkable result that the adiabatic current induced by a
time-dependent perturbation in a band is equal to the $q$-integral of
the Berry curvature $\Omega^n_{qt}$~\cite{thouless1983}.

\subsection{\label{sec:quant}Quantized adiabatic particle transport}

\begin{figure}
\includegraphics[width=8cm]{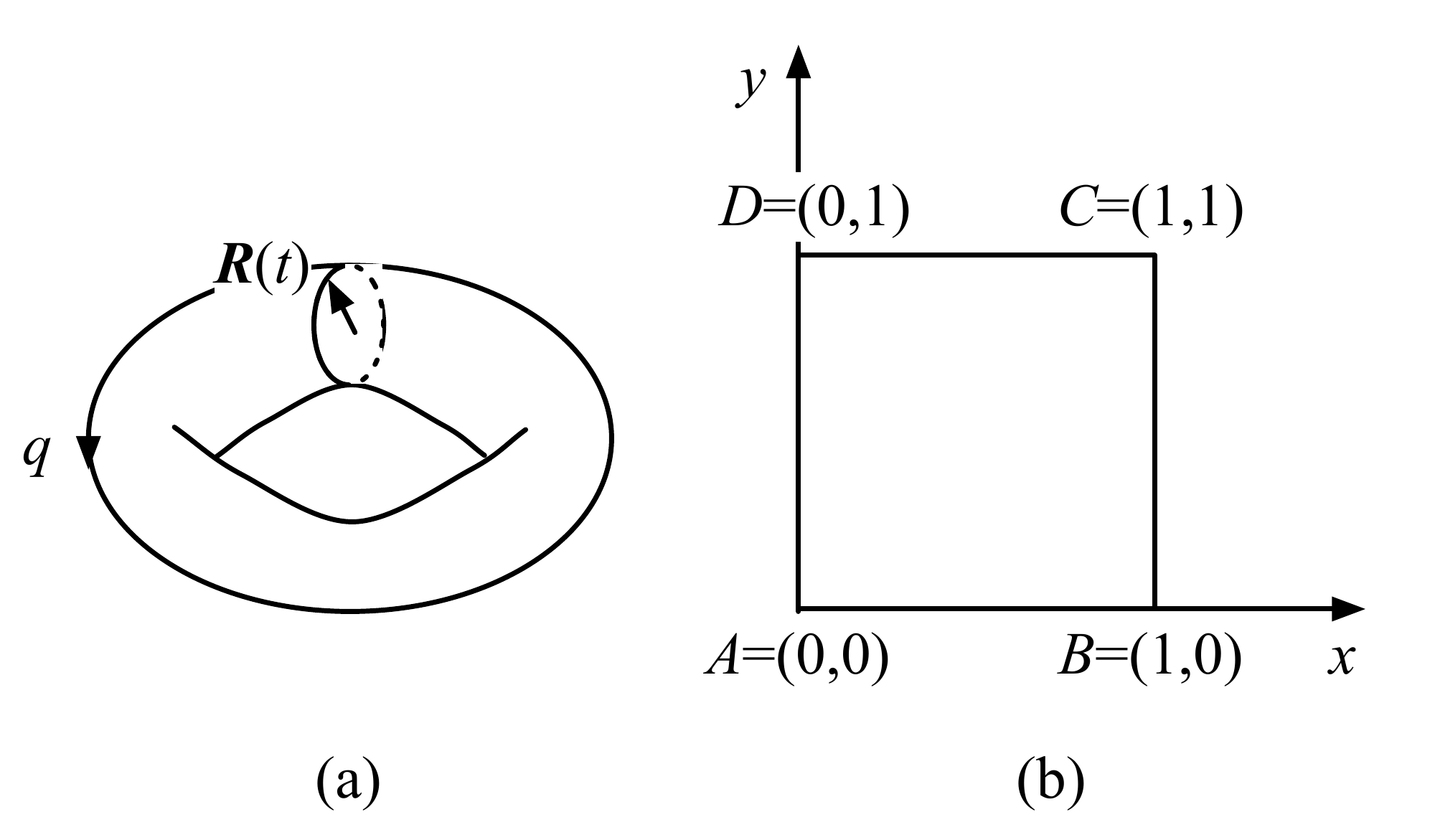}  
\caption{\label{fig:torus} (a) A torus with its surface parameterized
  by $(q, t)$.  The control parameter $\bm R(t)$ runs in circle along
  the $t$ direction.  (b) The equivalence of a torus: a rectangle with
  periodic boundary conditions: $\overline{AB} = \overline{DC}$ and
  $\overline{AD} = \overline{BC}$.  To use Stokes's theorem, we relax
  the boundary condition and allow the wave functions on parallel sides
  have different phases.}
\end{figure}

Next we consider the particle transport for the $n$th band over a time
cycle, given by
\begin{equation} \label{pump:chern}
c_n = -\int_0^T dt \intqq \Omega^n_{qt} \;.
\end{equation}
Since the Hamiltonian $H(q, t)$ is periodic in both $t$ and $q$, the
parameter space of $H(q, t)$ is a torus, schematically shown in
Fig.~\ref{fig:torus}(a).  By definition~\eqref{berry:phase1}, $2\pi
c_n$ is nothing but the Berry phase over the torus.

In Sec.~\ref{sec:spin}, we showed that the Berry phase over a closed
manifold, the surface of a sphere $S^2$ in that case, is quantized in
the unit of $2\pi$.  Here we prove that it is also true in the case of
a torus.  Our strategy is to evaluate the surface
integral~\eqref{pump:chern} using Stokes's theorem, which requires the
surface to be simply connected.  To do that, we cut the torus open and
transform it into a rectangle, as shown in Fig.~\ref{fig:torus}(b).
The basis function along the contour of the rectangle is assumed to be
single-valued.  Introduce $x = t/T$ and $y = q/2\pi$.  According to
Eq.~\eqref{berry:phase}, the Berry phase in Eq.~\eqref{pump:chern} can
be written into a contour integral of the Berry vector potential, i.e.,
\begin{equation} \label{pump:contour} \begin{split}
c = \frac{1}{2\pi}\Big\{ &\int_A^B dx\, \cal A_x(x, 0) 
+ \int_B^C dy\, \cal A_y(1, y) \\
&+ \int_C^D dx\,\cal A_x(x, 1) + \int_D^A dy\, \cal A_y(0, y) \Big\}\\
= \frac{1}{2\pi} \Bigl\{ &\int_0^1 dx\,
[\cal A_x(x, 0) - \cal A_x(x, 1)] \\
&- \int_0^1 dy\, [\cal A_y(0, y) - \cal A_y(1, y)] \Bigr\}\;,
\end{split} \end{equation}
where the band index $n$ is dropped for simplicity.  Let us consider
the integration over $x$.  By definition, $\cal A_x(x, y) =
\bracket{u(x, y)|i\nabla_x|u(x, y)}$.  Recall that $\ket{u(x, 0)}$ and
$\ket{u(x, 1)}$ describe physically equivalent states, therefore they
can only differ by a phase factor, i.e., $e^{i\theta_x(x)} \ket{u(x,
  1)} = \ket{u(x, 0)}$.  We thus have
\begin{equation}
\int_0^1 dx\, [\cal A_x(x, 0) - \cal A_x(x, 1)] 
= \theta_x(1)-\theta_x(0)\;.
\end{equation}
Similarly,
\begin{equation}
\int_0^1 dy\, [\cal A_y(0, y) - \cal A_y(1, y)]
= \theta_y(1)-\theta_y(0)\;,
\end{equation}
where $e^{i\theta_y(y)} \ket{u(y, 1)} = \ket{u(y, 0)}$.  The total
integral is
\begin{equation} \label{pump:chern2}
c = \frac{1}{2\pi} [ \theta_x(1) - \theta_x(0) 
+ \theta_y(0) - \theta_y(1) ] \;.
\end{equation}
On the other hand, using the phase matching relations at the four
corners $A$, $B$, $C$, and $D$,
\begin{align*}
e^{i\theta_x(0)} \ket{u(0, 1)} = \ket{u(0,0)} \;, \\
e^{i\theta_x(1)} \ket{u(1, 1)} = \ket{u(1,0)} \;, \\
e^{i\theta_y(0)} \ket{u(1, 0)} = \ket{u(0,0)} \;, \\
e^{i\theta_y(1)} \ket{u(1, 1)} = \ket{u(0,1)} \;,
\end{align*}
we obtain
\begin{equation}
\ket{u(0, 0)} = e^{i[\theta_x(1) - \theta_x(0) + \theta_y(0)
- \theta_y(1)]} \ket{u(0, 0)} \;.
\end{equation}
The single-valuedness of $\ket{u}$ requires that the exponent must be
an integer multiple of $2\pi$.  Therefore the transported particle
number $c$, given in Eq.~\eqref{pump:chern2}, must be quantized.  This
integer is called the first Chern number, which characterizes the
topological structure of the mapping from the parameter space $(q, t)$
to the Bloch states $\ket{u(q, t)}$.  Note that in our proof, we made
no reference to the original physical system; the quantization of the
Chern number is always true as long as the Hamiltonian is periodic in
both parameters.

An intuitive picture of the quantized particle transport is the
following.  If the periodic potential slides its position without
changing its shape, we expect that the electrons simply follow the
potential.  If the potential shifts one spatial period in the time
cycle, the particle transport should be equal to the number of filled
Bloch bands (double if the spin degeneracy is counted).  This follows
from the fact that there is on average one state per unit cell in each
filled band.

\subsubsection{Conditions for nonzero particle transport for cyclic motion}

We have shown that the adiabatic particle transport over a time period
takes the form of the Chern number and it is quantized.  However, the
exact quantization does not gurantee that the electrons will be
transported at the end of the cycle because zero is also an integer.
According to the discussion in Sec.~\ref{sec:spin}, the Chern number
counts the net number of monopoles enclosed by the surface.  Therefore
the number of transported electrons can be related to the number of
monopoles, which are degeneracy points in the parameter space.

To formulate this problem, we let the Hamiltonian depend on time
through a set of control parameters $\bm R(t)$, i.e.,
\begin{equation}
H(q, t) = H(q, \bm R(t)) \;, \qquad \bm R(t+T) = \bm R(t) \;.
\end{equation}
The particle transport is now given by, in terms of $\bm R$,
\begin{equation} \label{pump:chern3}
c_n = \frac{1}{2\pi} \oint dR_\alpha \int_\text{BZ} dq\,
\Omega^n_{qR_\alpha} \;.
\end{equation}
If $\bm R(t)$ is a smooth function of $t$, as it is usually the case
for physical quantities, then $\bm R$ must have at least two
components, say $R_1$ and $R_2$.  Otherwise the trajectory of $\bm
R(t)$ cannot trace out a circle on the torus [see
  Fig.~\ref{fig:torus}(a)].  To find the monopoles inside the torus,
we now relax the constraint that $R_1$ and $R_2$ can only move on the
surface and extend their domains inside the torus such that the
parameter space of $(q, R_1, R_2)$ becomes a toroid.  Thus, the
criterion for $c_n$ to be nonzero is that a degeneracy point must
occur somewhere inside the torus as one varies $q$, $R_1$ and $R_2$.
In the context of quasi one-dimensional ferroelectrics,
\citet{onoda2004} have discussed the situation where $\bm R$ has three
components, and showed how the topological structure in the $\bm R$
space affects the particle transport.

\subsubsection{\label{sec:twist}Many-body interactions and disorder}

In the above we have only considered band insulators of
non-interacting electrons.  However, in real materials both many-body
interactions and disorder are ubiquitous.  \citet{niu1984} studied
this problem and showed that in the general case the quantization of
particle transport is still valid as long as the system remains an
insulator during the whole process.

Let us consider a time-dependent Hamiltonian of an $N$-particle system
\begin{equation}
H(t) = \sum_i^N \Bigl[\frac{\hat{p}_i^2}{2m} + U(x_i, t)\Bigr]
+ \sum_{i > j}^N V(x_i - x_j) \;,
\end{equation}
where the one-particle potential $U(x_i, t)$ varies slowly in time and
repeats itself in period $T$.  Note that we have not assumed any
specific periodicity of the potential $U(x_i, t)$.  The trick is to
use the so-called twisted boundary condition by
requiring that the many-body wave function satisfies
\begin{equation} \label{pump:twist}
\Phi(x_1, \dots, x_i+L, \dots, x_N) = e^{i\kappa L}
\Phi(x_1, \dots, x_i, \dots, x_N) \;,
\end{equation}
where $L$ is the size of the system.  This is equivalent to
solving a $\kappa$-dependent Hamiltonian
\begin{equation} \label{pump:hh}
H(\kappa, t) = \exp(i\kappa\sum x_i) H(t) \exp(-i\kappa\sum x_i)
\end{equation}
with the strict periodic boundary condition
\begin{equation} \label{pump:bound}
\tilde{\Phi}(\kappa; x_1, \dots, x_i+L, \dots, x_N) = 
\tilde{\Phi}_\kappa(\kappa; x_1, \dots, x_i, \dots, x_N) \;.
\end{equation}
The Hamiltonian $H(\kappa, t)$ together with the boundary
condition~\eqref{pump:bound} describes a one-dimensional system placed
on a ring of length $L$ and threaded by a magnetic flux of
$(\hbar/e)\kappa L$~\cite{kohn1964}.  We note that the above
transformation~\eqref{pump:hh} with the boundary
condition~\eqref{pump:bound} is very similar to that of the
one-particle case, given by Eqs.~\eqref{zak:ham} and
\eqref{zak:bound}.

One can verify that the current operator is given by $\partial
H(\kappa, t)/\partial(\hbar\kappa)$.  For each $\kappa$, we can repeat
the same steps in Sec.~\ref{sec:current} by identifying $\ket{u_n}$ in
Eq.~\eqref{pump:un} as the many-body ground-state
$\ket{\tilde{\Phi}_0}$ and $\ket{u_{n'}}$ as the excited state.  We
have
\begin{equation} \label{pump:current} \begin{split}
j(\kappa) &= \dparh{\eps(\kappa)}{\kappa} 
- i \Bigl[\bracket{\dpar{\tilde{\Phi}_0}{\kappa}|
\dpar{\tilde{\Phi}_0}{t}} - \bracket{\dpar{\tilde{\Phi}_0}{t}|
\dpar{\tilde{\Phi}_0}{\kappa}} \Bigr] \\ 
&= \dparh{\eps(\kappa)}{\kappa} - \tilde{\Omega}_{\kappa t} \;.
\end{split} \end{equation}

So far the derivation is formal and we still cannot see why the
particle transport should be quantized.  The key step is achieved by
realizing that if the Fermi energy lies in a gap, then the current
$j(\kappa)$ should be insensitive to the boundary condition specified
by $\kappa$~\cite{niu1984,thouless1981}.  Consequently we can take the
thermodynamic limit and average $j(\kappa)$ over different boundary
conditions.  Note that $\kappa$ and $\kappa + 2\pi/L$ describe the
same boundary condition in Eq.~\eqref{pump:twist}.  Therefore the
parameter space for $\kappa$ and $t$ is a torus $T^2: \{0 < \kappa <
2\pi/L, 0 < t < T\}$.  The particle transport is given by
\begin{equation}
c = -\frac{1}{2\pi} \int_0^T dt \int_0^{2\pi/L} d\kappa\,
 \tilde{\Omega}_{\kappa t} \;,
\end{equation}
which, according to the previous discussion, is quantized.  

We emphasize that the quantization of the particle transport only
depends on two conditions:
\begin{enumerate}
\item The ground state is separated from the excited states in the
  bulk by a finite energy gap;
\item The ground state is non-degenerate.
\end{enumerate}
The exact quantization of the Chern number in the presence of
many-body interactions and disorder is very remarkable.  Usually,
small perturbations to the Hamiltonian results in small changes of
physical quantities.  However, the fact that the Chern number must be
an integer means that it can only be changed in a discontinous way and
does not change at all if the perturbation is small.  This is a
general outcome of the topological invariance.

Later we show that the same quantity also appears in the quantum Hall
effect.  The expression~\eqref{pump:current} of the induced current
also provides a many-body formulation for adiabatic transport.

\subsubsection{Adiabatic Pumping}

The phenomenon of adiabatic transport is sometimes called adiabatic
pumping because it can generate a dc current $I$ via periodic
variations of some parameters of the system, i.e.,
\begin{equation}
I = ec\nu \;,
\end{equation}
where $c$ is the Chern number and $\nu$ is the frequency of the
variation.  \citet{niu1990} suggested that the exact quantization of
the adiabatic transport can be used as a standard for charge current
and proposed an experimental realization in nanodevices, which could
serve as a charge pump.  It was later realized in the experimental
study of acoustoelectric current induced by a surface acoustic wave in
a one-dimensional channel in a GaAs-Al$_x$Ga$_{1-x}$
heterostructure~\cite{talyanskii1997}.  The same idea has led to the
proposal of a quantum spin pump in an antiferromagentic
chain~\cite{shindou2005a}.

Recently, much efforts have focused on adiabatic pumping in mesoscopic
systems~\cite{zheng2003,zhou1999,brouwer1998,sharma2001,mucciolo2002,avron2001}.
Experimentally, both charge and spin pumping have been observed in a
quantum dot system~\cite{switkes1999,watson2003}.  Instead of the wave
function, the central quantity in a mesoscopic system is the
scattering matrix.  \citet{brouwer1998} showed that the pumped charge
over a time period is given by
\begin{equation} \label{pump:Q}
Q(m) = \frac{e}{\pi} \int_A dX_1dX_2\, \sum_{\beta}\sum_{\alpha\in m}
\Im \dpar{S^*_{\alpha\beta}}{X_1} \dpar{S_{\alpha\beta}}{X_2} \;,
\end{equation}
where $m$ labels the contact, $X_1$ and $X_2$ are two external
parameters whose trace encloses the area $A$ in the parameter space,
$\alpha$ and $\beta$ labels the conducting channels, and
$S_{\alpha\beta}$ is the scattering matrix.  Although the physical
description of these open systems are dramatically different from the
closed ones, the concepts of gauge field and geometric phase can still
be applied.  The integrand in Eq.~\eqref{pump:Q} can be thought as the
Berry curvature $\Omega_{X_1X_2} = -2\Im\bracket{\partial_{X_1}
  u|\partial_{X_2} u}$ if we identify the inner product of the state
vector with the matrix product.  \citet{zhou2003} showed the pumped
charge (spin) and is essentially the Abelian (non-Abelian) geometric
phase associated with scattering matrix $S_{\alpha\beta}$.

\subsection{\label{sec:polar}Electric Polarization of Crystalline Solids}

Electric polarization is one of the fundamental quantities in
condensed matter physics, essential to any proper description of
dielectric phenomena of matter.  Despite its great importance, the
theory of polarization in crystals had been plagued by the lack of a
proper microscopic understanding.  The main difficulty lies in the
fact that in crystals the charge distribution is periodic in space,
for which the electric dipole operator is not well defined.  This
difficulty is most exemplified in covalent solids, where the electron
charges are continuously distributed between atoms.  In this case,
simple integration over charge density would give arbitrary values
depending on the choice of the unit cell~\cite{martin1972,martin1974}.
It has prompted the question whether the electric polarization can be
defined as a bulk property.  These problems are eventually solved by
the modern theory of polarization~\cite{resta1994,king-smith1993},
where it is shown that only the change in polarization has physical
meaning and it can be quantified by using the Berry phase of the
electronic wave function.  The resulting Berry-phase formula has been
very successful in first-principles studies of dielectric and
ferroelectric materials.  \citet{resta2007} reviewed recent progress
in this field.

Here we discuss the theory of polarization based on the concept of
adiabatic transport.  Their relation is revealed by elementary
arguments from macroscopic electrostatics~\cite{ortiz1994}.  We begin
with the relation
\begin{equation} \label{pump:polar}
\bm\nabla \cdot \bm P(\bm r) = -\rho(\bm r) \;,
\end{equation}
where $\bm P(\bm r)$ is the polarization density and $\rho(\bm r)$ is
the charge density.  Coupled with the continuity equation
\begin{equation}
\dpar{\rho(\bm r)}{t} + \bm\nabla \cdot \bm j = 0 \;,
\end{equation}
Eq.~\eqref{pump:polar} leads to
\begin{equation}
\bm \nabla \cdot (\dpar{\bm P}{t} - \bm j) = 0 \;.
\end{equation}
Therefore up to a divergence-free part,~\footnote{The divergence-free
  part of the current is usually given by the magnetization current.
  In a uniform system, such current vanishes identically in the bulk.
  \citet{hirst1997} gave an in-depth discussion on the separation
  between polarization and magnetization current.} the change in the
polarization density is given by
\begin{equation} \label{pump:pp}
\Delta P_\alpha = \int_0^T dt\, j_\alpha \;.
\end{equation}
The above equation can be interpreted in the following way: The
polarization difference between two states is given by the integrated
bulk current as the system adiabatically evolves from the initial
state at $t = 0$ to the final state at $t = T$.  This description
implies a time-dependent Hamiltonian $H(t)$, and the electric
polarization can be regarded as ``unquantized'' adiabatic particle
transport.  The above interpretation is also consistent with
experiments, as it is always the change of the polarization that has
been measured~\cite{resta2007}.

Obviously, the time $t$ in the Hamiltonian can be replaced by any
scalar that describes the adiabatic process.  For example, if the
process corresponds to a deformation of the crystal, then it makes
sense to use the parameter that characterizes the atomic displacement
within a unit cell.  For general purpose, we shall assume the
adiabatic transformation is parameterized by a scalar $\lambda(t)$
with $\lambda(0) = 0$ and $\lambda(T) = 1$.  It follows from
Eqs.~\eqref{pump:j} and \eqref{pump:pp} that
\begin{equation} \label{pump:p}
\Delta P_\alpha = e\sum_n\int_0^1 d\lambda \int_\text{BZ} 
\frac{d\bm q}{(2\pi)^d} \Omega^n_{q_\alpha \lambda} \;,
\end{equation}
where $d$ is the dimensionality of the system.  This is the
Berry-phase formula obtained by \citet{king-smith1993}.

In numerical calculations, a two-point version of Eq.~\eqref{pump:p}
that only involves the integration over $\bm q$ is commonly used to
reduce the computational load.  It is obtained under the periodic
gauge [see Eq.~\eqref{zak:bloch}]~\footnote{A more general phase
  choice is given by the path-independent gauge $\ket{u_n(\bm q,
    \lambda)} = e^{i[\theta(\bm q) + \bm G\cdot\bm r]} \ket{u_n(\bm q
    + \bm G, \lambda)}$, where $\theta(\bm q)$ is an arbitrary
  phase~\cite{ortiz1994}}.  The Berry curvature
$\Omega_{q_\alpha\lambda}$ is wirtten as $\partial_{q_\alpha}\cal
A_\lambda - \partial_\lambda \cal A_{q_\alpha}$.  Under the periodic
gauge, $A_\lambda$ is periodic in $q_\alpha$, and integration of
$\partial_{q_\alpha} \cal A_\lambda$ over $q_\alpha$ vanishes.  Hence
\begin{equation} \label{pump:p2}
\Delta P_\alpha = e\sum\int_\text{BZ} \frac{d\bm q}{(2\pi)^d}
\cal A^n_{q_\alpha}\Big|_{\lambda = 0}^1 \;.
\end{equation}
In view of Eq.~\eqref{pump:p2}, both the adiabatic transport and the
electric polarization can be regarded as the manifestation of Zak's
phase, given in Eq.~\eqref{zak:phase}.

However, a price must be paid to use the two-point formula, namely,
the polarization in Eq.~\eqref{pump:p2} is determined up to an
uncertainty quantum.  Since the integral~\eqref{pump:p2} does not
track the history of $\lambda$, there is no information on how many
cycles $\lambda$ has gone through.  According to our discussion on
quantized particle transport in Sec.~\ref{sec:quant}, for each cycle
an integer number of electrons are transported across the sample,
hence the polarization is changed by multiple of the quantum
\begin{equation} \label{pump:uncertain}
\frac{e \bm a}{\cal V_0} \;,
\end{equation}
where $\bm a$ is the Bravais lattice vector and $\cal
V_0$ is the volume of the unit cell.

Because of this uncertainty quantum, the polarization may be regarded
as a multi-valued quantity with each value differed by the quantum.
With this in mind, let us consider Zak's phase in a one-dimensional
system with inversion symmetry.  Now we know that Zak's phase is just
$2\pi/e$ times the polarization density $P$.  Under spatial inversion,
$P$ transforms to $-P$.  On the other hand, inversion symmetry
requires that $P$ and $-P$ describes the same state, which is only
possible if $P$ and $-P$ differ by multiple of the polarization
quantum $ea$.  Therefore $P$ is either 0 or $ea/2$ (modulo $ea$).  Any
other value of $P$ will break the inversion symmetry.  Consequently,
Zak's phase can only take the value $0$ or $\pi$ (modulo $2\pi$).

\citet{king-smith1993} further showed that, based on
Eq.~\eqref{pump:p2}, the polarization per unit cell can be defined as
the dipole moment of the Wannier charge density,
\begin{equation}
\bm P = -e\sum_n \int d\bm r\, \bm r|W_n(\bm r)|^2 \;,
\end{equation}
where $W_n(\bm r)$ is the Wannier function of the $n$th band,
\begin{equation}
W_n(\bm r - R) = \sqrt{N}\cal V_0 \int_\text{BZ}
\frac{d\bm q}{(2\pi)^3} e^{i\bm q\cdot (\bm r - \bm R)} 
u_{n\bm k}(\bm r) \;.
\end{equation}
In this definition, one effectively maps a band insulator into a
periodic array of localized distributions with truly quantized
charges.  This resembles an ideal ionic crystal where the polarization
can be understood in the classical picture of localized charges.  The
quantum uncertainty found in Eq.~\eqref{pump:uncertain} is reflected
by the fact that the Wannier center position is defined only up to a
lattice vector.

Before concluding, we point out that the polarization defined above is
clearly a bulk quantity as it is given by the Berry phase of the
ground state wave function.  A many-body formulation was developed by
\citet{ortiz1994} based on the work of \citet{niu1984}.

Recent development in this field falls into two categories.  On the
computational side, calculating polarization in finite electric fields
has been addressed, which has a deep influence on density functional
theory in extended systems~\cite{nunes1994,nunes2001,souza2002}.  On
the theory side, \citet{resta1998} proposed a quantum-mechanical
position operator for extended systems.  It was shown that the
expectation value of such an operator can be used to characterize the
phase transition between the metallic and insulating
state~\cite{resta1999,souza2000}, and is closely related to the
phenomenon of electron localization~\cite{kohn1964}.

\subsubsection{\label{sec:rice}The Rice-Mele model}

So far our discussion of the adiabatic transport and electric
polarization has been rather abstract.  We now consider a concrete
example: a one-dimensional dimerized lattice model described by the
following Hamiltonian
\begin{equation}
H = \sum_j (\frac{t}{2}+(-1)^j \frac{\delta}{2})
(c^\dag_j c_{j+1} + \text{h.c.}) + \Delta (-1)^j c^\dag_j c_{j+1} \;,
\end{equation}
where $t$ is the uniform hopping amplitude, $\delta$ is the
dimerization order, and $\Delta$ is a staggered sublattice potential.
It is the prototype Hamiltonian for a class of one-dimensional
ferroelectrics.  At half-filling, the system is a metal for $\Delta =
\delta = 0$, and an insulator otherwise.  \citet{rice1982} considered
this model in the study of solitons in polyenes.  It was later used to
study ferroelectricity~\cite{vanderbilt1993,onoda2004}.  If $\Delta =
0$ it reduces to the celebrated Su-Shrieffer-Heeger
model~\cite{su1979}.

Assuming periodic boundary condition, the Bloch representation of the
above Hamiltonian is given by $H(q) = \bm h(q) \cdot \bm\sigma$, where
\begin{equation}
\bm h = (t\cos\frac{qa}{2}, -\delta\sin\frac{qa}{2}, \Delta) \;.
\end{equation}
This is the two-level model we discussed in Sec.~\ref{sec:spin}.  Its
energy spectrum consists of two bands with eigenenergies $ \eps_\pm =
\pm (\Delta^2 + \delta^2 \sin^2\frac{qa}{2} + t^2
\cos^2\frac{qa}{2})^{1/2}$.  The degeneracy point occurs at
\begin{equation}
\Delta = 0 \;, \quad \delta = 0\;, \quad q = \pi/a \;.
\end{equation}
The polarization is calculated using the two-point
formula~\eqref{pump:p2} with the Berry connection given by
\begin{equation}
\cal A_q = \partial_q\phi \cal A_\phi + \partial_q\theta \cal A_\theta 
= \sin^2\frac{\theta}{2} \partial_q\phi \;,
\end{equation}
where $\theta$ and $\phi$ are the spherical angles of $\bm h$.

Let us consider the case of $\Delta = 0$.  In the parameter space of
$\bm h$, it lies in the $xy$-plane with $\theta = \pi/2$.  As $q$
varies from 0 to $2\pi/a$, $\phi$ changes from 0 to $\pi$ if $\delta <
0$ and 0 to $-\pi$ if $\delta > 0$.  Therefore the polarization
difference between $P(\delta)$ and $P(-\delta)$ is $ea/2$.  This is
consistent with the observation that the state with $P(-\delta)$ can
be obtained by shifting the state with $P(\delta)$ by half of the unit
cell length $a$.

\begin{figure}
\includegraphics[width=8cm]{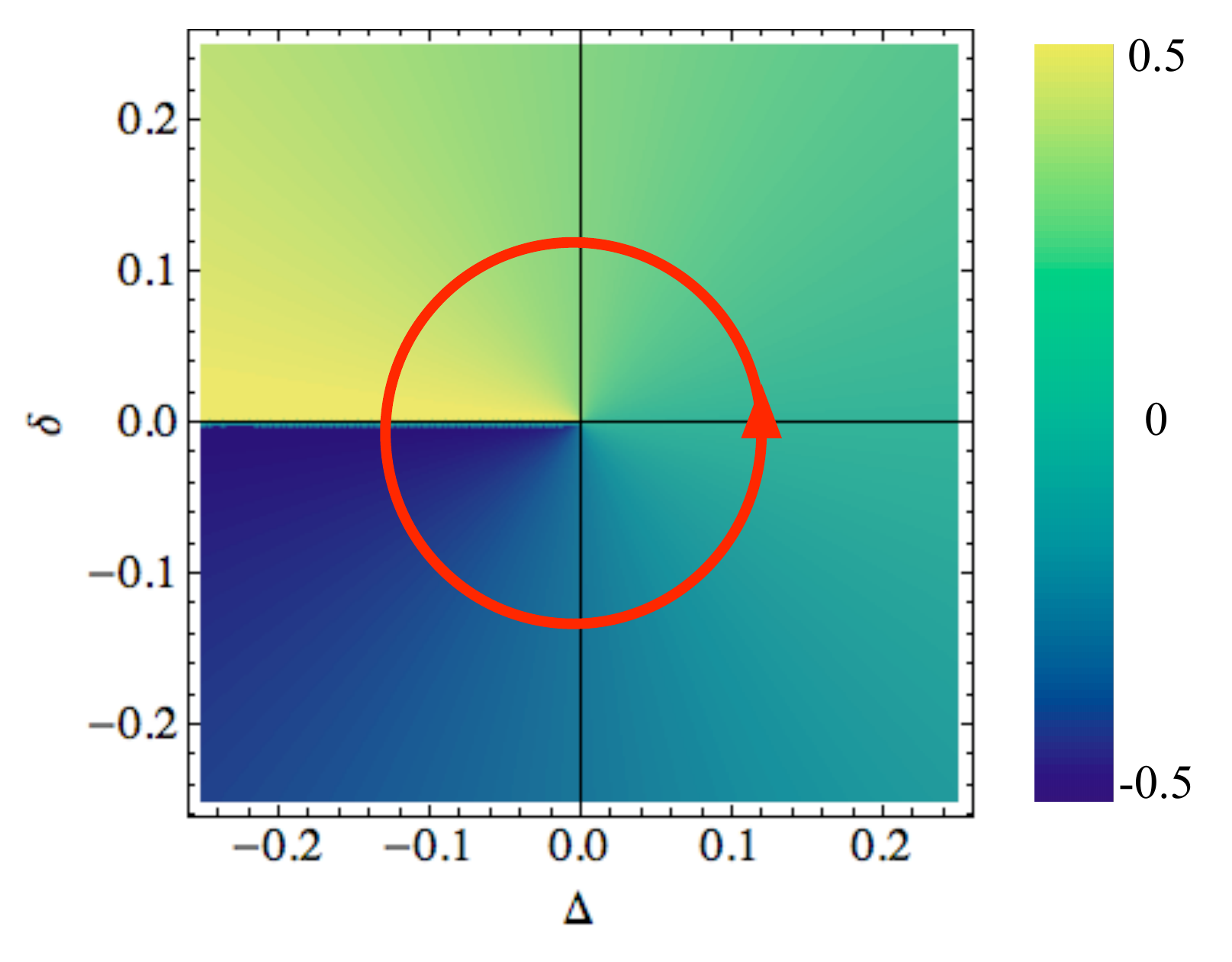}
\caption{\label{fig:rice}(color online). Polarization as a function of
  $\Delta$ and $\delta$ in the Rice-Mele model.  The units is $ea$
  with $a$ being the lattice constant.}
\end{figure}

Figure~\ref{fig:rice} shows the calculated polarization for arbitrary
$\Delta$ and $\delta$.  As we can see, if the system adiabatically
evolves along a loop enclosing the degeneracy point $(0,0)$ in the
$(\Delta, \delta)$ space, then the polarization will be changed by
$ea$, which means that if we allow $(\Delta, \delta)$ to change in
time along this loop, for example, $\Delta(t) = \Delta_0\sin(t)$ and
$\delta(t)=\delta_0 \cos(t)$, a quantized charge of $e$ is pumped out of
the system after one cycle.  On the other hand, if the loop does not
contain the degeneracy point, then the pumped charge is zero.


\section{\label{sec:hall}Electron dynamics in the presence of
  electric fields}

The dynamics of Bloch electrons under the perturbation of an electric
field is one of the oldest problems in solid state physics.  It is
usually understood that while the electric field can drive electron
motion in the momentum space, it does not appear in the electron
velocity; the latter is simply given by~\citep[for example,
  see][]{ashcroft-CM}
\begin{equation} \label{hall:oldv}
v_n(\bm q) = \dparh{\eps_n(\bm q)}{\bm q} \;.
\end{equation}
Through recent progress on the semiclassical dynamics of Bloch
electrons, it has been made increasingly clear that this description
is incomplete.  In the presence of an electric field, an electron can
acquire an anomalous velocity proportional to the Berry curvature of
the band~\cite{chang1995,chang1996,sundaram1999}.  This anomalous
velocity is responsible for a number of transport phenomena, in
particular various Hall effects, which we study in this section.

\subsection{Anomalous velocity}

Let us consider a crystal under the perturbation of a weak electric
field $\bm E$, which enters into the Hamiltonian through the coupling
to the electrostatic potential $\phi(\bm r)$.  However, a uniform $\bm
E$ means that $\phi(\bm r)$ varies linearly in space and breaks the
translational symmetry of the crystal such that Bloch's theorem cannot
be applied.  To go around this difficulty, one can let the electric
field enter through a uniform vector potential $\bm A(t)$ that changes
in time.  Using the Peierls substitution, the Hamiltonian is written
as
\begin{equation}
H(t) = \frac{[\hat{\bm p} + e\bm A(t)]^2}{2m} + V(\bm r) \;.
\end{equation}
This is the time-dependent problem we have studied in last section.
Transforming to the $\bm q$-space representation, we have
\begin{equation}
H(\bm q, t) = H(\bm q + \frac{e}{\hbar}\bm A(t)) \;.
\end{equation}
Introduce the gauge-invariant crystal momentum
\begin{equation} \label{hall:k}
\bm k = \bm q + \frac{e}{\hbar}\bm A(t) \;.
\end{equation}
The parameter-dependent Hamiltonian can be simply written as $H(\bm
k(\bm q, t))$.  Hence the eigenstates of the time-dependent
Hamiltonian can be labeled by a single parameter $\bm k$.  Moreover,
because $\bm A(t)$ preserves the translational symmetry, $\bm q$ is
still a good quantum number and is a constant of motion $\dot{\bm q} =
0$.  It then follows from Eq.~\eqref{hall:k} that $\bm k$ satisfies
the following equation of motion
\begin{equation}
\dot{\bm k} = -\frac{e}{\hbar} \bm E \;.
\end{equation}

Using the relation $\partial/\partial q_\alpha= \partial/\partial
k_\alpha$ and $\partial/\partial t = -(e/\hbar)E_\alpha
\partial/\partial k_\alpha$, the general formula~\eqref{pump:v}
for the velocity in a given state $\bm k$ becomes
\begin{equation} \label{hall:v}
\bm v_n(\bm k) = \dparh{\eps_n(\bm k)}{\bm k}
- \frac{e}{\hbar}\bm E \times \bm\Omega_n(\bm k) \;,
\end{equation}
where $\bm\Omega_n(\bm k)$ is the Berry curvature of the $n$th band:
\begin{equation}
\bm \Omega_n(\bm k) = i\bracket{\bm \nabla_{\bm k}u_n(\bm
  k)|\times|\bm\nabla_{\bm k} u_n(\bm k)} \;.
\end{equation}
We can see that, in addition to the usual band dispersion
contribution, an extra term previously known as an anomalous velocity
also contributes to $\bm v_n(\bm k)$.  This velocity is always
transverse to the electric field, which will give rise to a Hall
current.  Historically, the anomalous velocity has been obtained
by~\citet{adams1959,karplus1954,kohn1957}; however, its relation to
the Berry phase was not realized.  In Sec.~\ref{sec:em} we shall
rederive this term using a wave packet approach.

\subsection{Berry curvature: Symmetry considerations}

The velocity formula~\eqref{hall:v} reveals that, in addition to the
band energy, the Berry curvature of the Bloch bands is also required
for a complete description of the electron dynamics.  However, the
conventional formula, Eq.~\eqref{hall:oldv}, has had great success in
describing various electronic properties in the past.  It is thus
important to know under what conditions the Berry curvature term
cannot be neglected.

The general form of the Berry curvature $\bm\Omega_n(\bm k)$ can be
obtained via symmetry analysis.  The velocity formula~\eqref{hall:v}
should be invariant under time reversal and spatial inversion
operations if the unperturbed system has these symmetries.  Under time
reversal, $\bm v_n$ and $\bm k$ changes sign while $\bm E$ is fixed.
Under spatial inversion, $\bm v_n$, $\bm k$, and $\bm E$ changes sign.
If the system has time reversal symmetry, the symmetry condition on
Eq.~\eqref{hall:v} requires that
\begin{equation} \label{hall:T}
\bm\Omega_n(-\bm k) = -\bm\Omega_n(\bm k) \;.
\end{equation}
If the system has spatial inversion symmetry, then
\begin{equation}
\bm\Omega_n(-\bm k) = \bm\Omega_n(\bm k) \;.
\end{equation}
Therefore, for crystals with simultaneous time-reversal and spatial
inversion symmetry the Berry curvature vanish identically throughout
the Brillouin zone.  In this case Eq.~\eqref{hall:v} reduces to the
simple expression~\eqref{hall:oldv}.  However, in systems with either
broken time-reversal or inversion symmetries, their proper description
requires the use of the full velocity formula~\eqref{hall:v}.

\begin{figure}
\includegraphics[width=8cm]{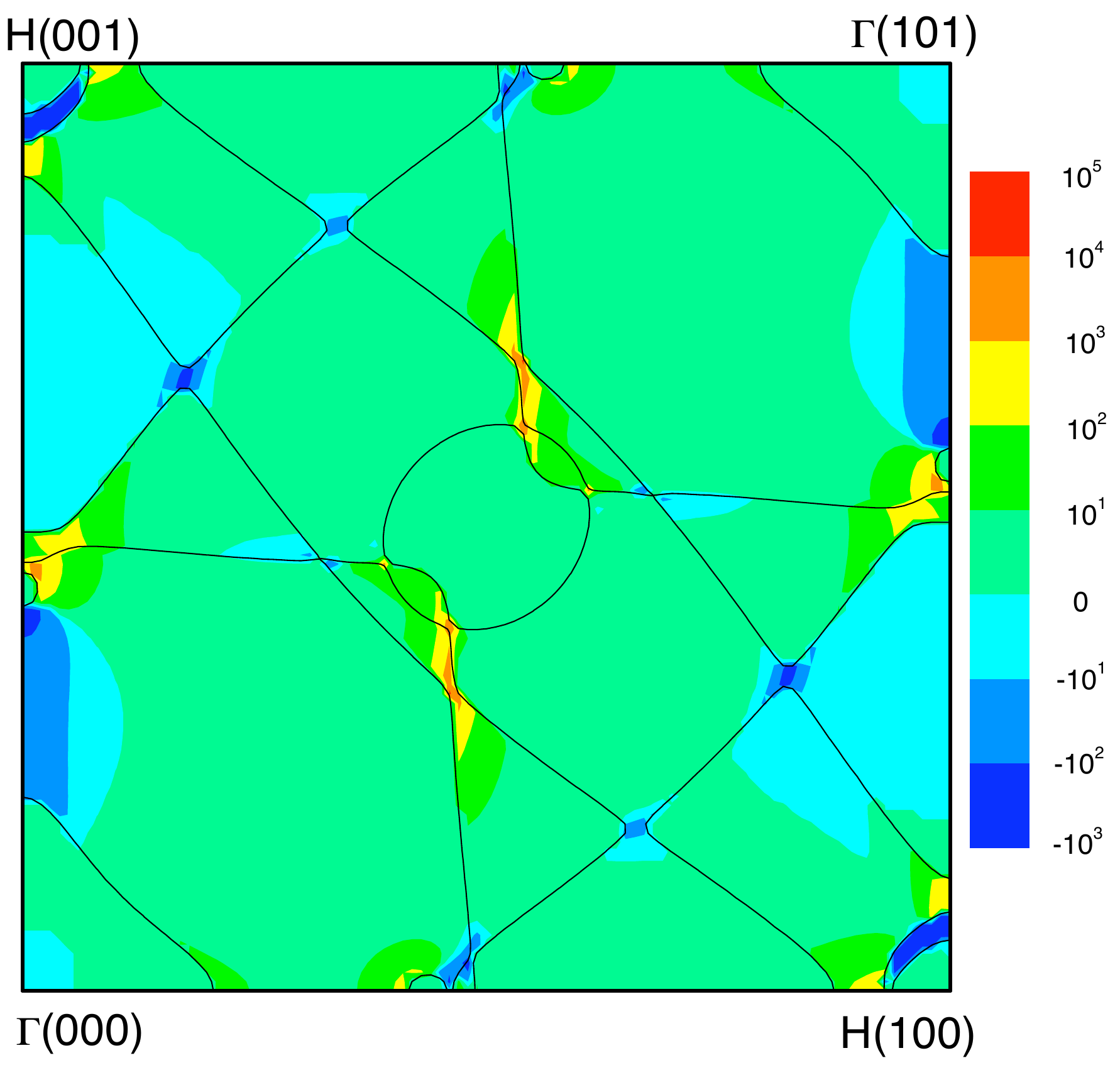}
\caption{\label{fig:curv_Fe}(color online). Fermi surface in
  (010) plane (solid lines) and Berry curvature $-\Omega_z(\bm k)$ in
  atomic units (color map) of fcc Fe.  From~\onlinecite{yao2004}.}
\end{figure}

\begin{figure}
\begin{tabular}{c}
\resizebox{7cm}{!}{\includegraphics{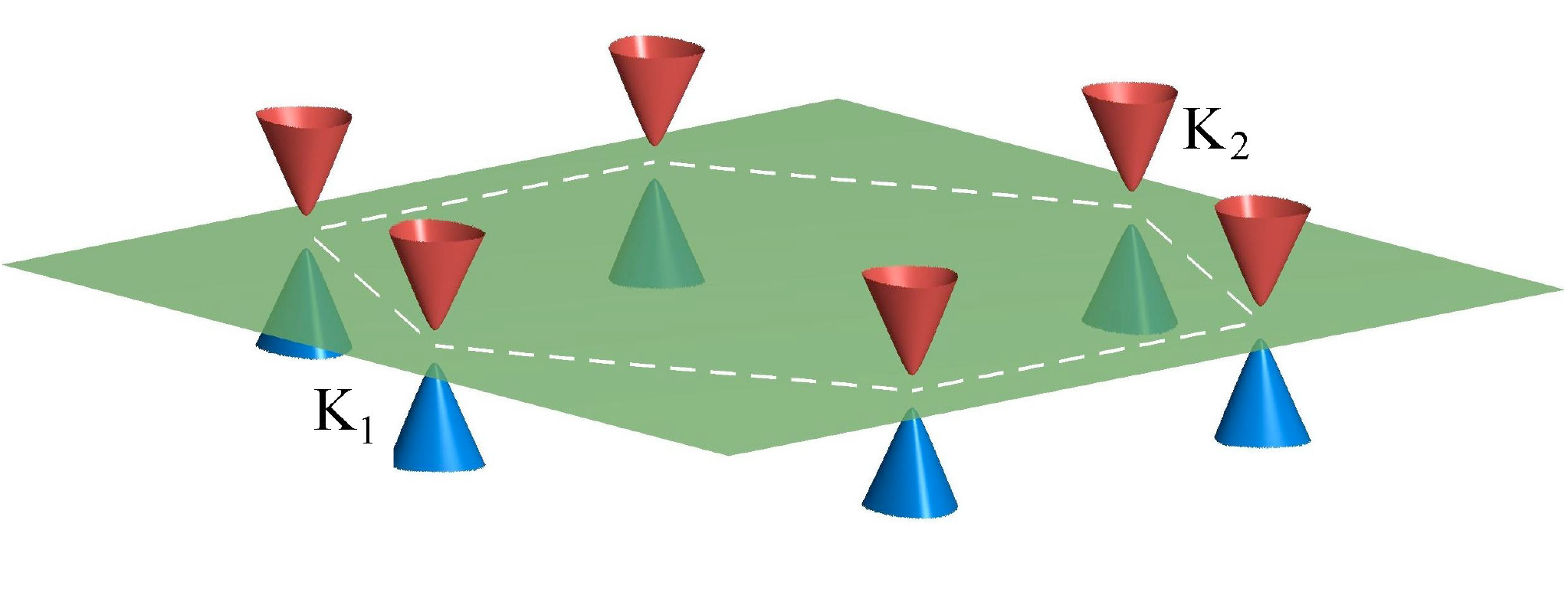}} \\
\resizebox{8cm}{!}{\includegraphics{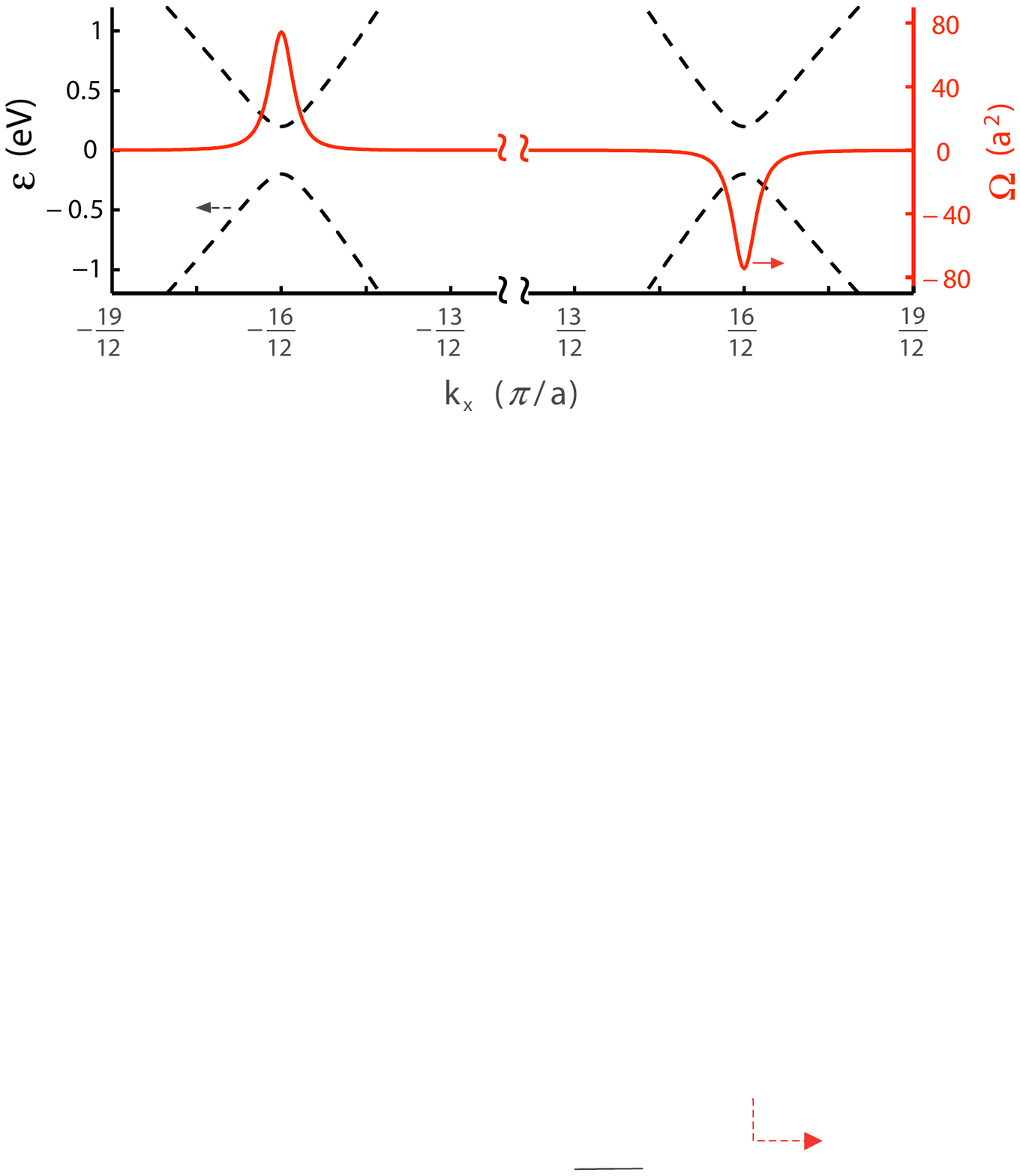}} \\
\end{tabular}
\caption{\label{fig:curv_graphene}(color online).  Energy bands
  (top panel) and Berry curvature of the conduction band (bottom
  panel) of a graphene sheet with broken inversion symmetry.  The
  first Brillouin zone is outlined by the dashed lines, and two
  inequivalent valleys are labeled as K$_1$ and K$_2$.  Details are
  presented in \onlinecite{xiao2007}.}
\end{figure}

There are many important physical systems where both symmetries are
not simultaneously present.  For example, in the presence of ferro- or
antiferro-magnetic ordering the crystal breaks the time-reversal
symmetry.  Figure~\ref{fig:curv_Fe} shows the Berry curvature on the
Fermi surface of fcc Fe.  As we can see, the Berry curvature is
negligible in most areas in the momentum space and displays very sharp
and pronounced peaks in regions where the Fermi lines (intersection of
the Fermi surface with (010) plane) have avoided crossings due to
spin-orbit coupling.  The monopole structure has been identified in
other materials as well~\cite{fang2003}.  Another example is provided
by single-layered graphene sheet with staggered sublattice potential,
which breaks the inversion symmetry~\cite{zhou2007}.
Figure~\ref{fig:curv_graphene} shows the energy band and Berry
curvature of this system.  The Berry curvature at valley $K_1$ and
$K_2$ have opposite signs due to time-reversal symmetry.  We note that
as the gap approaches zero, the Berry phase acquired by an electron
during one circle around the valley becomes exactly $\pm\pi$.  This
Berry phase of $\pi$ has been observed in intrinsic graphene
sheet~\cite{novoselov2005,zhang2005}.

\subsection{\label{sec:hallc}The quantum Hall effect}

The quantum Hall effect was discovered by \citet{klitzing1980}.  They
found that in a strong magnetic field the Hall conductivity of a
two-dimensional electron gas is \emph{exactly} quantized in the units
of $e^2/h$.  The exact quantization was subsequently explained by
\citet{laughlin1981} based on gauge invariance, and was later related
to a topological invariance of the energy
bands~\cite{avron1983,thouless1982,niu1985}.  Since then it has
blossomed into an important research field in condensed matter
physics.  In this subsection we shall focus only on the quantization
aspect of the quantum Hall effect using the formulation developed so
far.

Let us consider a two-dimensional band insulator.  It follows from
Eq.~\eqref{hall:v} that the Hall conductivity of the system is
given by
\begin{equation} \label{hall:qhe}
\sigma_{xy} = \frac{e^2}{\hbar} \int_\text{BZ} \frac{d^2k}{(2\pi)^2}
\Omega_{k_x k_y} \;,
\end{equation}
where the integration is over the entire Brillouin.  Once again we
encounter the situation where the Berry curvature is integrated over a
closed manifold.  Here $\sigma_{xy}$ is just the Chern number in the
units of $e^2/h$, i.e.,
\begin{equation}
\sigma_{xy} = n \frac{e^2}{h} \;.
\end{equation}
Therefore the Hall conductivity is quantized for a two-dimensional
band insulator of non-interacting electrons.

Historically, the quantization of the Hall conductivity in a crystal
was first shown by \citet{thouless1982} for magnetic Bloch bands (see
also Sec.~\ref{sec:mbb}.  It was shown that, due to the magnetic
translational symmetry, the phase of the wave function in the magnetic
Brillouin zone carries a vortex and leads to a \emph{nonzero}
quantized Hall conductivity~\cite{kohmoto1985}.  However, from the
above derivation it is clear that a magnetic field is not necessary
for the quantum Hall effect to occur as the condition is a nonzero
Chern number of the band. \citet{haldane1988} constructed a
tight-binding model on a honeycomb lattice which displays the quantum
Hall effect with zero net flux per unit cell.  Another model utilizing
the spin-orbit interaction in a semiconductor quantum well was
recently proposed~\cite{qi2006a,liu2008}.  The possibility of
realizing the quantum Hall effect without a magnetic field is very
attractive in device design.

\begin{figure}
\includegraphics[width=6cm]{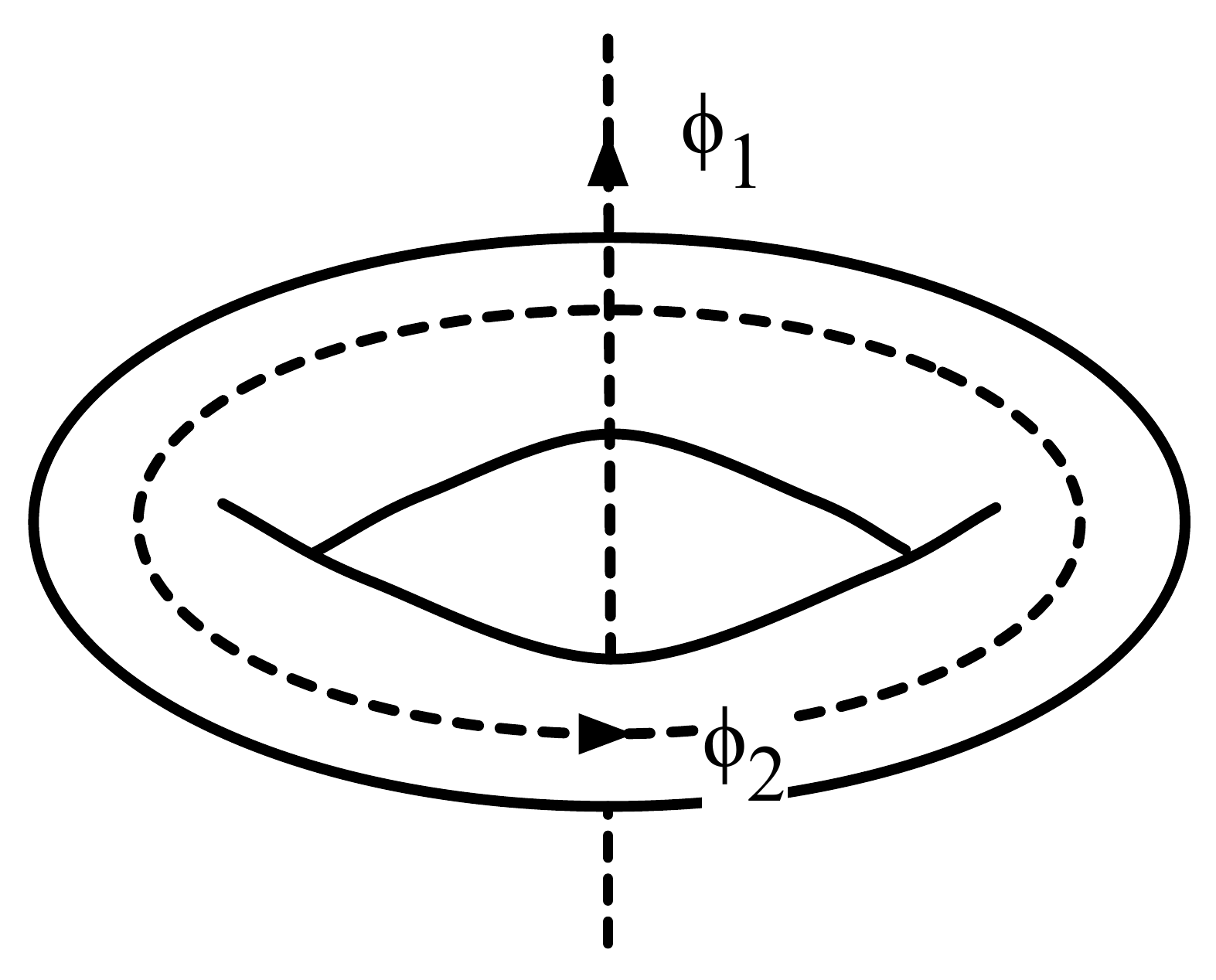}
\caption{\label{fig:twist}Magnetic flux going through the holes of the
  torus.}
\end{figure}

\citet{niu1985} further showed that the quantized Hall conductivity in
two-dimensions is robust against many-body interactions and disorder.
Their derivation involves the same technique discussed in
Sec.~\ref{sec:twist}.  A two-dimensional many-body system is placed on
a torus by assuming periodic boundary conditions in both directions.
One can then thread the torus with magnetic flux through its holes
(Fig.~\ref{fig:twist}) and make the Hamiltonian $H(\phi_1, \phi_2)$
depend on the flux $\phi_1$ and $\phi_2$.  The Hall conductivity is
calculated using the Kubo formula
\begin{equation} \label{hall:kubo}
\sigma_H = ie^2\hbar\sum_{n > 0} \frac{\bracket{\Phi_0|v_1|\Phi_n}
\bracket{\Phi_n|v_2|\Phi_0} - (1\leftrightarrow 2)}
{(\eps_0 - \eps_n)^2} \;,
\end{equation}
where $\Phi_n$ is the many-body wave function with $\ket{\Phi_0}$ the
ground state.  In the presence of the flux, the velocity operator is
given by $v_i = \partial H(\kappa_1,
\kappa_2)/\partial(\hbar\kappa_i)$ with $\kappa_i =
(e/\hbar)\phi_i/L_i$ and $L_i$ the dimensions of the system.  We
recognize that Eq.~\eqref{hall:kubo} is the summation
formula~\eqref{berry:sum} for the Berry curvature
$\Omega_{\kappa_1\kappa_2}$ of the state $\ket{\Phi_0}$.  The
existence of a bulk energy gap guarantees that the Hall conductivity
remains unchanged after thermodynamic averaging, which is given by
\begin{equation}
\sigma_H = \frac{e^2}{\hbar}\int_0^{2\pi/L_1}d\kappa_1 \int_0^{2\pi/L_2}
d\kappa_2 \, \Omega_{\kappa_1\kappa_2} \;.
\end{equation}
Note that the Hamiltonian $H(\kappa_1, \kappa_2)$ is periodic in
$\kappa_i$ with period $2\pi/L_i$ because the system returns to its
original state after the flux is changed by a flux quantum $h/e$ (and
$\kappa_i$ changed by $2\pi/L_i$).  Therefore the Hall conductivity is
quantized even in the presence of many-body interaction and disorder.
Due to the high precision of the measurement and the robustness of the
quantization, the quantum Hall resistance is now used as the primary
standard of resistance.

The geometric and topological ideas developed in the study of the
quantum Hall effect has a far-reaching impact on modern condensed
matter physics.  The robustness of the Hall conductivity suggests that
it can be used as a topological invariance to classify many-body
phases of electronic states with a bulk energy gap~\cite{avron1983}:
states with different topological orders (Hall conductivities in the
quantum Hall effect) cannot be adiabatically transformed into each
other; if that happens, a phase transition must occur.  It has
important applications in strongly correlated electron systems, such
as the fractional quantum Hall effect~\cite{wen1990}, and most
recently, the topological quantum computing~\citep[for a review,
  see][]{nayak2008}.

\subsection{{\label{sec:ahe}}The anomalous Hall Effect}

Next we discuss the anomalous Hall effect, which refers to the
appearance of a large spontaneous Hall current in a ferromagnet in
response to an electric field alone~\cite{hallbook}.  Despite its
century-long history and importance in sample characterization, the
microscopic mechanism of the anomalous Hall effect has been a
controversial subject~\citep[for a brief review on the history of the
  anomalous Hall effect, see][]{sinova2004}.  In the past, three
mechanisms have been identified: the intrinsic
contribution~\cite{karplus1954,luttinger1958}, and the extrinsic
contributions from the skew~\cite{smit1958} and side-jump
scattering~\cite{berger1970}.  It was later realized that the
scattering-independent intrinsic contribution comes from the Berry-phase
supported anomalous velocity.  It will be our primary interests here.

The intrinsic contribution to the anomalous Hall effect can be
regarded as an ``unquantized'' version of the quantum Hall effect.
The Hall conductivity is given by
\begin{equation} \label{hall:ahe}
\sigma_{xy} = \frac{e^2}{\hbar} \int \frac{d\bm k}{(2\pi)^d}
f(\eps_{\bm k}) \Omega_{k_xk_y} \;,
\end{equation}
where $f(\eps_{\bm k})$ is the Fermi-Dirac distribution function.
However, unlike the quantum Hall effect, the anomalous Hall effect
does not require a nonzero Chern number of the band; for a band with
zero Chern number, the local Berry curvature can be nonzero and give
rise to a nonzero anomalous Hall conductivity.

To appreciate the intrinsic contribution, let us consider a
generic Hamiltonian with spin-orbit (SO) split bands~\cite{onoda2006}
\begin{equation}
H = \frac{\hbar^2k^2}{2m} + \lambda (\bm k \times \bm\sigma) \cdot \bm
e_z - \Delta \sigma_z \;.
\end{equation}
where $2\Delta$ is the SO split gap in the energy spectrum $\eps_\pm =
\hbar^2k^2/2m \pm \sqrt{\lambda^2k^2 + \Delta^2}$, and $\lambda$ gives
a linear dispersion in the absence of $\Delta$.  This model also
describes spin-polarized two-dimensional electron gas with Rashba SO
coupling, with $\lambda$ being the SO coupling strength and $\Delta$
the exchange field~\cite{culcer2003}.  The Berry curvature is given
by, using Eq.~\eqref{berry:R1R2},
\begin{equation}  \label{ahe:curv}
\Omega_{\pm} = \mp \frac{\lambda^2\Delta}
{2(\lambda^2k^2 + \Delta^2)^{3/2}} \;.
\end{equation}
The Berry curvatures of the two energy bands have opposite sign, and
is highly concentrated around the gap (In fact, the Berry curvature
has the same form of the Berry curvature in one valley of the
graphene, shown in Fig.~\ref{fig:curv_graphene}).  One can verify that
the integration of the Berry curvature of a full band,
$2\pi\int_0^\infty qdq\,\Omega_\pm$, is $\pm\pi$ for the upper and
lower bands, respectively.

\begin{figure}
\includegraphics[width=7cm]{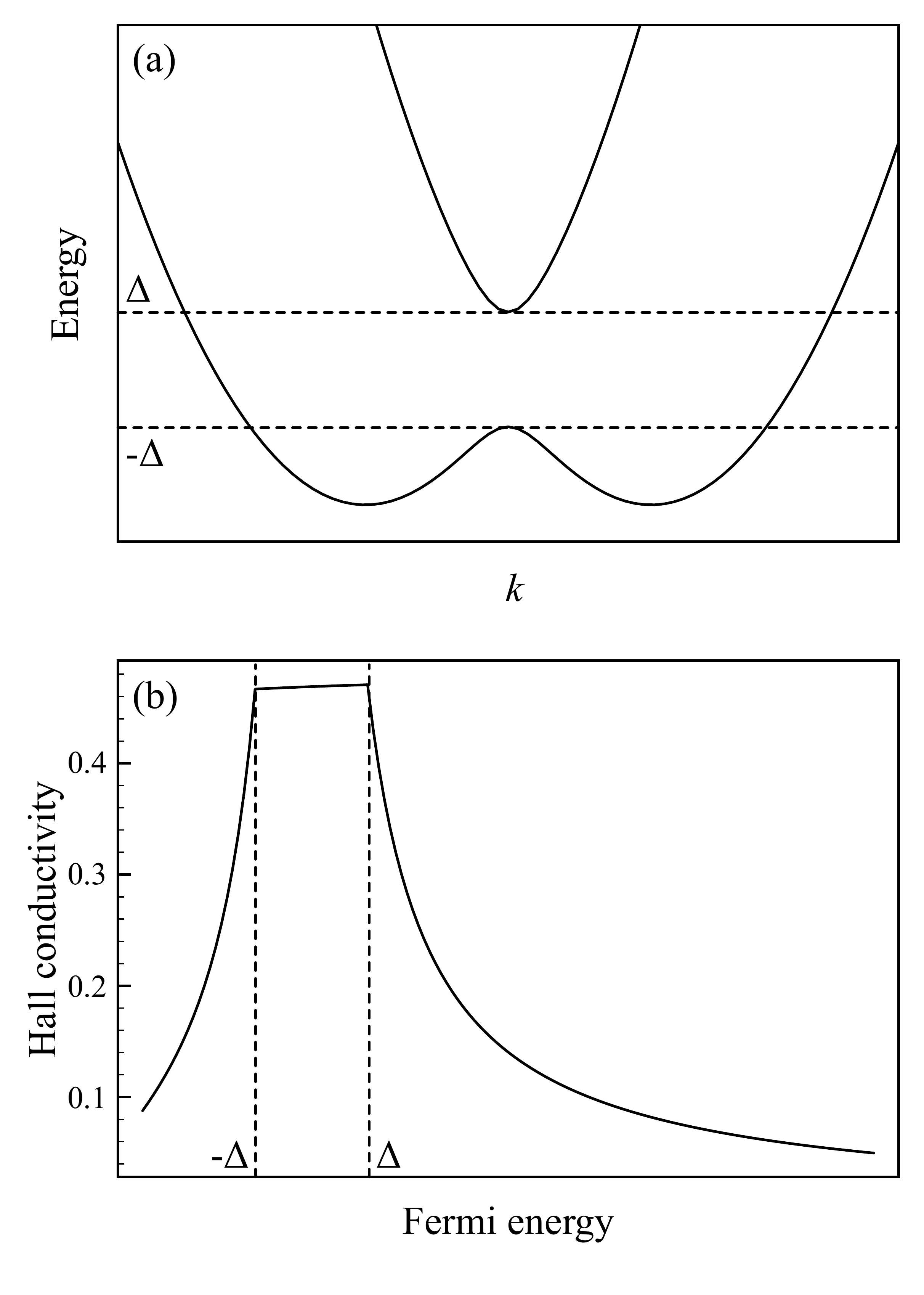}
\caption{\label{fig:ahe} (a) Energy dispersion of spin-split bands.
  (b) The Hall conductivity $-\sigma_{xy}$ in the units of $e^2/h$ as
  a function of Fermi energy.}
\end{figure}

Figure~\ref{fig:ahe} shows the band dispersion, and the intrinsic Hall
conductivity, Eq.~\eqref{hall:ahe}, as the Fermi energy sweeps across
the SO split gap.  As we can see, when the Fermi energy $\eps_F$ is in
the gap region, the Hall conductivity reaches its maximum value (about
$-e^2/2h$).  If $\eps_F < -\Delta$, the states with energies just
below $-\Delta$, which contribute most to the Hall conductivity, are
empty.  If $\eps_F > \Delta$, contributions from upper and lower bands
cancel each other, and the Hall conductivity decreases quickly as
$\eps_F$ moves away from the bad gap.  It is only when $-\Delta <
\eps_F < \Delta$, the Hall conductivity is resonantly
enhanced~\cite{onoda2006}.

\subsubsection{\label{sec:halld1}Intrinsic vs. extrinsic contributions}

The above discussion does not take into account the fact that, unlike
insulators, in metallic systems electron scattering can be important
in transport phenomena.  Two contributions to the anomalous Hall
effect arises due to scattering: (i) the skew scattering that refers
to the asymmetric scattering amplitude with respect to the scattering
angle between the incoming and outgoing electron
waves~\cite{smit1958}, and (ii) the side jump which is a sudden shift
of the electron coordinates during scattering~\cite{berger1970}.  In a
more careful analysis, a systematic study of the anomalous Hall effect
based on the semiclassical Boltzmann transport theory has been carried
out~\cite{sinitsyn2008}.  The basic idea is to solve the following
Boltzmann equation:
\begin{equation} \begin{split}
&\dpar{g_{\bm k}}{t} - e\bm E \cdot \dparh{\eps}{\bm k} \dpar{f}{\eps}\\
&\quad = \sum_{\bm k'}\omega_{\bm k\bm k'} \Bigl[g_{\bm k} - g_{\bm k'}
- \dpar{f}{\eps} e\bm E \cdot \delta \bm r_{\bm k\bm k'} \Bigr] \;,
\end{split} \end{equation}
where $g$ is the non-equilibrium distribution function, $\omega_{\bm
  k\bm k'}$ represents the asymmetric skew scattering, and $\delta \bm
r_{\bm k\bm k'}$ describes the side-jump of the scattered electrons.
The Hall conductivity is the sum of different contributions
\begin{equation}
\sigma_H = \sigma_H^\text{in} + \sigma_H^\text{sk} +
\sigma_H^\text{sj} \;,
\end{equation}
where $\sigma_H^\text{in}$ is the intrinsic contribution given by
Eq.~\eqref{hall:ahe}, $\sigma_H^\text{sk}$ is the skew scattering
contribution, which is proportional to the relaxation time $\tau$, and
$\sigma_H^\text{sj}$ is the side jump contribution, which is
independent of $\tau$.  Note that in addition to Berger's original
proposal, $\sigma_H^\text{sj}$ also includes two other contributions:
the intrinsic skew-scattering and anomalous distribution
function~\cite{sinitsyn2008}.

An important question is to identify the dominant contribution to the
AHE among these mechanisms.  If the sample is clean and the
temperature is low, the relaxation time $\tau$ can be extremely large,
and the skew scattering is expected to dominate.  On the other hand,
in dirty samples and at high temperatures, $\sigma_H^\text{sk}$
becomes small compared to both $\sigma_H^\text{in}$ and
$\sigma_H^\text{sj}$.  Because the Berry-phase contribution
$\sigma_H^\text{in}$ is independent of scattering, it can be readily
evaluated using first-principles methods or effective Hamiltonians.
Excellent agreement with experiments has been demonstrated in
ferromagnetic transition metals and
semiconductors~\cite{fang2003,yao2004,xiao2006,yao2007,jungwirth2002},
which leaves little room for the side jump contribution.

\begin{figure}
\includegraphics[width=8cm]{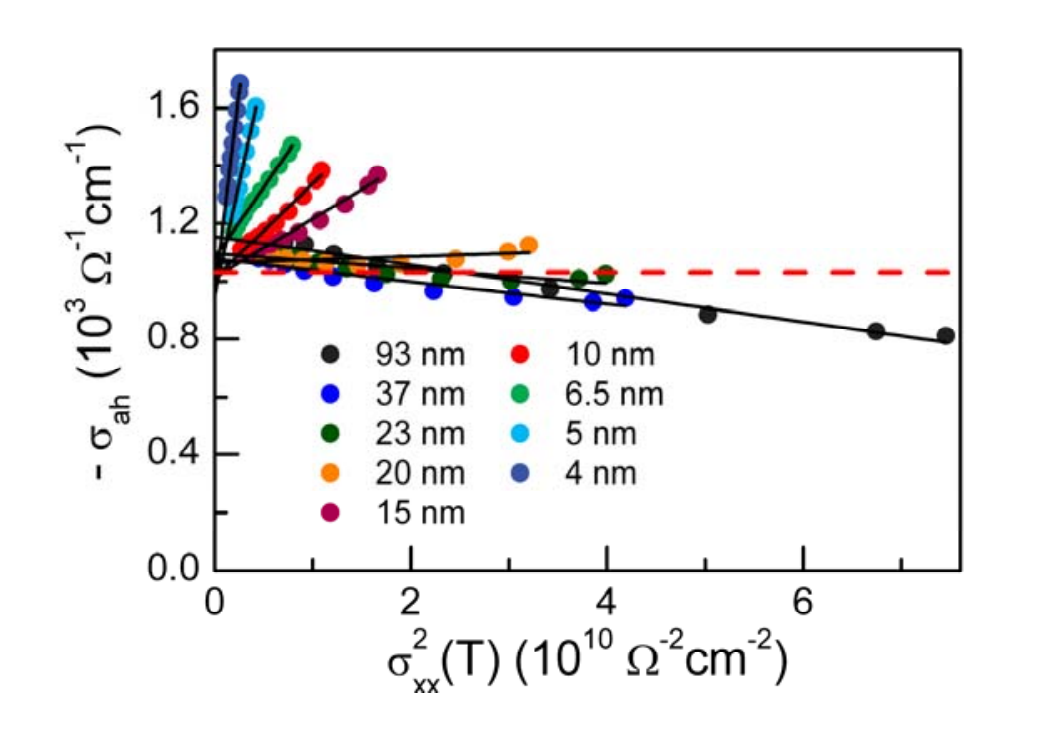}
\caption{\label{fig:ahe_jin} (color online) $\sigma_{ah}$ vs. $\sigma_{xx}(T)^2$ in
  Fe thin films with different film thickness over the temperature
  range 5-300 K.  From~\onlinecite{jin2009}.}
\end{figure}

In addition, a number of experimental results also gave favorable
evidence for the dominance of the intrinsic
contribution~\cite{lee2004,zeng2006,sales2006,mathieu2004,chun2007}.
In particular, \citet{jin2009} recently measured the anomalous Hall
conductivity in Fe thin films.  By varying the film thickness and the
temperature, they are able to control various scattering process such
as the impurity scattering and the phonon scattering.
Figure~\ref{fig:ahe_jin} shows their measured $\sigma_{ah}$ as a function
of $\sigma_{xx}(T)^2$.  We can see that although $\sigma_{ah}$ in
different thin films and at different temperatures shows a large
variation at finite $\sigma_{xx}$, they converge to a single point as
$\sigma_{xx}$ approaches zero, where the impurity-scattering induced
contribution should be washed out by the phonon scattering and only
the intrinsic contribution survives.  It turns out that this converged
value is very close to the bulk $\sigma_{ah}$ of Fe, which confirms
the dominance of the intrinsic contribution in Fe.

In addition to the semiclassical
approach~\cite{sinitsyn2008,sinitsyn2005}, there are a number of works
based on a full quantum mechanical
approach~\cite{inoue2006,onoda2006,kato2007,onoda2008,onoda2002,dugaev2005,nozieres1973,sinitsyn2007}.
In both approaches, the Berry-phase supported intrinsic contribution
to the anomalous Hall effect has been firmly established.

\subsubsection{Anomalous Hall conductivity as a Fermi surface property}

An interesting aspect of the intrinsic contribution to the anomalous
Hall effect is that the Hall conductivity, Eq.~\eqref{hall:ahe}, is
given as an integration over all occupied states below the Fermi
energy.  It seems to be against the spirit of the Landau Fermi liquid
theory, which states that the transport property of an electron system
is determined by quasiparticles at the Fermi energy.  This issue was
first raised by \citet{haldane2004}, and he showed that the Hall
conductivity can be written, in the units of $e^2/2\pi h$, as the
Berry phase of quasiparticles on the Fermi surface, up to a multiple
of $2\pi$.  Therefore the intrinsic Hall conductivity is also a Fermi
surface property.  This observation suggests that the Berry phase on
the Fermi surface should be added as a topological ingredient to the
Landau Fermi liquid theory.

For simplicity, let us consider a two-dimensional system.  We assume
there is only one partially filled band.  Write the Berry curvature in
terms of the Berry vector potential and integrate Eq.~\eqref{hall:ahe}
by part; one finds
\begin{equation}
\sigma_{xy}^\text{2D} = \frac{e^2}{\hbar}
\int \frac{d^2\bm k}{2\pi}
(\dpar{f}{k_y} \cal A_{k_x} - \dpar{f}{k_x} \cal A_{k_y}) \;,
\end{equation}
Note that the Fermi distribution function $f$ is a step
function at the Fermi energy.  If we assume the Fermi surface is a
closed loop in the Brillouin zone, then
\begin{equation}
\sigma^\text{2D}_{xy} = \frac{e^2}{2\pi h} \oint d\bm k \cdot \bm{\cal
  A}_{\bm k} \;.
\end{equation}
The integral is nothing but the Berry phase along the Fermi circle in
the Brillouin zone. The three-dimensional case is more complicated;
\citet{haldane2004} showed that the same conclusion can be reached.

\citet{wang2007} has implemented Haldane's idea in first-principles
calculations of the anomalous Hall conductivity.  From a computational
point of view, the advantage lies in that the integral over the Fermi
sea is converted to a more efficient integral on the Fermi surface.
On the theory side, \citet{shindou2006,shindou2008} derived an
effective Boltzmann equation for quasiparticles on the Fermi surface
using the Keldysh formalism, where the Berry phase of the Fermi
surface is defined in terms of the quasiparticle Green functions,
which nicely fits into the Landau Fermi liquid theory.

\subsection{\label{sec:valley}The valley Hall effect}

A necessary condition for the charge Hall effect to manifest is the
broken time-reversal symmetry of the system.  In this subsection we
discuss another type of Hall effect which relies on inversion symmetry
breaking, and survives in time-reversal invariant systems.

We shall use graphene as our prototype system.  The band structure of
intrinsic graphene has two degenerate and inequivalent Dirac points at
the corners of the Brillouin zone, where the conduction and valance
bands touch each other, forming a valley structure.  Because of their
large separation in momentum space, the intervalley scattering is
strongly suppressed~\cite{gorbachev2007,morpurgo2006,morozov2006},
which makes the valley index a good quantum number.  Interesting
valley-dependent phenomena, which concerns about the detection and
generation of valley polarization, are being actively
explored~\cite{rycerz2007,akhmerov2007,xiao2007,yao2008}.

The system we are interested in is graphene with broken inversion
symmetry.  \citet{zhou2007} have recently reported the observation of
a band gap opening in epitaxial graphene, attributed to the inversion
symmetry breaking by the substrate potential.  In addition, in biased
graphene bilayer, inversion symmetry can be explicitly broken by the
applied interlayer voltage~\cite{ohta2006,mccann2006,min2007}.  It is
this broken inversion symmetry that allows a valley Hall effect.
Introducing the valley index $\tau_z = \pm 1$ which labels the two
valleys, we can write the valley Hall effect as
\begin{equation}
\bm j^v = \sigma^v_H \hat{z} \times \bm E \;,
\end{equation}
where $\sigma^v_H$ is the valley Hall conductivity, and the valley
current $\bm j^v = \bracket{\tau_z \bm v}$ is defined as the average
of the valley index $\tau_z$ times the velocity operator $\bm v$.
Under time reversal, both the valley current and electric field are
invariant ($\tau_z$ changes sign because the two valleys switch when
the crystal momentum changes sign).  Under spatial inversion, the
valley current is still invariant but the electric field changes sign.
Therefore, the valley Hall conductivity can be nonzero when the
inversion symmetry is broken, even if time reversal symmetry remains.

In the tight-binding approximation, the Hamiltonian of a single
graphene sheet can be modeled with a nearest-neighbor hopping energy
$t$ and a site energy difference $\Delta$ between sublattices.  For
relatively low doping, we can resort to the low-energy description
near the Dirac points.  The Hamiltonian is given
by~\cite{semenoff1984}
\begin{equation}
H = \frac{\sqrt{3}}{2} at (q_x \tau_z \sigma_x + q_y \sigma_y) +
\frac{\Delta}{2} \sigma_z \;,
\end{equation}
where $\bm\sigma$ is the Pauli matrix accounting for the sublattice
index, and $\bm q$ is measured from the valley center $\bm K_{1,2} =
(\pm 4\pi/3a)\hat{\bm x}$ with $a$ being the lattice constant.  The
Berry curvature of the conduction band is given by
\begin{equation}
\Omega(\bm q) = \tau_z \frac{3a^2t^2\Delta}
{2(\Delta^2 + 3q^2a^2t^2)^{3/2}} \;.
\end{equation}
Note that the Berry curvatures in two valleys have opposite sign, as
required by time-reversal symmetry.  The energy spectrum and the Berry
curvature are already shown in Fig.~\ref{fig:curv_graphene}.  Once the
structure of the Berry curvature is revealed, the valley Hall effect
becomes transparent: upon the application of an electric field,
electrons in different valleys will flow to opposite transverse edges,
giving rise to a net valley Hall current in the bulk.

We remark that as $\Delta$ goes to zero, the Berry curvature vanishes
everywhere except at the Dirac points where it diverges.  Meanwhile,
The Berry phase around the Dirac points becomes exactly $\pm\pi$ (also see Sec.~\ref{sec:qzc}).

As we can see, the valley transport in systems with broken inversion
symmetry is a very general phenomenon.  It provides a new and standard
pathway to potential applications of valleytronics, or valley-based
electronic applications, in a broad class of
semiconductors~\cite{gunawan2006,xiao2007,yao2008}.


\section{\label{sec:wave}Wave packet: Construction and Properties}

So far, our discussion has focused on crystals under time-dependent
perturbations, and we have shown that the Berry phase will manifest
itself as an anomalous term in the electron velocity.  However, in
general situations the electron dynamics can be also driven by
perturbations that vary in space.  In this case, the most useful
description is provided by the semiclassical theory of Bloch electron
dynamics, which describes the motion of a narrow wave packet obtained
by superposing the Bloch states of a band~\citep[see, for example,
  Chap. 12 of][]{ashcroft-CM}.  The current and next
sections are devoted to the study of the wave packet dynamics, where
the Berry curvature naturally appears in the equations of motion.

In this section we discuss the construction and the general properties
of the wave packet.  Two quantities emerges in the wave packet
approach, i.e., the orbital magnetic moment of the wave packet and the
dipole moment of a physical observable.  For their applications, we
consider the problems of orbital magnetization and anomalous
thermoelectric transport in ferromagnets.

\subsection{Construction of the wave packet and its orbital moment}

We assume the perturbations are sufficiently weak such that
transitions between different bands can be neglected. i.e., the
electron dynamics is confined within a single band.  Under this
assumption, we construct a wave packet using the Bloch functions
$\ket{\psi_n(\bm q)}$ from the $n$th band:
\begin{equation}
\ket{W_0} = \int d\bm q\, w(\bm q, t) \ket{\psi_n(\bm q)} \;.
\end{equation}
There are two requirements on the envelope function $w(\bm q, t)$.
Firstly, $w(\bm q, t)$ must have a sharp distribution in the Brillouin
zone such that it makes sense to speak of the wave vector $\bm q_c$ of
the wave packet, given by
\begin{equation} \label{wave:kc1}
\bm q_c = \int d\bm q\, \bm q |w(\bm q, t)|^2 \;.
\end{equation}
To first order, $|w(\bm q, t)|^2$ can be approximated by $\delta(\bm q
- \bm q_c)$ and one has
\begin{equation} \label{wave:kc}
\int d\bm q\, f(\bm q)|w(\bm q,t)|^2 \approx f(\bm q_c) \;,
\end{equation}
where $f(\bm q)$ is an arbitrary function of $\bm q$.
Equation~\eqref{wave:kc} is very useful in evaluating various
quantities related to the wave packet.  Secondly, the wave packet has
to be narrowly localized around its center of mass, denoted by $\bm
r_c$, in the real space, i.e.,
\begin{equation}
\bm r_c = \bracket{W_0|\bm r|W_0} \;.
\end{equation}
Using Eq.~\eqref{wave:kc} we obtain
\begin{equation} \label{wave:rc}
\bm r_c = -\dpar{}{\bm q_c} \arg w(\bm q_c, t) + \bm{\cal A}^n_{\bm q}
(\bm q_c) \;,
\end{equation}
where $\bm{\cal A}^n_{\bm q} = i\bracket{u_n(\bm q)|\bm\nabla_{\bm
    q}|u_n(\bm q)}$ is the Berry connection of the $n$th band defined
using $\ket{u_n(\bm q)} = e^{-i\bm k\cdot\bm r}\ket{\psi_n(\bm q)}$.

In principle, more dynamical variables, such as the width of the wave
packet in both the real space and momentum space, can be added to
allow a more elaborate description of the time evolution of the wave
packet.  However, in short period the dynamics is dominated by the
motion of the wave packet center, and Eqs.~\eqref{wave:kc1} and
\eqref{wave:rc} are sufficient requirements.

When more than one band come close to each other, or even become
degenerate, the single-band approximation breaks down and the wave
packet must be constructed using Bloch functions from multiple bands.
\citet{culcer2005,shindou2005} developed the multi-band formalism for
electron dynamics, which will be presented in Sec.~\ref{sec:nab}.  For
now, we will focus on the single-band formulation and drop the band
index $n$ for simple notation.

The wave packet, unlike a classical point particle, has a finite
spread in real space.  In fact, since it is constructed using an
incomplete basis of the Bloch functions, the size of the wave packet
has a nonzero lower bound~\cite{marzari1997}.  Therefore, a wave
packet may possess a self-rotation around its center of mass, which
will in turn give rise to an orbital magnetic moment.  By calculating
the angular momentum of a wave packet directly, one finds~\cite{chang1996}
\begin{equation} \label{wave:m} \begin{split}
\bm m(\bm q) &= -\frac{e}{2}\bracket{W_0|(\bm r - \bm r_c) \times \bm
  j|W_0} \\ &= -i\frac{e}{2\hbar}\bracket{\bm\nabla_{\bm q}u|
  \times[H(\bm q) - \eps(\bm q)]|\bm\nabla_{\bm q}u} \;,
\end{split} \end{equation}
where $H(\bm q) = e^{-i\bm q\cdot\bm r}H e^{i\bm q\cdot \bm r}$ is the
$\bm q$-dependent Hamiltonian.  This shows that the wave packet of a
Bloch electron generally rotates around its mass center and carries an
orbital magnetic moment in addition to its spin moment.

We emphasize that the orbital moment is an intrinsic property of the
band.  Its final expression, Eq.~\eqref{wave:m}, does not depend on
the actual shape and size of the wave packet, and only depends on the
Bloch functions.  Under symmetry operations, the orbital moment
transforms exactly like the Berry curvature.  Therefore unless the
system has both time-reversal and inversion symmetry, $\bm m(\bm q)$
is in general nonzero.  Information of the orbital moment can be
obtained by measuring magnetic circular dichroism spectrum of a
crystal~\cite{souza2008,yao2008}.

This orbital moment behaves exactly like the electron spin.  For
example, upon the application of a magnetic field, the orbital moment
will couple to the field through a Zeeman-like term $-\bm m(\bm q)
\cdot \bm B$.  If one construct a wave packet using only the positive
energy states (the electron branch) of the Dirac Hamiltonian, its
orbital moment in the non-relativistic limit is exactly the Bohr
magneton (Sec.~\ref{sec:nab}).  For Bloch electrons, the orbital
moment can be related to the electron $g$-factor~\cite{yafet1963}.
Let us consider a specific example.  For the graphene model with
broken-inversion symmetry, discussed in Sec.~\ref{sec:valley}, the
orbital moment of the conduction band is given by~\cite{xiao2007}
\begin{equation}
m(\tau_z, \bm q) = 
\tau_z \frac{3ea^2\Delta t^2}{4\hbar(\Delta^2+3q^2a^2t^2)} \;.
\end{equation}
So orbital moments in different valleys have opposite signs, as
required by time-reversal symmetry.  Interestingly, the orbital moment
at exactly the band bottom takes the following form
\begin{equation}
m(\tau_z) = \tau_z \mu^*_B, \qquad
\mu^*_B = \frac{e\hbar}{2m^*} \;,
\end{equation}
where $m^*$ is the effective mass at the band bottom.  The close
analogy with the Bohr magneton for the electron spin is transparent.
In realistic situations, the moment can be 30 times larger than the
spin moment, and can be used as an effective way to detect and
generate the valley polarization~\cite{xiao2007,yao2008}.

\subsection{\label{sec:om}Orbital magnetization}

A closely related quantity to the orbital magnetic moment is the
orbital magnetization in a crystal.  Although this phenomenon has been
known for a long time, our understanding of orbital magnetization in
crystals has remained in a primitive stage.  In fact, there was no
proper way to calculate this quantity until recently when the Berry
phase theory of orbital magnetization is
developed~\cite{xiao2005,thonhauser2005,shi2007}.  Here we provide a
rather pictorial derivation of the orbital magnetization based on the
wave packet approach.  This derivation gives an intuitive picture of
different contributions to the total orbital magnetization.

The main difficulty of calculating the orbital magnetization is
exactly the same one we faced when calculating the electric
polarization: the magnetic dipole $e\bm r \times \bm p$ is not defined
in a periodic system.  For a wave packet this is not a problem because
it is localized in space.  As we showed in last subsection, each
wave packet carries an orbital moment.  Thus, it is tempting to
conclude that the orbital magnetization is simply the thermodynamic
average of the orbital moment.  As it turns out, this is only part of
the contribution.  There is another contribution due to the
center-of-mass motion of the wave packet.

\begin{figure}
\includegraphics[width=8.5cm]{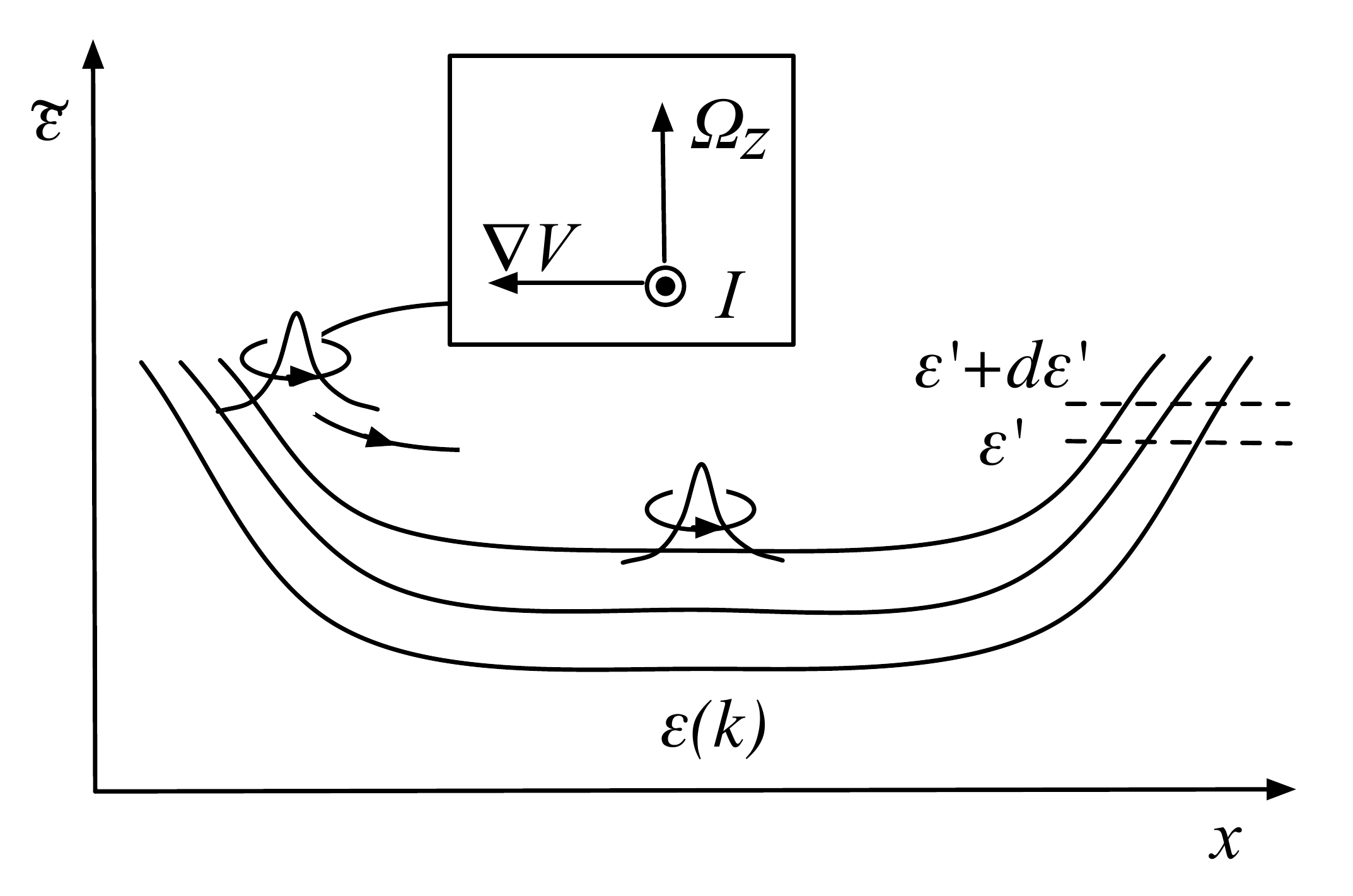}
\caption{\label{fig:energy}Electron energy $\tilde{\eps}$ in a slowly
  varying confining potential $V(\bm r)$.  In addition to the
  self-rotation, wave packets near the boundary will also move along
  the boundary due to the potential $V$.  Level spacings between
  different bulk $\bm q$-states are exaggerated; they are continuous
  in the semiclassical limit.  The insert shows directions of the
  Berry curvature, the effective force, and the current carried by a
  wave packet on the left boundary.}
\end{figure}

For simplicity, let us consider a finite system of electrons in a
two-dimensional lattice with a confining potential $V(\bm r)$.  We
further assume that the potential $V(\bm r)$ varies slowly at atomic
length scale such that the wave packet description of the electron is
still valid on the boundary.  In the bulk where $V(\bm r)$ vanishes,
the electron energy is just the bulk band-energy; near the boundary,
it will be tilted up due to the increase of $V(\bm r)$.  Thus to a
good approximation, we can write the electron energy as
\begin{equation}
\teps(\bm r, \bm q) = \eps(\bm q) + V(\bm r) \;.
\end{equation}
The energy spectrum in real space is sketched in
Fig.~\ref{fig:energy}.

Before proceeding further, we need generalize the velocity
formula~\eqref{hall:v}, which is derived in the presence of an
electric field.  In our derivation the electric field enters through a
time-dependent vector potential $\bm A(t)$ so that we can avoid the
technical difficulty of calculating the matrix element of the position
operator.  However, the electric field may be also given by the
gradient of the electrostatic potential.  In both cases, the velocity
formula should stay the same because it should be gauge invariant.
Therefore, in general a scalar potential $V(\bm r)$ will induce a
transverse velocity of the following form
\begin{equation} \label{wave:v}
\frac{1}{\hbar} \bm\nabla V(\bm r) \times \bm\Omega(\bm q) \;.
\end{equation}
This generalization will be justified in Sec.~\ref{sec:gen}

Now consider a wave packet in the boundary region.  It will feel a
force $\bm\nabla V(\bm r)$ due to the presence of the confining
potential.  Consequently, according to Eq.~\eqref{wave:v} the electron
acquires a transverse velocity, whose direction is parallel with the
boundary (Fig.~\ref{fig:energy}).  This transverse velocity will lead
to a boundary current (of the dimension ``current density $\times$
width'' in 2D) given by
\begin{equation} \label{wave:current}
I = \frac{e}{\hbar} \int dx \int \frac{d\bm q}{(2\pi)^2}\, \dtot{V}{x} 
f(\eps(\bm q) + V) \Omega_z(\bm q) \;,
\end{equation}
where $x$ is in the direction perpendicular to the boundary, and the
integration range is taken from deep into the bulk to outside the
sample.  Recall that for a current $I$ flowing in a closed circuit
enclosing a sufficiently small area $A$, the circuit carries a
magnetic moment given by $I \cdot A$.  Therefore the magnetization
(magnetic moment per unit area) has the magnitude of the current $I$.
Integrating Eq.~\eqref{wave:current} by part, we obtain
\begin{equation} \label{wave:mcurrent}
M_f = \frac{1}{e}\int d\eps\, f(\eps) \sigma_{xy}(\eps) \;,
\end{equation}
where $\sigma_{xy}(\eps)$ is the zero-temperature Hall conductivity
for a system with Fermi energy $\eps$:
\begin{equation} \label{wave:sigma}
\sigma_{xy}(\eps) = \frac{e^2}{\hbar} \int\frac{d\bm q}{(2\pi)^d}
\Theta(\eps-\eps(\bm q)) \Omega_z(\bm q) \;.
\end{equation}
Since the boundary current corresponds to the global movement of the
wave packet center, we call this contribution the ``free current''
contribution, whereas the orbital moment are due to ``localized''
current.  The total magnetization thus is
\begin{equation} \label{wave:mag}
M_z = \int\frac{d\bm q}{(2\pi)^d} f(\bm q) m_z(\bm q) + \frac{1}{e}
\int d\eps\, f(\eps) \sigma_{xy}(\eps) \;.
\end{equation}
The orbital magnetization has two different contributions: one is from
the self-rotation of the wave packet, and the other is due to the
center-of-mass motion.

The reader may still have one question in mind: the above derivation
relies on the existence of a confining potential, which seems to
contradict the fact that the orbital magnetization is a bulk property.
This is a wrong assertion as the final expression,
Eq.~\eqref{wave:mag}, is given in terms of the bulk Bloch functions
and does not depend on the boundary condition.  Here, the boundary is
merely a tool to expose the ``free current'' contribution because in a
uniform system, the magnetization current always vanishes in the bulk.
Finally, in more rigorous approaches~\cite{xiao2005,shi2007} the
boundary is not needed and the derivation is based on a pure bulk
picture.  It is similar to the quantum Hall effect, which can be
understood in terms of either the bulk states~\cite{thouless1982} or
the edge states~\cite{halperin1982}.

\subsection{\label{sec:dipole}Dipole moment}

The finite size of the wave packet not only allows an orbital magnetic
moment, but also leads to the concept of the dipole moment associated
with an operator.

\begin{figure}[b]
\includegraphics[width=8cm]{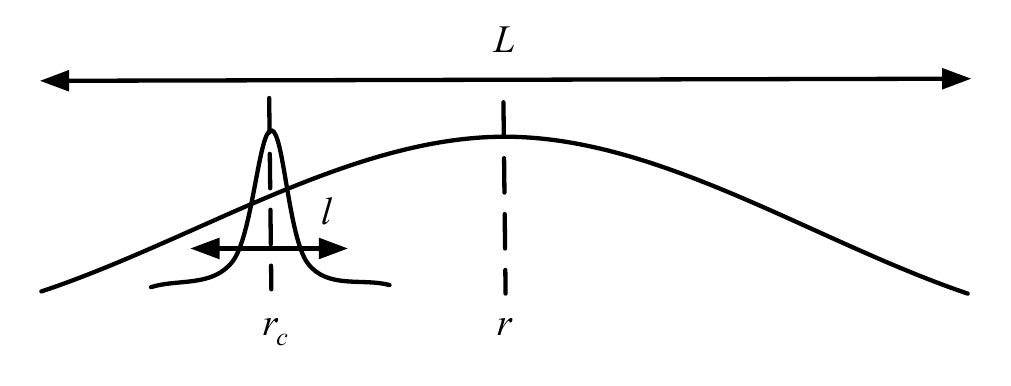}
\caption{\label{fig:sampling} Sampling function and a wave packet at $\bm
  r_c$.  The width $L$ of the sampling function is sufficiently small
  so that it can be treated as a $\delta$-function at the macroscopic
  level and is sufficiently big so that it contains a large number of
  wave packets of width $l$ inside its range.  Eq.~\eqref{wave:op} is
  indeed a microscopic average over the distance $L$ around the point
  $\bm r$.  See Section 6.6 in \citet{jackson-em} for an analogy
  in macroscopic electromagnetism.}
\end{figure}

The dipole moment appears naturally when we consider the thermodynamic
average of a physical quantity, with its operator denoted by $\hat{\cal O}$.
In the wave packet approach, it is given by
\begin{equation} \label{wave:op}
\cal O(\bm r) = \int\frac{d\bm r_c d\bm q_c}{(2\pi)^3} g(\bm r_c,
\bm q_c) \bracket{W|\hat{\cal O}\delta(\bm r-\hat{\bm r})|W} \;,
\end{equation}
where $g(\bm r, \bm q)$ is the distribution function,
$\bracket{W|\cdots|W}$ denotes the expectation in the wave packet
state, and $\delta(\bm r-\hat{\bm r})$ plays the role as a sampling
function, as shown in Fig.~\ref{fig:sampling}.  An intuitive way to
view Eq.~\eqref{wave:op} is to think the wave packets as small
molecules, then Eq.~\eqref{wave:op} is the quantum mechanical version
of the familiar coarse graining process which averages over the length
scale larger than the size of the wave packet.  A multi-pole expansion
can be carried out.  But for most purposes, the dipole term is enough.
Expand the $\delta$-function to first order of $\hat{\bm r}-\bm r_c$:
\begin{equation}
\delta(\bm r - \hat{\bm r}) = \delta(\bm r - \bm r_c) 
- (\hat{\bm r} - \bm r_c)
\cdot \bm\nabla \delta(\bm r - \bm r_c) \;.
\end{equation}
Inserting it into Eq.~\eqref{wave:op} yields
\begin{equation} \label{wave:d} \begin{split}
\cal O(\bm r) &= \int\frac{d\bm q}{(2\pi)^3}g(\bm r, \bm q)
\bracket{W|\hat{\cal O}|W}\bigr|_{\bm r_c = \bm r} \\
&\quad - \bm\nabla 
\cdot \int\frac{d\bm q}{(2\pi)^3} g(\bm r, \bm q) \bracket{W|
\hat{\cal O}(\hat{\bm r} - \bm r_c)|W}\bigr|_{\bm r_c = \bm r} \;.
\end{split} \end{equation}
The first term is what one would obtain if the wave packet is treated
as a point particle.  The second term is due to the finite size of the
wave packet.  We can see that the bracket in the second integral has
the form of a dipole of the operator $\cal O$, defined by
\begin{equation} \label{wave:dipole}
\bm P_{\hat{\cal O}} = \bracket{W|\hat{\cal O}(\hat{\bm r} - \bm r_c)|W} \;,
\end{equation}
The dipole moment of an observable is a general consequence of the
wave packet approach and must be included in the semiclassical
theory.  Its contribution appears only when the system is inhomogeneous.

\begin{figure}
\includegraphics[width=5.5cm]{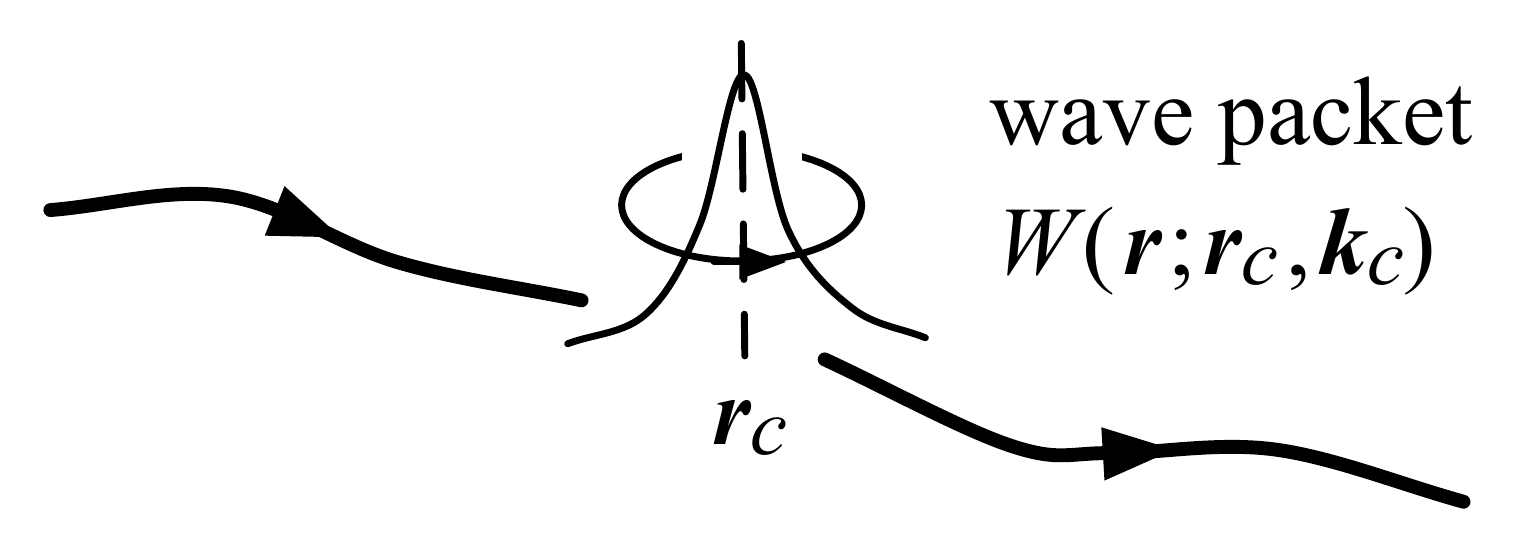}
\caption{\label{fig:wavepacket}The wave packet description of a charge
  carrier whose center is $(\bm r_c, \bm q_c)$.  A wave packet
  generally possesses two kinds of motion: the center of mass motion
  and the self-rotation around its center. From
  \onlinecite{xiao2006}.}
\end{figure}

In particular, we find:
\begin{enumerate}
\item If $\hat{\cal O} = e$, then $\bm P_e = 0$.  This is consistent
  with the fact that the charge center coincides with the mass center
  of the electron.
\item If $\hat{\cal O} = \hat{\bm v}$, one finds the expression for
  the \emph{local} current:
\begin{equation} \label{wave:local}
\bm j^L = \int\frac{d\bm q}{(2\pi)^3}g(\bm r, \bm q)\dot{\bm r} \\
+ \bm\nabla \times \int \frac{d\bm q}{(2\pi)^3} g(\bm r, \bm q) 
\bm m(\bm q) \;.
\end{equation}
We will explain the meaning of \emph{local} later.  Interestingly,
this is the second time we encounter the quantity $\bm m(\bm q)$, but
in an entirely different context.  The physical meaning of the second
term becomes transparent if we make reference to the self-rotation of
the wave packet.  The self-rotation can be thought as localized
circuit.  Therefore if the distribution is not uniform, those
localized circuit will contribute to the local current $\bm j^L$ (See
Fig.~\ref{fig:wavepacket}).
\item If $\hat{\cal O}$ is the spin operator $\hat{s}$, then
  Eq.~\eqref{wave:dipole} gives the spin dipole
\begin{equation}
\bm P_s = \bracket{u|s(i\dpar{}{\bm q} - \bm{\cal A}_{\bm q})|u} \;.
\end{equation}
It shows that in general the spin center and the mass center do not
coincide, which is usually due to the spin-orbit
interaction.  The time derivative of the spin dipole contributes to
the total spin current~\cite{culcer2004}.
\end{enumerate}

\subsection{Anomalous thermoelectric transport}

As an application of the above concepts, we consider the problem of
anomalous thermoelectric transport in ferromagnets, which refers to
the appearance of a Hall current driven by statistical forces, such as
the gradient of temperature and chemical potential~\cite{hallbook}.
Similar to the anomalous Hall effect, there are also intrinsic and
extrinsic contributions, and we will focus on the former.

A question immediately arises when one tries to formulate this problem.
Recall that in the presence of an electric field, the electron
acquires an anomalous velocity proportional to the Berry curvature,
which gives rise to a Hall current.  In this case, the driving force
is of mechanical nature: it exists on the microscopic level and can be
described by a perturbation to the Hamiltonian for the carriers.  On
the other hand, transport can be also driven by statistical force.
However, the statistical force manifests on the macroscopic level and
makes sense only through the statistical distribution of the carriers.
Since there is no force acting directly on individual particles, the
obvious cause for the Berry phase assisted transport is eliminated.
This conclusion would introduce a number of basis contradictions to
the standard transport theory.  First, a chemical potential gradient
would be distinct from the electric force, violating the basis for the
Einstein relation.  Second, a temperature gradient would not induce an
intrinsic charge Hall current, violating the Mott relation.  Finally,
it is also unclear whether the Onsager relation is satisfied or not.

It turns out the correct description of anomalous thermoelectric
transport in ferromagnets requires the knowledge of both the magnetic
moment and orbital magnetization.  Firstly, as we showed in
Eq.~\eqref{wave:local}, the local current is given by
\begin{equation}
\bm j^L = \int \frac{d\bm q}{(2\pi)^d} g(\bm r, \bm q) \dot{\bm r} 
+ \bm\nabla \times \int \frac{d\bm q}{(2\pi)^d}
f(\bm r, \bm q) \bm m(\bm q) \;,
\end{equation}
where in the second term we have replaced the distribution function
$g(\bm r, \bm q)$ with the \emph{local} Fermi-Dirac function $f(\bm r,
\bm q)$, which is sufficient for a first-order calculation.  Secondly,
in ferromagnetic systems, it is important to discount the contribution
from the magnetization current.  It was argued that the magnetization
current cannot be measured by conventional transport
experiments~\cite{cooper1997}.  Therefore the transport current is
given by
\begin{equation}
\bm j = \bm j^L - \bm\nabla \times M(\bm r) \;.
\end{equation}
Using Eq.~\eqref{wave:mag}, one finds
\begin{equation} \label{wave:jh}
\bm j = \int\frac{d\bm q}{(2\pi)^d} g(\bm r, \bm q) \dot{\bm r} 
-\frac{1}{e} \bm\nabla \times \int d\eps f(\eps)
\sigma^\text{AH}_z(\eps) \;.
\end{equation}

Equation~\eqref{wave:jh} is the most general expression for the
transport current.  We notice that the contribution from the orbital
magnetic moment $\bm m(\bm q)$ cancels out.  This agrees with the
intuitive picture we developed in Sec.~\ref{sec:om}, i.e., the orbital
moment is due to the self-rotation of the wave packet, therefore it is
localized and cannot contribute to transport (see
Fig.~\ref{fig:wavepacket}).

In the presence of a statistical force, there are two ways for a Hall
current to occur.  The asymmetric scattering will have an effect on
the distribution $g(\bm r, \bm q)$, which is obtained from the
Boltzmann equation~\cite{berger1972}.  This results in a transverse
current in the first term of Eq.~\eqref{wave:jh}.  In addition to
that, there is an intrinsic contribution comes from the orbital
magnetization, which is the second term of Eq.~\eqref{wave:jh}.  Note
that the spatial dependence enters through $T(\bm r)$ and $\mu(\bm r)$
in the distribution function.  It is straightforward to verify that
for the intrinsic contribution to the anomalous thermoelectric
transport, both the Einstein relation and Mott relation still
hold~\cite{xiao2006,onoda2008}.  Hence, the measurement of this type
of transport, such as the anomalous Nernst effect, can give further
insight of the intrinsic mechanism of the anomalous Hall effect.
Great experimental efforts have been put along this line.  The
intrinsic contribution has been verified in
CuCr$_{2}$Se$_{4-x}$Br$_{x}$~\cite{lee2004,lee2004a},
La$_{1-x}$Sr$_x$CoO$_3$~\cite{miyasato2007}, Nd$_2$Mo$_2$O$_7$ and
Sm$_2$Mo$_2$O$_7$~\cite{hanasaki2008}, Ga$_{1-x}$Mn$_x$~\cite{pu2008}.

Equation~\eqref{wave:jh} is not limited to transport driven by
statistical forces.  As we shall show later, at the microscopic level
the mechanical force generally has two effects: it can drive the
electron motion directly, and appears in the expression for $\dot{\bm
  r}$; it can also make the electron energy and the Berry curvature
spatially dependent, hence also manifest in the second term in
Eq.~\eqref{wave:jh}.  The latter provides another route for the Berry
phase to enter the transport problems in inhomogeneous situations,
which can be caused by a non-uniform distribution function, or a
spatially-dependent perturbation, or both.


\section{\label{sec:em}Electron dynamics in the presence of
  electromagnetic fields}

In last section we discussed the construction and general properties
of a wave packet.  Now we are set to study its dynamics under external
perturbations.  The most common perturbations to a crystal is the
electromagnetic fields.  The study of the electron dynamics under such
perturbations dates back to Bloch, Peierls, Jones, and Zener in the
early 1930s, and is continued by \citet{slater1949},
\citet{luttinger1951}, \citet{adams1952}, \citet{karplus1954},
\citet{kohn1957}, \citet{adams1959}, \citet{blount1962},
\citet{brown1967}, \citet{zak1977}, \citet{rammal1990},
\citet{wilkinson1996}.  In this section we present the semiclassical
theory based on the wave packet approach~\cite{chang1995,chang1996}.

\subsection{Equations of motion}

In the presence of electromagnetic fields, the Hamiltonian is given by
\begin{equation}
H = \frac{[\bm p + e\bm A(\bm r)]^2}{2m} + V(\bm r) - e\phi(\bm r) \;,
\end{equation}
where $V(\bm r)$ is the periodic lattice potential, and $\bm A(\bm r)$
and $\phi(\bm r)$ are the electromagnetic potentials.  If the length
scale of the perturbations is much larger than the spatial spread of
the wave packet, the approximate Hamiltonian that the wave packet
``feels'' may be obtained by linearizing the perturbations about the
wave packet center $\bm r_c$ as
\begin{align}
H &\approx H_c + \Delta H \;, \\
H_c &= \frac{[\bm p + e\bm A(\bm r_c)]^2}{2m} + V(\bm r) - e\phi(\bm
r_c) \;, \\
\Delta H &= \frac{e}{2m}\{\bm A(\bm r) - \bm A(\bm r_c), \bm p\}
- e\bm E \cdot (\bm r - \bm r_c) \;,
\end{align}
where $\{,\}$ is the anticommutator.  Naturally, we can then construct
the wave packet using the eigenstates of the local Hamiltonian $H_c$.
The effect of a uniform $\bm A(\bm r_c)$ is to add a phase to the
eigenstates of the unperturbed Hamiltonian.  Therefore the wave packet
can be written as
\begin{equation}
\ket{W(\bm k_c, \bm r_c)} = e^{-ie/\hbar \bm A(\bm r_c) \cdot \bm r}
\ket{W_0(\bm k_c, \bm r_c)} \;,
\end{equation}
where $\ket{W_0}$ is the wave packet constructed using the unperturbed
Bloch functions.

The wave packet dynamics can be obtained from the time-dependent
variational principles~\cite{kramer1981}.  The basic recipe is to
first obtain the Langragian from the following equation,
\begin{equation}
L = \bracket{W|i\hbar\dpar{}{t} - H|W}
\end{equation}
then obtain the equations of motion using the Euler equations.
Straightforward calculation shows that $\bracket{W|i\hbar\dpar{}{t}|W}
= e\dot{\bm A} \cdot \bm R_c - \hbar\dpar{}{t} \arg w(\bm k_c, t)$.
For the wave packet energy, we have $\bracket{W|\Delta H|W} = -\bm
m(\bm k) \cdot \bm B$.  This is expected as we already showed that the
wave packet carries an orbital magnetic moment $\bm m(\bm k)$ that
will couple to the magnetic field.  Using Eq.~\eqref{wave:rc}, we find
the Langragian is given by, up to some unimportant total
time-derivative terms (dropping the subscript $c$ on $\bm r_c$ and
$\bm k_c$),
\begin{equation}\label{em:Lag}
L = \hbar\bm k\cdot\dot{\bm r} - \eps_M(\bm k) + e\phi(\bm r)
- e \dot{\bm r} \cdot \bm
A(\bm r, t) + \hbar\dot{\bm k} \cdot \bm{\cal A_n(\bm k)} \;,
\end{equation}
where $\eps_M(\bm k) = \eps_0(\bm k) - \bm B \cdot \bm m(\bm k)$ with
$\eps_0(\bm k)$ being the unperturbed band energy.  The equations of
motion is
\begin{subequations} \label{em:EOM}
\begin{align}
\dot{\bm r} &= \dparh{\eps_M(\bm k)}{\bm k} - \dot{\bm k} \times
\bm\Omega(\bm k) \;, \label{em:EOM1} \\
\hbar \dot{\bm k} &= -e\bm E - e\dot{\bm r} \times \bm B \;.
\end{align}
\end{subequations}
Compared to the conventional equations of motion for Bloch
electrons~\cite{ashcroft-CM}, there are two differences: (1) The
electron energy is modified by the orbital magnetic moment; (2) the
electron velocity gains an extra velocity term proportional to the
Berry curvature.  As we can see, in the case of only an electric
field, Eq.~\eqref{em:EOM1} reduces to the anomalous velocity formula
\eqref{hall:v} we derived before.

\subsection{Modified density of states}

The Berry curvature not only modifies the electron dynamics, but also
has a profound effect on the electron density of states in the phase
space~\cite{xiao2005}.

Recall that in solid state physics, the expectation value of an
observable, in the Bloch representation, is given by
\begin{equation} \label{em:sum}
\sum_{n\bm k} f_{n\bm k} \bracket{\psi_{nk}|\hat{O}|\psi_{nk}} \;,
\end{equation}
where $f_{n\bm k}$ is the distribution function.  In the semiclassical
limit, the sum is converted to an integral in the $\bm k$-space,
\begin{equation}
\sum_{\bm k} \to \frac{1}{V} \int \frac{d\bm k}{(2\pi)^d} \;,
\end{equation}
where $V$ is the volume, and $(2\pi)^d$ is the density of states,
i.e., number of states per unit $\bm k$-volume.  From a classical
point of view, the constant density of states is guaranteed by the
Liouvilles theorem, which states that the volume element is a
conserved quantity during the time evolution of the
system.~\footnote{The actual value of this constant volume for a
  quantum state, however, can be determined only from the quantization
  conditions in quantum mechanics.}  However, as we shall show below,
this is no longer the case for the Berry-phase modified dynamics.

The time evolution of a volume element $\Delta V = \Delta\bm r
\Delta\bm k$ is given by
\begin{equation}
\frac{1}{\Delta V}\dpar{\Delta V}{t} = \bm\nabla_{\bm r} \cdot
\dot{\bm r} + \bm\nabla_{\bm k} \cdot \dot{\bm k} \;.
\end{equation}
Insert the equations of motion~\eqref{em:EOM} into the above
equation.  After some algebra, we find
\begin{equation}
\Delta V = \frac{V_0}{(1 + (e/\hbar)\bm B\cdot\bm \Omega)} \;.
\end{equation}
The fact that the Berry curvature is generally $\bm k$-dependent and
the magnetic fields is $\bm r$-dependent implies that the phase-space
volume $\Delta V$ changes during time evolution of the state variables
$(\bm r, \bm k)$.

Although the phase space volume is no longer conserved, it is a local
function of the state variables and has nothing to do with the
history of time evolution.  We can thus introduce a modified density
of states
\begin{equation} \label{em:dos}
D(\bm r, \bm k) = \frac{1}{(2\pi)^d}(1 + \frac{e}{\hbar}\bm
B\cdot\bm\Omega)
\end{equation}
such that the number of states in the volume element, $D_n(\bm r, \bm
k)\Delta V$, remains constant in time.  Therefore, the correct
semiclassical limit of the sum in Eq.~\eqref{em:sum} is
\begin{equation}
O(\bm R) = \int d\bm k\,D(\bm r, \bm k) \bracket{O\delta(
\hat{\bm r}- \bm R)}_W \;,
\end{equation}
where $\bracket{\cdots}_W$ is the expectation value in a wave
packet, which could includes the dipole contribution due to the finite
size of the wave packet (See Sec.~\ref{wave:dipole}).  In a uniform
system it is simply given by
\begin{equation}
O = \int d\bm k\, D(\bm k) f(\bm k) O(\bm k)
\end{equation}

We emphasize that although the density of states is no longer a
constant, the dynamics itself is still Hamiltonian.  The modification
comes from the fact that the dynamical variables, $\bm r$ and $\bm k$,
are no longer canonical variables, and the density of states can be
regarded as the phase space
measure~\cite{duval2006,xiao2006a,bliokh2006,duval2006a}.  A more
profound reason for this modification has its quantum
mechanical origin in non-commutative quantum mechanics, discussed in
Sec.~\ref{sec:qz}.

In the following we discuss two direct applications of the modified
density of states in metals and in insulators.

\subsubsection{\label{sec:fermi}Fermi volume}

We show that the Fermi volume can be changed linearly by a magnetic
field when the Berry curvature is nonzero.  Assume zero temperature,
the electron density is given by
\begin{equation} \label{em:density}
n_e = \int \frac{d\bm k}{(2\pi)^d}\Bigl(1 + \frac{e}{\hbar}\bm B
\cdot \bm\Omega\Bigr) \Theta(\eps_F - \eps)
\end{equation}
We work in the canonical ensemble by requiring the electron number
fixed, therefore, to first order of $\bm B$, the Fermi energy must be
changed by
\begin{equation} \label{em:ef}
\delta V_F = -\int \frac{d\bm k}{(2\pi)^d}
\frac{e}{\hbar} \bm B\cdot \bm \Omega \;.
\end{equation}

It is particularly interesting to look at insulators, where the
integration is limited to the Brillouin zone.  Then the electron must
populate a higher band if $\int_\text{BZ} d\bm k\,\bm B\cdot\bm\Omega$
is negative.  When this quantity is positive, holes must appear at the
top of the valance bands.  Discontinuous behavior of physical
properties in a magnetic field is therefore expected for band
insulators with a nonzero integral of the Berry curvatures (Chern
numbers).

\subsubsection{Streda Formula}

In the context of the quantum Hall effect, \citet{streda1982} derived
a formula relating the Hall conductivity to the field derivative of
the electron density at a fixed chemical potential
\begin{equation}
\sigma_{xy} = -e\Bigl(\dpar{n_e}{B_z}\Bigr)_\mu \;.
\end{equation}
There is a simple justification of this relation by a thermodynamic
argument by considering the following adiabatic process in two
dimensions. A time dependent magnetic flux generates an electric field
with an emf around the boundary of some region; and the Hall current
leads to a net flow of electrons across the boundary and thus a change
of electron density inside.   Note that this argument is valid only
for insulators because in metals the adiabaticity would break down.  Using
Eq.~\eqref{em:density} for an insulator, we obtain, in 2D
\begin{equation}
\sigma_{xy} = - \frac{e^2}{\hbar}\int_\text{BZ} \frac{d\bm
  k}{(2\pi)^2} \sigma_{xy} \;.
\end{equation}
This is what \citet{thouless1982} obtained using the Kubo formula.
The fact that the quantum Hall conductivity can be derived using the
modified density of states further confirms the necessity to introduce
this concept.

\subsection{Orbital magnetization: Revisit}

We have discussed the orbital magnetization using a rather pictorial
derivation in Sec.~\ref{sec:om}.  Here we derive the formula again by
using the field-dependent density of states~\eqref{em:dos}.

The equilibrium magnetization density can be obtained from the grand
canonical potential, which, within first order in the magnetic field,
may be written as
\begin{equation} \label{grand}
\begin{split}
F &= -\frac{1}{\beta} \sum_{\bm k}
\log(1 + e^{-\beta(\eps_M - \mu)}) \\
&= -\frac{1}{\beta} \int\frac{d\bm k}{(2\pi)^d} (1 + \frac{e}{\hbar}\bm B \cdot
\bm\Omega) \log(1 + e^{-\beta(\eps_M - \mu)}) \;,
\end{split}
\end{equation}
where the electron energy $\eps_M = \eps(\bm k) - \bm m(\bm k) \cdot
\bm B$ includes a correction due to the orbital magnetic moment $\bm
m(\bm k)$.  The magnetization is then the field derivative at fixed
temperature and chemical potential, $\bm M = -(\partial F/\partial \bm
B)_{\mu,T}$, with the result
\begin{equation} \label{magnetization}
\begin{split}
\bm M(\bm r) &= \int\frac{d\bm k}{(2\pi)^d}\, f(\bm k) \bm m(\bm k)
\\ &\quad + \frac{1}{\beta} \int\frac{d\bm k}{(2\pi)^d}\, \frac{e}{\hbar}
\bm\Omega(\bm k) \log(1 + e^{-\beta(\eps-\mu)}) \;.
\end{split}
\end{equation}
Integration by parts of the second term will give us the exact formula
obtained in Eq.~\eqref{wave:mag}.  We have thus derived a general
expression for the equilibrium orbital magnetization density, valid at
zero magnetic field but at arbitrary temperatures.  From this
derivation we can clearly see that the orbital magnetization is indeed
a bulk property.  The center-of-mass contribution identified before
comes from the Berry-phase correction to the electron density of
states.

Following the discussions on band insulators in our first example in
Sec.~\ref{sec:fermi}, there will be a discontinuity of the orbital
magnetization if the integral of the Berry curvature over the
Brillouin zone, or the anomalous Hall conductivity, is non-zero and
quantized. Depending on the direction of the field, the chemical
potential $\mu_0$ in the above formula should be taken at the top of
the valence bands or the bottom of the conduction bands.  The size of
the discontinuity is given by the quantized anomalous Hall
conductivity times $E_g/e$, where $E_g$ is the energy gap.

Similar formula for insulators with zero Chern number has been
obtained by~\citet{thonhauser2005,ceresoli2006} using the Wannier
function approach, and has been numerically confirmed for a
tight-binding model.  Recently, \citet{shi2007} provided a full
quantum mechanical derivation of the formula, and showed that it is
valid in the presence of electron-electron interaction, provided the
one-electron energies and wave functions are calculated
self-consistently within the framework of the exact current and
spin-density functional theory~\cite{vignale1988}.

The appearance of the Hall conductivity is not a coincidence.  Let us
consider an insulator.  The free energy is given by
\begin{equation}
dF = -MdB - nd\mu - SdT \;.
\end{equation}
Using the Maxwell relation, we have
\begin{equation} \label{em:maxwell}
\sigma_H = -e\Bigl(\dpar{n}{B}\Bigr)_{\mu, T} = -e \Bigl(\dpar{M}{\mu}
\Bigr)_{B,T} \;.
\end{equation}
On the hand, the zero-temperature formula of the magnetization for an
insulator is given by
\begin{equation}
\bm M = \int_\text{BZ} \frac{d\bm k}{(2\pi)^3} \{ \bm m(\bm k) + \frac{e}{\hbar}
(\mu - \eps)\bm \Omega\} \;.
\end{equation}
Inserting it into Eq.~\eqref{em:maxwell} gives us once again the
quantized Hall conductivity.

\subsection{Magnetotransport}

The equations of motion~\eqref{em:EOM} and the density of
states~\eqref{em:dos} gives us a complete description of the
electron dynamics in the presence of electromagnetic fields.  In this
subsection we apply these results to the problem of magnetotransport.
For simple notations, we set $e = \hbar = 1$ and introduce the
shorthand $[d\bm k] = d\bm k/(2\pi)^d$.

\subsubsection{Cyclotron period}

Semiclassical motion of a Bloch electron in a uniform magnetic field
is important to understand various magneto-effects in solids.  In this
case, the equations of motion reduce to
\begin{subequations}  \label{eom-B}
\begin{align}
\bar{D}(\bm k) \dot{\bm r} &= \bm v + (\bm v \cdot \bm \Omega) \bm B \;, \\
\bar{D}(\bm k) \dot{\bm k} &= -\bm v \times \bm B \;, \label{em:EOM3}
\end{align}
\end{subequations}
where $\bar{D}(\bm k) = D(\bm k)/(2\pi)^d = 1 + (e/\hbar)\bm B\cdot\bm
\Omega$.

We assume the field is along the $z$-axis.  From the second equation
of \eqref{eom-B} we can see that motion in $\bm k$-space is confined
in the $xy$-plane and is completely determined once the energy $\eps$
and the $z$ component of the wave vector $k_z$ is given.  Let us
calculate the period of the cyclotron motion.  The time for the wave
vector to move from $\bm k_1$ to $\bm k_2$ is
\begin{equation}
t_2 - t_1 = \int^{t_2}_{t_1} \rmd t = \int^{\bm k_2}_{\bm k_1}
\frac{\rmd k}{|\dot{\bm k}|} \;.
\end{equation}
From the equations of motion \eqref{eom-B} we have
\begin{equation}
|\dot{\bm k}| = \frac{B|\bm v_\perp|}{\bar{D}(\bm k)} =
\frac{B|(\partial\eps/\partial\bm k)_\perp|}{\hbar \bar{D}(\bm k)} \;.
\end{equation}
On the other hand, the quantity $(\partial\eps/\partial\bm k)_\perp$
can be written as $\Delta \eps/\Delta \bm k$, where $\Delta\bm k$
denotes the vector in the plane connecting points on neighboring
orbits of energy $\eps$ and $\eps+\Delta\eps$, respectively. Then
\begin{equation}
t_2 - t_1 = \frac{\hbar}{B} \int^{\bm k_2}_{\bm k_1}
\frac{\bar{D}(\bm k) \Delta\bm k\,\rmd k}{\Delta \eps} \;.
\end{equation}
Introducing the 2D electron density for given $\eps$ and $k_z$
\begin{equation} \label{em:2DD}
n_2(\eps,k_z) = \iint_{k_z, \eps(\bm k) < \eps}
\frac{\bar{D}(\bm k) \,\rmd k_x\rmd k_y}{(2\pi)^2} \;,
\end{equation}
the period of a cyclotron motion can be written as
\begin{equation} \label{em:T}
T = (2\pi)^2 \frac{\hbar}{B} \dpar{n_2(\eps, k_z)}{\eps} \;.
\end{equation}
We thus recovered the usual expression for the cyclotron period, with
the 2D electron density, Eq.~\eqref{em:2DD}, defined with the modified
density of states.

In addition, we note that there is a velocity term proportional to
$\bm B$ in Eq.~\eqref{eom-B}, which seems to suggest there will be a
current along the field direction.  We show that after averaging over
the distribution function, this current is actually zero.  The current
along $\bm B$ is given by
\begin{equation}
\begin{split}
\bm j_B &= -e \bm B \int[\rmd\bm k] f\bm v \cdot \bm \Omega \\
&= -\frac{e}{\hbar} \bm B \int[\rmd\bm k]\bm \nabla_{\bm k} F
\cdot \bm \Omega \\
&= -\frac{e}{\hbar} \bm B \Bigl( \int[\rmd\bm k] \bm\nabla_{\bm k}
(F\bm\Omega) - \int[\rmd\bm k] F\bm\nabla_{\bm k} \cdot\bm\Omega\Bigr) \;,
\end{split}
\end{equation}
where $F(\eps) = -\int^\infty_\eps f(\eps')\rmd\eps'$ and $f(\eps) =
\partial F/\partial \eps$.  The first term vanishes
\footnote{For any periodic function $F(\bm k)$ with the periodicity of
a reciprocal Bravais lattice, the following identity holds for
integrals taken over a Brillouin zone, $\int_\text{BZ} \rmd\bm k\, \bm
\nabla_{\bm k} F(\bm k) = 0$.  To see this, consider $I(\bm k') =
\int\rmd\bm k\,F(\bm k + \bm k')$.  Because $F(\bm k)$ is periodic in
$\bm k$, $I(\bm k')$ should not depend on $\bm k'$.  Therefore, $\bm
\nabla_{\bm k'} I(\bm k') = \int\rmd\bm k, \bm \nabla_{\bm k'} F(\bm k
+ \bm k') = \int\rmd\bm k \bm \nabla_{\bm k} F(\bm k + \bm k') = 0$.
Setting $\bm k' = 0$ gives the desired expression.  This is also true if
$\bm F(\bm k)$ is a vector function.}
and if there is no magnetic monopole in $\bm k$-space, the second term
also vanishes.  In above calculation we did not consider the change of
the Fermi surface.  Since it always comes in the form $(\partial
f/\partial \mu) \delta \mu = -(\partial f/\partial \eps) \delta \mu$
we can use the same technique to prove that the corresponding current
also vanishes.

\subsubsection{The high field limit}

We now consider the magnetotransport at the so-called high field
limit, i.e., $\omega_c\tau \gg 1$, where $\omega_c = 2\pi/T$ is the
cyclotron frequency and $\tau$ is the relaxation time.  We consider
configuration where the electric and magnetic fields are perpendicular
to each other, i.e., $\bm E = E\hat{\bm x}$, $\bm B = B \hat{\bm z}$
and $\bm E \cdot \bm B = 0$.

In the high fied limit, $\omega_c\tau \gg 1$, the electron can finish
several turns between two successive collisions.  We can then assume
all orbits are closed.  According to the theorem of adiabatic
drifting~\cite{niu2001}, an originally closed orbit remains closed for
weak perturbations, so that
\begin{equation}
0 = \bracket{\dot{\bm k}} = \bm E + \bracket{\dot{\bm r}} \times \bm B \;.
\end{equation}
Or
\begin{equation}
\bracket{\dot{\bm r}}_\perp = \frac{\bm E \times \bm B}{B^2} \;.
\end{equation}
The Hall current is simply the sum over $\bracket{\dot{\bm r}}_\perp$
of occupied states:
\begin{equation} \begin{split}
\bm j_H &= -e \frac{\bm E \times \bm B}{B^2} \int[\rmd\bm k] f(\bm k)
(1 + \bm B \cdot \bm \Omega) \\
&= -e\frac{\bm E \times \bm B}{B^2}
\int[\rmd\bm k] f(\bm k)\bar{D}(\bm k) \;.
\end{split} \end{equation}
Therefore in the high field limit we reach the remarkable conclusion:
the total current in crossed electric and magnetic fields is the Hall
current as if calculated from free electron model
\begin{equation}
\bm j = -e \frac{\bm E \times \bm B}{B^2} n \;,
\end{equation}
and it has no dependence on the relaxation time $\tau$.  This result
ensures that even in the presence of anomalous Hall effect, the high
field Hall current gives the ``real'' electron density.


Let us now consider the hole-like band.  The Hall current is obtained
by substracting the contribution of holes from that of the filled
band, which is given by $-e\bm E \times \int[d\bm k]\bm\Omega$.
Therefore
\begin{equation}
\bm j^\text{hole} = e\frac{\bm E \times \bm B}{B^2} \int[\rmd\bm k]
\bar{D}(\bm k) [1 - f(\bm k)] - e\bm E \times \int[\rmd\bm k] \bm \Omega \;.
\end{equation}
So for the hole-like band, there is an additional term in the current
expression proportional to the Chern number (the second integral) of
the band.

\subsubsection{The Low Field Limit}

Next we consider the magnetotransport at the low field limit, i.e.,
$\omega_c\tau \ll 1$.  In particular, we show that the Berry phase
induce a linear magnetoresistance.  By solving the Boltzmann equation,
one finds that the diagonal element of the conductivity is given by
\begin{equation} \label{em:sigmaxx}
\sigma_{xx} = -e^2\int[\rmd\bm k] \tau\dpar{f_0}{\eps}
\frac{v_x^2}{\bar{D}(\bm k)} \;.
\end{equation}
This is just the zeroth order expansion based on $\omega_c\tau$.
There are four places in this expression depending on $B$.  (1) There
is an explicit $B$-dependence in $\bar{D}(\bm k)$.  (2) The electron
velocity $v_x$ is modified by the orbital magnetic moment:
\begin{equation}
v_x = \frac{1}{\hbar} \dpar{(\eps_0 - m_zB)}{k_x} = v_x^{(0)}
- \frac{1}{\hbar} \dpar{m_z}{k_x} B \;.
\end{equation}
(3) There is also a modification to the Fermi energy, given by
Eq.~\eqref{em:ef}.  (4) The relaxation time $\tau$ can also
depend on $B$.  In the presence of the Berry curvature, the collision
term in the Boltzmann equation is given by
\begin{equation}
\dpar{f}{t}\Bigr|_\text{coll} = - \int[\rmd\bm k'] \bar{D}(\bm k')
W_{\bm k\bm k'}[f(\bm k) - f(\bm k')] \;,
\end{equation}
where $W_{\bm k\bm k'}$ is the transition probability from $\bm k'$ to
$\bm k$ state.  In the relaxation-time approximation we make the
assumption that a characteristic relaxation time exists so that
\begin{equation}
\frac{f-f_0}{\tau} = \bar{D}(\bm k) \int[\rmd\bm k']
\frac{\bar{D}(\bm k')}{\bar{D}(\bm k)} W_{\bm k\bm k'}[f(\bm k)
- f(\bm k')] \;.
\end{equation}
If we assume $\bm\Omega(\bm k)$ is smooth and $W_{\bm k\bm k'}$ is
localized, the relaxation time can be approximated by
\begin{equation}
\tau = \frac{\tau_0}{\bar{D}(\bm k)} \approx \tau_0
\Bigl(1 - \frac{e}{\hbar} \bm B \cdot \bm \Omega \Bigr) \;.
\end{equation}
More generally, we can always expand the relaxation time to first
order of $(e/\hbar) \bm B \cdot \bm \Omega$,
\begin{equation}
\tau = \tau_0 + \tau_1 \frac{e}{\hbar} \bm B \cdot \bm \Omega \;,
\end{equation}
where $\tau_1$ should be regarded as a fitting parameter within this
theory.

Expand expression~\eqref{em:sigmaxx} to first order of $B$, and
take the spherical band approximation, we obtain
\begin{equation} \begin{split}
\sigma_{xx}^{(1)} &= e^2\tau_0B \Bigl[\int[\rmd\bm k] \dpar{f_0}{\eps}
\Bigl(\frac{2e\Omega_z}{\hbar}v_x^2
+ \frac{2}{\hbar} \dpar{m_z}{k_x}v_x \Bigr) \\
&\quad -
\frac{e}{\hbar} \bracket{(\mathbf M)^{-1}_{xx}}_{k_F}
\int[\rmd\bm k] f \Omega_z \Bigr]\;,
\end{split} \end{equation}
where $\mathbf M$ is the effect mass tensor.  The zero-field
conductivity takes the usual form
\begin{equation}
\sigma_{xx}^{(0)} = -e^2\tau_0\int[\rmd\bm k]\dpar{f_0}{\eps}v_x^2 \;.
\end{equation}
The ratio $-\sigma_{xx}^{(1)}/\sigma_{xx}^{(0)}$ will then give us the
linear magnetoresistance.


\section{\label{sec:gen}Electron dynamics in the presence of general perturbations}

In this section we present the general theory of electron dynamics in
slowly perturbed crystals~\cite{sundaram1999,shindou2005,panati2003}.
As expected, the Berry curvature enters into the equations of motion
and modifies the density of states.  The difference is that one needs
to introduce the Berry curvature defined in the extended parameter
space $(\bm r, \bm q, t)$.  Two physical applications are considered:
electron dynamics in deformed crystals, and adiabatic current induced
by inhomogeneity.

\subsection{Equations of motion}

We consider a slowly perturbed crystal whose Hamiltonian can be
expressed in the following form
\begin{equation}
H[\bm r, \bm p; \beta_1(\bm r, t), ... \beta_g(\bm r, t)] \;,
\end{equation}
where $\{\beta_i(\bm r, t)\}$ are the modulation function
characterizing the perturbations.  They may represent either gauge
potentials of electromagnetic fields, atomic displacements, charge or
spin density waves, helical magnetic structures, or compositional
gradients.  Following the same procedure as we have done in last
section, we expand the Hamiltonian around the wave packet center, and
obtain
\begin{align}
H &= H_c + \Delta H \;, \\
H_c &= H[\bm r, \bm p; \{\beta_i(\bm r_c, t)\}] \;, \\
\Delta H &= \sum_i \bm\nabla_{\bm r_c} \beta_i(\bm r_c, t) \cdot
\{(\bm r - \bm r_c), \dpar{H}{\beta_i}\} \;.
\end{align}
Since the local Hamiltonian $H_c$ maintains periodicity of the
unperturbed crystal, its eigenstates take the Bloch form
\begin{equation}
H_c(\bm r_c, t) \ket{\psi_{\bm q}(\bm r_c, t)}
= \eps_c(\bm r_c, \bm q, t) \ket{\psi_{\bm q}(\bm r_c, t)} \;,
\end{equation}
where $\bm q$ is the Bloch wave vector and $\eps_c(\bm r_c, \bm q, t)$
is the band energy.  Here we have dropped the band index $n$ for
simple notations.

Following the discussion in Sec.~\ref{sec:zak}, we switch to the Bloch
Hamiltonian $H_c(\bm q, \bm r_c, t) = e^{-i\bm q \cdot\bm r} H_c(\bm
r_c, t) e^{i\bm q\cdot \bm r}$, whose eigenstate is the periodic part
of the Bloch function, $\ket{u(\bm q, \bm r_c, t)} = e^{-i\bm q\cdot
  \bm r}\ket{\psi(\bm q, \bm r_c, t)}$.  The Berry vector potentials
can be defined for each of the coordinates of the parameter space
$(\bm q, \bm r_c, t)$; for example,
\begin{equation}
\mathcal A_t = \bracket{u|i\partial_t|u} \;.
\end{equation}

After constructing the wave packet using the local Bloch functions
$\ket{\psi_{\bm q}(\bm r_c, t)}$, one can apply the time-dependent
variational principle to find the Langragian governing the dynamics of
the wave packet:
\begin{equation} \label{gen:L}
L = -\eps + \bm q_c \cdot \dot{\bm r}_c + \dot{\bm q}_c \cdot
\bm{\mathcal A}_{\bm q} + \dot{\bm r}_c \cdot \bm{\mathcal A}_{\bm r}
+ \mathcal A_t \;,
\end{equation}
Note that the wave packet energy $\eps = \eps_c + \Delta\eps$ has a
correction $\Delta\eps$ from $\Delta H$,
\begin{equation} \label{gen:deltae}
\Delta\eps = \bracket{W|\Delta H|W} = -\Im \bracket{\dpar{u}{\bm r_c}|
\cdot (\eps_c - H_c)|\dpar{u}{\bm q}} \;.
\end{equation}
From the Lagrangian~\eqref{gen:L} we obtain the following equations
of equation:
\begin{subequations} \label{gen:eom} \begin{align}
\dot{\bm r}_c &= \dpar{\eps}{\bm q_c} - (\tensor{\Omega}_{\bm q\bm r}
\cdot \dot{\bm r}_c + \tensor{\Omega}_{\bm q\bm q} \cdot \dot{\bm
  q}_c) - \bm\Omega_{\bm qt} \;, \\
\dot{\bm q}_c &= -\dpar{\eps}{\bm r_c} + (\tensor{\Omega}_{\bm r\bm r}
\cdot \dot{\bm r}_c + \tensor{\Omega}_{\bm x\bm q} \cdot \dot{\bm
  q}_c) + \bm\Omega_{\bm rt} \;
\end{align} \end{subequations}
where $\Omega$'s are the Berry curvatures.  For example,
\begin{equation}
(\tensor{\Omega}_{\bm q\bm r})_{\alpha\beta} = \partial_{q_\alpha}
\mathcal A_{r_\beta} - \partial_{r_\beta} \mathcal A_{k_\alpha} \;.
\end{equation}
In the following we will also drop the subscript $c$ on $\bm r_c$ and
$\bm q_c$.

The form of the equations of motion is quite symmetrical with respect
to $\bm r$ and $\bm q$, and there are Berry curvatures between every
pair of phase space variables plus time.  The term $\bm\Omega_{\bm
  qt}$ was identified as the adiabatic velocity vector in
Sec.~\ref{sec:pump}.  In fact, if the perturbation is uniform in space
(has the same period as the unperturbed crystal) and only varies in
time, all the spatial derivatives vanish; we obtain
\begin{equation} \label{gen:eom1}
\dot{\bm r} = \dpar{\eps}{\bm q} - \bm\Omega_{\bm qt} \;, \qquad
\dot{\bm q} = 0 \;.
\end{equation}
The first equation is the velocity formula~\eqref{pump:v} obtained in
Sec.~\ref{sec:pump}.  The term $\tensor{\Omega}_{\bm q\bm q}$ was
identified as the Hall conductivity tensor.  In the presence of
electromagnetic perturbations, we have
\begin{equation}
H = H_0[\bm q + e\bm A(\bm r)] - e\phi(\bm r, t) \;.
\end{equation}
Hence the local basis can be written as $\ket{u(\bm r_c, \bm q)} =
\ket{u(\bm k)}$, where $\bm k = \bm q + e\bm A(\bm r)$.  One can
verify that by using the chain rule $\partial_{q_\alpha} =
\partial_{k_\alpha}$ and $\partial_{r_\alpha} = (\partial_{r_\alpha}
A_\beta) \partial_{k_\beta}$, $\Delta\eps$ given in
Eq.~\eqref{gen:deltae} becomes $-\bm m(\bm k) \cdot \bm B$, and the
equations of motion~\eqref{gen:eom} reduces to Eq.~\eqref{em:EOM}.
The physics of quantum adiabatic transport and the quantum and
anomalous Hall effect can be described from a unified point of view.
The Berry curvature $\bm\Omega_{\bm rt}$ plays a role like the electric
force.  The antisymmetric tensor $\tensor{\Omega}_{\bm r\bm r}$ is
realized in terms of the magnetic field in the Lorenz force and is
also seen in the singular form ($\delta$-function like distribution)
of dislocations in a deformed crystal~\cite{bird1988}.  Finally, the
Berry curvature between $\bm r$ and $\bm q$ can be realized in
deformed crystals as a quantity proportional to the strain and the
electronic mass renormalization in the crystal~\cite{sundaram1999}.

\subsection{Modified density of states}

The electron density of states is also modified by the Berry
curvature.  Let us consider the time-independent case.  To better
appreciate the origin of this modification, we introduce the phase
space coordinates $\bm\xi = (\bm r, \bm q)$.  The equations of motion
can be written as
\begin{equation} \label{gen:hh}
\Gamma_{\alpha\beta} \dot{\xi}_\beta = \nabla_{\xi_\alpha} \eps \;,
\end{equation}
where $\tensor{\Gamma} = \tensor{\Omega} - \tensor{J}$ is an
antisymmetric matrix with
\begin{equation}
\tensor{\Omega} = \begin{pmatrix} \tensor{\Omega}_{\bm r\bm r} &
\tensor{\Omega}_{\bm r\bm q} \\ \tensor{\Omega}_{\bm q\bm r} &
\tensor{\Omega}_{\bm q\bm q} \end{pmatrix} \;, \qquad
\tensor{J} = \begin{pmatrix} 0 & \tensor{I} \\ -\tensor{I} & 0
\end{pmatrix} \;.
\end{equation}
According to standard theory of Hamiltonian
dynamics~\cite{arnold1978}, the density of states, which is
proportional to the phase space measure is given by
\begin{equation} \label{gen:DOS}
D(\bm r, \bm q) = \frac{1}{(2\pi)^d}
\sqrt{\det(\tensor{\Omega} - \tensor{J})} \;.
\end{equation}
One can show that in the time-dependent case $D(\bm r, \bm q)$ has the
same form.

Let us consider the following situations.  (i) If the perturbation is
electromagnetic field, by the variable substitution $\bm k = \bm q +
e\bm A(\bm r)$, Eq.~\eqref{gen:DOS} reduces to Eq.~\eqref{em:dos}.
(ii) In many situations we are aiming at a first-order calculation in
the spatial gradient.  In this case, the density of states given by
\begin{equation} \label{gen:dos2}
D = \frac{1}{(2\pi)^d} (1 + \Tr \tensor{\Omega}_{\bm q\bm r}) \;.
\end{equation}

Note that if the Berry curvature vanishes, Eq.~\eqref{gen:hh} becomes
the canonical equations of motion for Hamiltonian dynamics, and $\bm
r$ and $\bm q$ are called canonical variables.  The density of states
is a constant in this case.  The presence of the Berry curvature
renders the variables non-canonical and, as a consequence, modifies
the density of states.  The non-canonical variables is a common
feature of Berry-phase participated dynamics~\cite{littlejohn1991}.

\begin{figure}
\includegraphics[width=8cm]{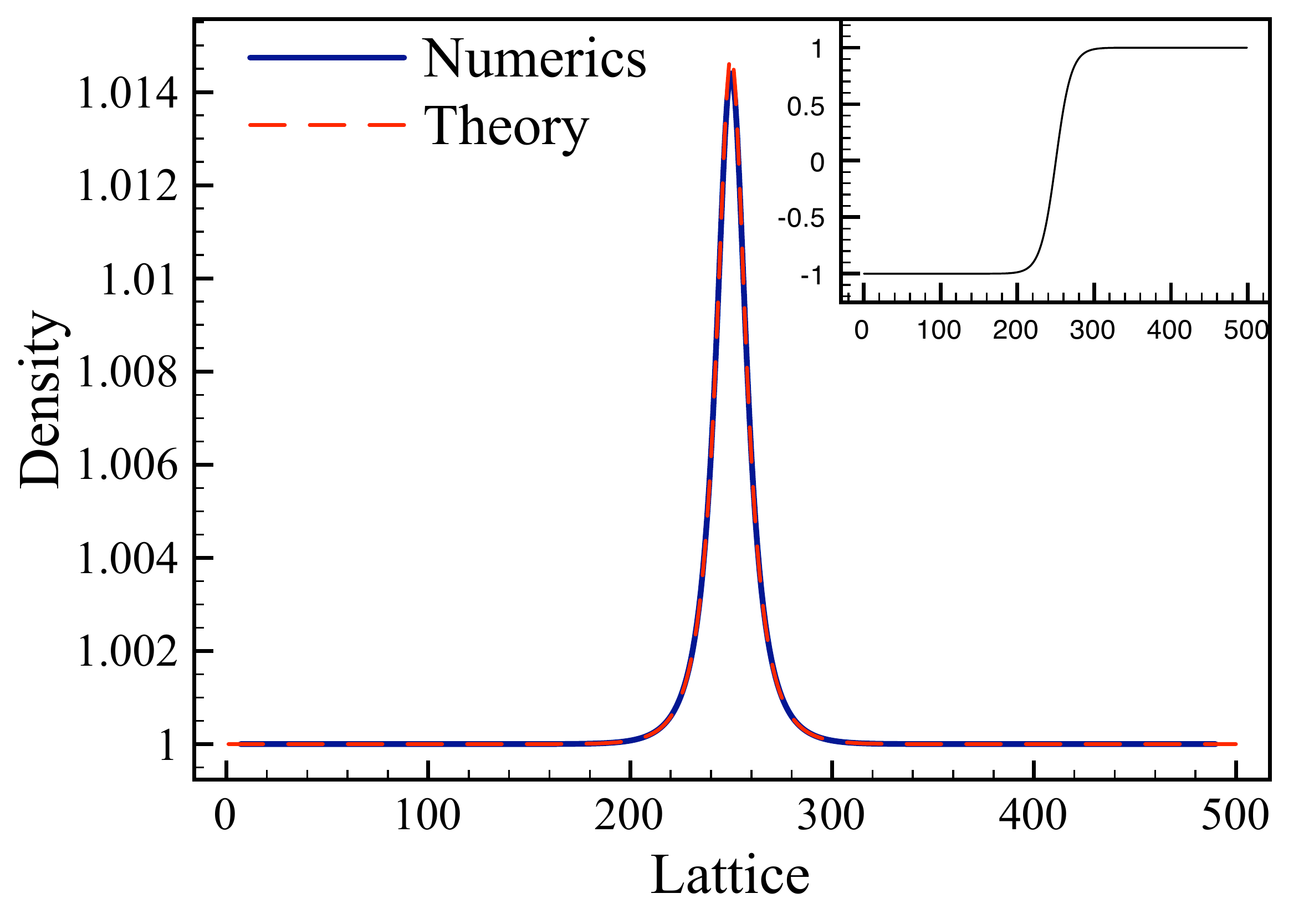}
\caption{\label{fig:dimer} (color online) Electron density of the Rice-Mele model
  with a spatial varying dimerization parameter.  The parameters we
  used are $\Delta = 0.5$, $t = 2$, and $\delta = \tanh(0.02x)$.
  Inset: The profile of $\delta(x)$.  From~\citet{xiao2009}.}
\end{figure}

To demonstrate the modified density of states, we again consider the
Rice-Mele model discussed in Sec.~\ref{sec:rice}.  This time we
introduce the spatial dependence by letting the dimerization parameter
$\delta(x)$ vary in space.  Using Eq.~\eqref{berry:R1R2} we find
\begin{equation}
\Omega_{qx} = \frac{\Delta t \sin^2\frac{q}{2} \partial_x \delta}
{4(\Delta^2 + t^2\cos^2\frac{q}{2} + \delta^2
  \sin^2\frac{q}{2})^{3/2}} \;.
\end{equation}
At half filling, the system is an insulator and its electron density
is given by
\begin{equation} \label{gen:ne}
n_e = \int_{-\pi}^{\pi} \frac{dq}{2\pi}\, \Omega_{qx} \;.
\end{equation}
We let $\delta(x)$ have a kink in its profile.  Such a domain wall is
known to carry fractional charge~\cite{su1979,rice1982}.
Figure~\ref{fig:dimer} shows the calculated electron density using
Eq.~\eqref{gen:ne} together with numerical result obtained by direct
diagonalization of the tight-binding Hamiltonian.  These two results are
virtually indistinguishable in the plot, which confirms the
Berry-phase modification to the density of states.

\subsection{Deformed Crystal}

In this subsection we present a general theory of electron dynamics in
crystals with deformation~\cite{sundaram1999}, which could be caused
by external pressure, defects in the lattice, or interfacial strain.

Let us set up the basic notations for this problem.  Consider a
deformation described by the atomic displacement $\{\bm u_l\}$.  We
denote the deformed crystal potential as $V(\bm r; \{\bm R_l + \bm
u_l\})$, where $\bm R_l$ is the atomic position with $l$ labeling the
  atomic site.  Introducing a smooth displacement field $\bm u(\bm r)$
  such that $\bm u(\bm R_l + \bm u_l) = \bm u_l$, the Hamiltonian can
  be written as
\begin{equation}
H = \frac{p^2}{2m} + V[\bm r - \bm u(\bm r)] + s_{\alpha\beta}(\bm r)
\cal V_{\alpha\beta}[\bm r - \bm u(\bm r)] \;,
\end{equation}
where $s_{\alpha\beta} = \partial u_\alpha/\partial r_\beta$ is the
unsymmetrized strain, and $\cal V_{\alpha\beta}[\bm r - \bm u(\bm r)]
= \sum_l [\bm R_l + \bm u(\bm r) - \bm r]_\beta(\partial V/\partial
R_{l\alpha})$ is a gradient expansion of the crystal potential.  The
last term, being proportional to the strain, can be treated
perturbatively.  The local Hamiltonian is given by
\begin{equation}
H_c = \frac{p^2}{2m} + V[\bm r - \bm u(\bm r_c)] \;,
\end{equation}
with its eigenstates $\ket{\psi_{\bm q}(\bm r - \bm u(\bm r_c))}$.

To write down the equations of motion, two pieces of information are
needed.  One is the gradient correction to the electron energy, given
in Eq.~\eqref{gen:deltae}.  It is found that~\cite{sundaram1999}
\begin{equation}
\Delta\eps = s_{\alpha\beta} D_{\alpha\beta}(\bm q) \;,
\end{equation}
where
\begin{equation}
D_{\alpha\beta} = m[v_\alpha v_\beta - \bracket{\hat{v}_\alpha
    \hat{v}_\beta}] + \bracket{\cal V_{\alpha\beta}} \;, \\
\end{equation}
with $\bracket{\cdots}$ the expectation value of the enclosed
operators in the Bloch state, and $\hat{v}_\alpha$ is the velocity operator.
Note that in the free electron limit $(V \to 0)$ this quantity
vanishes.  Which is expected since a wave packet should not feel the
effect of a deformation of the lattice in the absence of
electron-phonon coupling.  The other is the Berry curvature, which is
derived from the Berry vector potentials.  For deformed crystals, in
addition to $\bm{\cal A}_{\bm q}$, there are two other vector
potentials
\begin{equation}
\bm{\cal A}_{\bm r} = f_\alpha \dpar{u_\alpha}{\bm r} \;, \qquad
\bm{\cal A}_t = f_\alpha \dpar{u_\alpha}{t} \;,
\end{equation}
with
\begin{equation}
\bm f(\bm q) = \frac{m}{\hbar} \dpar{\eps}{\bm q} - \hbar\bm q\;.
\end{equation}
It then leads to the following Berry curvatures
\begin{equation} \begin{split}
&\Omega_{q_\alpha r_\beta} = \dpar{u_\gamma}{r_\beta}
\dpar{f_\gamma}{q_\alpha} \;, \qquad
\Omega_{k_\alpha t} = -\dpar{u_\gamma}{t} \dpar{f_\gamma}{q_\alpha}
\;, \\ &\Omega_{r_\alpha r_\beta} = \Omega_{x_\alpha t} = 0 \;.
\end{split} \end{equation}
With the above information we just need plug in the electron energy as
well as the Berry curvatures into Eq.~\eqref{gen:eom} to obtain the
equations of motion.

We first consider a one-dimensional insulator with lattice constant
$a$ and is under a uniform strain with a new lattice constant $a +
\delta a$, i.e., $\partial_x u= \delta a/ a$.  Assuming one electron
per unit cell, the electron density goes from $1/a$ to
\begin{equation}
\frac{1}{a + \delta a} = \frac{1}{a} ( 1 - \frac{\delta a}{a}) \;.
\end{equation}
On the other hand, we can also directly calculate the change of the
electron density using the modified density of
states~\eqref{gen:dos2}, which gives
\begin{equation}
\int_0^{2\pi/a} \frac{dq}{2\pi} \Omega_{qx} = - \frac{\delta a}{a^2} \;.
\end{equation}
From a physical point of view, it says an insulator under a uniform
strain remains an insulator.

The above formalism is also applicable to dislocation strain fields,
which are well defined except in a region of a few atomic spacings
around the lin of dislocation.  Outside this region, the displacement
field $\bm u(\bm r)$ is a smooth but multiple-valued function.  On
account of this multiple-valuedness, a wave packet of incident wave
vector $\bm q$ taken around the line of dislocation acquires a Berry
phase
\begin{equation}
\gamma = \oint_c d\bm r \cdot \bm{\cal A}_{\bm r} =
\oint_c d\bm u \cdot f(\bm k) \approx \bm b \cdot \bm f(\bm k) \;,
\end{equation}
where $\bm b = \oint dr_\alpha \partial\bm u/\partial r_\alpha$ is
known as the Burgers vector.  What we have here is a situation similar
to the Aharonov-Bohm effect~\cite{aharonov1959}, with the dislocation
playing the role of the solenoid, and the Berry curvature
$\bm\Omega_{\bm r\bm r}$ the role of the magnetic field.
\citet{bird1988} showed that this Berry phase can affect the electron
diffraction pattern of a deformed crystal.

The above discussion only touches a few general ideas of the Berry
phase effect in deformed crystals.  With the complete information of
the equations of motion, the semiclassical theory provides a powerful
tool to investigate the effects of deformation on electron dynamics
and equilibrium properties.

\subsection{Polarization induced by inhomogeneity}

In Sec.~\ref{sec:polar} we have discussed the Berry phase theory of
polarization in crystalline solids, based on the basic idea that the
polarization is identical to the integration of the adiabatic current
flow in the bulk.  There the system is assumed to be periodic and the
perturbation depends only on time (or any scalar for that matter).  In
this case, it is straightforward to obtain the polarization based on
the equations of motion~\eqref{gen:eom1}.  However, in many physical
situations the system is in an inhomogeneous state and the electric
polarization strongly depends on the inhomogeneity.  Examples include
flexoelectricity where a finite polarization is produced by a strain
gradient~\cite{tagantsev1986,tagantsev1991}, and multiferroic
materials where the magnetic ordering varies in space and induces a
polarization~\cite{fiebig2002,kimura2003,hur2004,cheong2007}.

Let us consider an insulating crystal with an order parameter that
varies slowly in space.  We assume that, at least at the mean-field
level, the system can be described by a perfect crystal under the
influence of an external field $\bm h(\bm r)$.  If, for example, the
order parameter is the magnetization, then $\bm h(\bm r)$ can be
chosen as the exchange field that yields the corresponding spin
configuration.  Our goal is to calculate the electric polarization to
first order in the spatial gradient as the field $\bm h(\bm r)$ is
gradually turned on.  The Hamiltonian thus takes the form $H[\bm h(\bm
  r); \lambda]$ where $\lambda$ is the parameter describing the
turning on process.  \citet{xiao2008} showed that the first order
contribution to the polarization can be classified into two
categories: the perturbative contribution due to the correction to the
wave function, and the topological contribution which is from the
dynamics of the electrons.

Let us first consider the perturbation contribution, which is
basically a correction to the polarization formula obtained by
\citet{king-smith1993} for a uniform system.  The perturbative
contribution is obtained by evaluating the Berry curvature
$\Omega_{qt}$ in Eq.~\eqref{pump:p} to first order of the gradient.
Remember that we always expand the Hamiltonian into the form $H = H_c
+ \Delta H$, and choose the eigenfunctions of $H_c$ as our expansion
basis.  Hence to calculate the Berry curvature to first order of the
gradient, one needs to know the form of the wave function perturbed by
$\Delta H$.  It has been extensively discussed in the case of an
electric field~\cite{souza2002,nunes2001}.

The topological contribution is of different nature.  Starting from
Eq.~\eqref{gen:eom} and making use of the modified density of
states~\eqref{gen:dos2}, one finds the adiabatic current induced by
inhomogeneity is give by
\begin{equation} \label{gen:j2}
j^{(2)}_\alpha = e \int_\text{BZ} \frac{d\bm q}{(2\pi)^d}
\Bigl(\Omega^{qq}_{\alpha\beta}\Omega^{r\lambda}_\beta +
\Omega^{qr}_{\beta\beta} \Omega^{q\lambda}_\alpha -
\Omega^{qr}_{\alpha\beta} \Omega^{q\lambda}_\beta\Bigr) \;.
\end{equation}
We can see that this current is explicitly proportional to the spatial
gradient.  Comparing this equation with Eq.~\eqref{pump:j} reveals a
very elegant structure: the zeroth order contribution,
Eq.~\eqref{pump:j}, is given as an integral of the first Chern form,
while the first order contribution, Eq.~\eqref{gen:j2}, is given as an
integral of the second Chern form.  Similar result has been obtained
by~\citet{qi2008}.

The polarization is obtained by integrating the current.  As usually
in the case of multiferroics, we can assume the order parameter is
periodic in space (but in general incommensurate with the crystal
lattice).  A two-point formula can be written down~\footnote{So far we
only considered the Abelian Berry case.  The non-Abelian reslut is
obtained by replacing the Chern-Simons form with its non-Abelian form}
\begin{equation} \label{gen:p2}
P^{(2)}_\alpha = \frac{e}{\cal V}
\int d \bm r \int_\text{BZ} \frac{d\bm q}{(2\pi)^d}
\Bigl( \cal A^q_\alpha \nabla^r_\beta \cal A^q_\beta + \cal A^q_\beta
\nabla^q_\alpha A^r_\beta + \cal A^r_\beta \nabla^q_\beta \cal
A^q_\alpha \Bigr) \Big|_0^1 \;,
\end{equation}
where $\cal V$ is the volume of the periodic structure of the order
parameter.  Again, due to the loss of tracking of $\lambda$, there is
an uncertain quantum which is the second Chern number.  If we assume
the order parameter has period $l_y$ in the $y$-direction, the
polarization quantum in the $x$-direction is given by
\begin{equation}
\frac{e}{l_ya_z} \;,
\end{equation}
where $a$ is the lattice constant.

\citet{kunz1986} has discussed the charge pumping in incommensurate
potentials and he showed that in general the charge transport is
quantized and is given in the form of Chern numbers, which is
consistent with what we have derived.

The second Chern form demands that the system must be two-dimensional
or higher, otherwise the second Chern form vanishes.  It allows us to
determine the general form of the induced polarization.  Consider a
two-dimensional minimal model with $\bm h(\bm r)$ having two
components. If we write $H[\bm h(\bm r);\lambda]$ as $H[\lambda\bm
  h(\bm r)]$, i.e., $\lambda$ acts like a switch, the polarization can
be written as
\begin{equation}  \label{gen:pp}
\bm P^{(2)} = \frac{e}{\cal V} \int d\bm r
\chi[(\bm\nabla \cdot \bm h)\bm h - (\bm h \cdot \bm\nabla) \bm h] \;.
\end{equation}
The coefficient $\chi$ is given by
\begin{equation}
\chi = \frac{e}{8}\int_\text{BZ} \frac{d\bm q}{(2\pi)^2} \int_0^1 d\lambda \,
\epsilon_{abcd} \Omega_{ab}\Omega_{cd} \;,
\end{equation}
where the Berry curvature is defined on the parameter space $(\bm q,
\bm h)$, and $\epsilon_{abcd}$ is the Levi-Civita antisymmetric
tensor.

\citet{xiao2008} also showed how the two-point formula can be implemented in
numerical calculations using a discretized version~\cite{kotiuga1989}.

\subsubsection{Magnetic field induced polarization}

Using the above formalism, \citet{essin2008} considered the
polarization induced by a magnetic field, which can be regarded as
arising from a spatial inhomogeneity in the vector potential $\bm
A(\bm r)$.  Consider $\bm A = By\hat{\bm z}$ with $\bm B = h/ea_zl_y
\hat{\bm x}$.  Note that magnetic flux over the supercell $a_z \times
l_y$ in the $x$-direction is exactly $h/e$, therefore the system is
periodic in the $y$-direction with period $l_y$.  According to our
discussion in Sec.~\ref{sec:em}, the effect of a magnetic field can be
counted by the Peierls substitution, $k_z \to k_z + eBy/\hbar$, hence
$\nabla_y = (eB/\hbar)\nabla_{kz}$.  The induced polarization is given
by
\begin{equation}
P_x = \frac{\theta e^2}{2\pi h} B \;,
\end{equation}
with
\begin{equation}
\theta = \frac{1}{2\pi}\int_\text{BZ} d\bm
  k \eps_{\alpha\beta\gamma} \TR[\cal A_\alpha
  \partial_\beta \cal A_\gamma - i\frac{2}{3}\cal A_\alpha \cal
  A_\beta \cal A_\gamma] \;. 
\end{equation}
Recall that $\bm P = \partial \mathcal E /\partial \bm E$, such a
magnetic-field-induced polarization implies that there is an
electromagnetic coupling of the form
\begin{equation}
\Delta\cal L_\text{EM} = \frac{\theta e^2}{2\pi h}\bm E \cdot \bm B \;.
\end{equation}
This coupling, labeled ``axion electrodynamics'', was discussed by
\citet{Wilczek1987}.  When $\theta = \pi$, the corresponding insulator
is known as a 3D $Z_2$ topological insualtor~\cite{qi2008}.

\subsection{Spin Texture}

So far our discussion has focused on the physical effects of the Berry
curvature in the momentum space ($\bm\Omega_{\bm k\bm k}$), or in the
mixed space of the momentum coordinates and some other parameters
($\bm\Omega_{\bm k\bm r}$ and $\bm\Omega_{\bm kt}$).  In this
subsection we discuss the Berry curvatures which originate only from
the nontrivial real space configuration of the system.

One of such systems is magnetic materials with domain walls or spin
textures.  Let us consider a ferromagnetic thin films described by the
following Hamiltonian
\begin{equation}
H = \frac{\bm p^2}{2m} - J \hat{\bm n}(\bm r, t) \cdot \bm\sigma \;,
\end{equation}
where the first term is the bare Hamiltonian for a conduction
electron, and the second term is the $s$-$d$ coupling between the
conduction electron and the local $d$-electron spin along the
direction $\hat{\bm n}(\bm r, t)$ with $J$ being the coupling
strength.  Note that we have allowed the spin texture to vary in both
space and time.  The simple momentum-dependence of the Hamiltonian
dictates that all $\bm k$-dependent Berry curvatures vanish.

Because of the strong $s$-$d$ coupling, we adopt the adiabatic
approximation which states that the electron spin will follow the
local spin direction during its motion.  Then the spatial variation of
local spin textures gives rise to the Berry curvature field
\begin{equation}
\bm \Omega_{\bm r\bm r} = \frac{1}{2} \sin\theta (\bm\nabla \theta
\times \bm\nabla \phi) \;,
\end{equation}
where $\theta$ and $\phi$ are the spherical angles specifying the
direction of $\hat{\bm n}$.  According to Eqs.~\eqref{gen:eom}, this field
acts on the electrons as an effective magnetic field.  In addition,
the time-dependence of the spin texture also gives rises to
\begin{equation}
\bm\Omega_{\bm rt} =
\frac{1}{2}\sin\theta(\partial_t\phi\bm\nabla\theta - \partial_t\theta
\bm\nabla\phi) \;.
\end{equation}
Similarly, $\bm\Omega_{\bm rt}$ acts on the electrons as an effective
electric field.  This is in analogy with a moving magnetic field
($\bm\Omega_{\bm r\bm r}$) generating an electric field
($\bm\Omega_{\bm rt}$).

The physical consequences of these two fields are obvious by analogy
with the electromagnetic fields.  The Berry curvature $\bm\Omega_{\bm
  r\bm r}$ will drive a Hall current, just like the ordinary Hall
effect~\cite{ye1999,bruno2004}.  Unlike the anomalous Hall effect
discussed in Sec.~\ref{sec:ahe}, this mechanism for a nonvanishing
Hall effect does not require the spin-orbit coupling, but does need a
topologically nontrivial spin texture, for example, a skyrmion.  On
the other hand, for a moving domain wall in a thin magnetic wire, the
Berry curvature $\bm\Omega_{rt}$ will induce an electromotive force,
which results in a voltage difference between the two ends.  This
Berry curvature induced emf has has been experimentally measured
recently~\cite{yang2009}.


\section{\label{sec:qz}Quantization of electron dynamics}

In previous sections, we have reviewed several Berry phase effects
in solid state systems. Berry curvature often appears as a result
of restricting (or projecting) the extent of a theory to its
subspace. In particular, the Berry curvature plays a crucial role
in the semiclassical dynamics of electrons, which is valid under
the one-band approximation. In the following, we will explain how
could the semiclassical formulation be re-quantized. This is
necessary, for example, in studying the quantized Wannier-Stark
ladders from the Bloch oscillation, or the quantized Landau levels
from the cyclotron orbit \cite{ashcroft-CM}. The re-quantized
theory is valid in the same subspace of the semiclassical theory.
It will become clear that, the knowledge of the Bloch energy, the
Berry curvature, and the magnetic moment in the semiclassical
theory constitute sufficient information for building the
re-quantized theory.  In this section, we focus on the following
methods of quantization: the Bohr-Sommerfeld quantization and the
canonical quantization.

\subsection{\label{sec:qza}Bohr-Sommerfeld quantization}

A method of quantization is a way to select quantum mechanically
allowed states out of a continuum of classical states. This is
often formulated using the generalized coordinates $q_i$ and their
conjugate momenta $p_i$. The Bohr-Sommerfeld quantization requires
the action integral for each set of the conjugate variables to
satisfy
\begin{equation}\label{qza:bs0}
\oint_{C_i}p_i d
q_i=\left(m_i+\frac{\nu_i}{4}\right)h,i=1,\cdots,d,
\end{equation}
where $C_i$ are closed loops in the phase space with dimension
$2d$, $m_i$ are integers, and $\nu_i$ are the so-called Maslov
indices, which are usually integers.

However, since the choices of conjugate variables are not unique,
the Bohr-Sommerfeld quantization method may give inconsistent
quantization rules. This is known to happen in the case of an
isotropic three-dimensional harmonic oscillator \cite{tabor1989}.
This problem can be fixed if, instead of Eq.~\eqref{qza:bs0}, one
uses the following integrals,
\begin{equation}
\oint_{C_k}\sum_{\ell=1}^d p_\ell d
q_\ell=\left(m_k+\frac{\nu_k}{4}\right)h,k=1,\cdots,d,
\end{equation}
where $C_k$ are the periodic orbits on invariant tori. An
invariant torus is a torus with constant action in the phase
space. The revised rule above is often called the Einstein-
Brillouin-Keller (EBK) quantization \cite{tabor1989}.

In the wavepacket formulation of Bloch electrons, both ${\bf r}_c$
and ${\bf q}_c$ are treated as generalized coordinates. With the
Lagrangian in Eq.~\eqref{em:Lag}, one can find their conjugate
momenta $\partial L/\partial {\dot{\bf r}}_c$ and $\partial
L/\partial {\dot{\bf q}}_c$, which are equal to $\hbar{\bf q}_c$
and $\hbar\langle u|i\partial u/\partial{\bf
q}_c\rangle=\hbar{\cal A}$ respectively \cite{sundaram1999}. The
quantization condition for an orbit with constant energy thus
becomes,
\begin{equation}\label{qza:bs}
\oint_C {\bf q}_c\cdot d{\bf
r}_c=2\pi\left(m+\frac{\nu}{4}-\frac{\Gamma_C}{2\pi}\right),
\end{equation}
where $\Gamma_C\equiv\oint_C{\cal A}\cdot d{\bf q}_c$ is the Berry
phase of a constant-energy orbit $C$ (also see
\citet{kuratsuji1985,wilkinson1984a}). Since the Berry phase is
path dependent, one may need to solve the equation
self-consistently to obtain the quantized orbits.

Before applying the Bohr-Sommerfeld quantization in the following
subsections, we would like to point out two disadvantages of this
method. First, the value of the Maslov index is not always
evident. For example, for an one-dimensional particle bounded by
two walls, its value would depend on the slopes of the walls
\cite{houten1989}. In fact, a non-integer value may give a more
accurate prediction of the energy levels \cite{friedrich1996}.
Second, this method fails if the trajectory in phase space is not
closed, or if the dynamic system is chaotic so that no invariant
torus exists. On the other hand, the method of canonical
quantization in Sec.~\ref{sec:qzd} does not have these problems.

\subsection{\label{sec:qzb}Wannier-Stark ladder}

Consider an electron moving in a one-dimensional periodic lattice
with lattice constant $a$. Under a weak uniform electric field
${\bf E}$, according to the semiclassical equations of motion, the
quasi-momentum of an electron wavepacket is simply (see
Eq.~\eqref{em:EOM})
\begin{equation}
\hbar{\bf q}_c(t)=-e{\bf E}t.
\end{equation}
It takes the time $T_B=h/(eEa)$ for the electron to traverse the
first Brillouin zone. Therefore, the angular frequency of the
periodic motion is $\omega_B=eEa/\hbar$.  This is the so-called
Bloch oscillation \cite{ashcroft-CM}.

Similar to a simple harmonic oscillator, the energy of the
oscillatory electron is quantized in multiples of $\hbar\omega_B$.
However, unlike the former, the Bloch oscillator has no zero-point
energy (that is, the Maslov index is zero). These equally spaced
energy levels are called the Wannier-Stark ladders. Since the
Brillouin zone is periodic, the electron orbit is closed.
According to the Bohr-Sommerfeld quantization, one has
\begin{equation}\label{qzb:ws}
\oint_{C_m}{\bf r}_c\cdot d{\bf
q}_c=-2\pi\left(m-\frac{\Gamma_{C_m}}{2\pi}\right).
\end{equation}
For a simple one-dimensional lattice with inversion symmetry, the
Berry phase $\Gamma_{C_m}$ can only have two values, 0 or $\pi$
\cite{zak1989a}, as discussed earlier in Sec.~\ref{sec:polar}.

Starting from Eq.~\eqref{qzb:ws}, it is not difficult to find the
average position of the electron,
\begin{equation}
\langle r_c\rangle_m=a\left(m-\frac{\Gamma_C}{2\pi}\right),
\end{equation}
where we have neglected the subscript $m$ in $\Gamma_{C_m}$ since
all of the paths in the same energy band have the same Berry phase
here. Such average positions $\langle r_c\rangle_m$ are the
average positions of the Wannier function \cite{vanderbilt1993}.
Due to the Berry phase, they are displaced from the positive ions
located at $am$.

In Sec.~\ref{sec:polar}, the electric polarization is derived
using the theory of adiabatic transport.  It can also be obtained
from the expectation value of the position operator directly.
Because of the charge separation mentioned above, the
one-dimensional crystal has a polarization $\Delta
P=e\Gamma_c/2\pi$ (compared to the state without charge
separation), which is the electric dipole per unit cell. This is
consistent with the result in Eq.~\eqref{pump:p2}.

After time average, the quantized energies of the electron
wavepacket are,
\begin{eqnarray}
\langle {\cal E}\rangle_m &=&\langle\epsilon(q_c)\rangle-eE\langle r_c\rangle_m\nonumber\\
&=&\epsilon_0-eEa\left(m-\frac{\Gamma_C}{2\pi}\right),
\end{eqnarray}
which are the energy levels of the Wannier-Stark ladders.

Two short comments are in order: First, beyond the one-band
approximation, there exist Zener tunnellings between Bloch bands.
Therefore, the quantized levels are not stationary states of the
system. They should be understood as resonances with finite
life-times \cite{gluck1999}. Second, the fascinating phenomenon of
Bloch oscillation is not commonly observed in laboratory for the
following reason: In an usual solid, the electron scattering time
is shorter than the period $T_B$ by several orders of magnitude.
Therefore, the phase coherence of the electron is destroyed within
a tiny fraction of a period. Nonetheless, with the help of a
superlattice that has a much larger lattice constant, the period
$T_B$ can be reduced by two orders of magnitude, which could make
the Bloch oscillation and the accompanying Wannier-Stark ladders
detectable \cite{mendes1993}. Alternatively, the Bloch oscillation
and Wannier-Stark ladders can also be realized in an optical
lattice \cite{wilkinson1996,salomon1996}, in which the atom can be
coherent over a long period of time.

\subsection{\label{sec:qzc}de Haas-van Alphen oscillation}

When an uniform ${\bf B}$ field is applied to a solid, the
electron would execute a cyclotron motion in both the $r$-space
and the $k$-space. From Eq.~\eqref{em:EOM3}, it is not difficult
to see that an orbit $C$ in $k$-space lies on the intersection of
a plane perpendicular to the magnetic field and the
constant-energy surface \cite{ashcroft-CM}. Without quantization,
the size of an orbit is determined by the initial energy of the
electron and can be varied continuously. One then applies the
Bohr-Sommerfeld quantization rule, as Onsager did, to quantize the
size of the orbit \cite{onsager1952}. That is, only certain orbits
satisfying the quantization rule are allowed. Each orbit
corresponds to an energy level of the electron (i.e., the Landau
level). Such a method remains valid in the presence of the Berry
phase.

With the help of the semiclassical equation (see
Eq.~\eqref{em:EOM}),
\begin{equation}
\hbar\dot{\bf k}_c=-e\dot{\bf r}_c\times{\bf B},
\end{equation}
the Bohr-Sommerfeld condition in Eq.~\eqref{qza:bs} can be written
as (notice that $\hbar{\bf q}_c=\hbar{\bf k}_c-e{\bf A}$, and
$\nu=2$),
\begin{equation}\label{qzc:r-orbit}
\frac{\bf B}{2}\cdot\oint_{C_m}{\bf r}_c\times d{\bf
r}_c=\left(m+\frac{1}{2}-\frac{\Gamma_{C_m}}{2\pi}\right)\phi_0,
\end{equation}
where $\phi_0\equiv h/e$ is the flux quantum. The integral on the
left-hand side is simply the magnetic flux enclosed by the
real-space orbit (allowing a drift along the ${\bf B}$-direction).
Therefore, the enclosed flux has to jump in steps of the flux
quantum (plus a Berry phase correction).

Similar to the Bohr atom model, in which the electron has to form
a standing wave, here the total phase acquired by the electron
after one circular motion also has to be integer multiples of
$2\pi$. Three types of phases contribute to the total phase: (a),
The Aharonov-Bohm phase: an electron circulating a flux quantum
picks up a phase of $2\pi$. (b), The phase lag of $\pi$ at each
turning point (there are two of them). This explains why the
Maslov index is two. (c), The Berry phase intrinsic to the solid.
Therefore, Eq.~\eqref{qzc:r-orbit} simply says that the summation
of these three phases should be equal to $2\pi m$.

The orbit in $k$-space can be obtained by re-scaling the $r$-space
orbit in Eq.~\eqref{qzc:r-orbit} with a linear factor of
$\lambda_B^2$, followed by a rotation of ninety degrees, where
$\lambda_B\equiv\sqrt{\hbar/eB}$ is the magnetic length
\cite{ashcroft-CM}. Therefore, one has
\begin{equation}\label{qzc:k-orbit}
\frac{\hat{\bf B}}{2}\cdot\oint_{C_m}{\bf k}_c\times d{\bf
k}_c=2\pi\left(m+\frac{1}{2}-\frac{\Gamma_{C_m}}{2\pi}\right)\frac{eB}{\hbar}.
\end{equation}
The size of the orbit, combined with the knowledge of the electron
energy $E(\bf k)_c=\epsilon({\bf k}_c)-{\bf M}\cdot{\bf B}$, help
determining the quantized energy levels. For an electron with a
quadratic energy dispersion (before applying the magnetic field),
these levels are equally spaced. However, with the Berry phase
correction, which are usually different for different orbits, the
energy levels are no longer uniformly distributed. This is related
to the discussion in Sec.~\ref{sec:em} on the relation between the
density of states and the Berry curvature \cite{xiao2005}.

As a demonstration, we apply the quantization rule to graphene and
calculate the energies of Landau levels near the Dirac point.
Before applying a magnetic field, the energy dispersion near the
Dirac point is linear, $E({\bf k })=\hbar v_F k$. It is known
that, if the energy dispersion near a degenerate point is linear,
then the cyclotron orbit will acquire a Berry phase
$\Gamma_C=\pi$, independent of the shape of the orbit
\cite{blount1962c}. As a result, the $1/2$ on the right hand side
of Eq.~\eqref{qzc:k-orbit} is cancelled by the Berry phase term.
According to Eq.~\eqref{qzc:k-orbit}, the area of a cyclotron
orbit is thus $\pi k^2=2\pi |m|\frac{eB}{\hbar}, m\in Z$. From
which one can easily obtain the Landau level energy
$E_m=v_F\sqrt{2eB\hbar m }$. The experimental observation of a
quantum Hall plateau at zero energy is thus a direct consequence
of the Berry phase \cite{novoselov2005,novoselov2006,zhang2005}.

In addition to point degeneracy, other types of degeneracy in
momentum space can also be a source of the Berry phase. For
example, the effect of the Berry phase generated by a line of band
contact on magneto-oscillations is studied in
\citet{mikitik1999,mikitik2004}.

The discussion so far is based on the one-band approximation. In
reality, the orbit in one band would couple with the orbits in
other bands. As a result, the Landau levels are broadened into
mini-bands \cite{wilkinson1984b}. Similar situation occurs in a
magnetic Bloch band, which is the subject of Sec.~\ref{sec:mbb}.

\subsection{\label{sec:qzd}Canonical quantization (Abelian case)}

In addition to the Bohr-Sommerfeld quantization, an alternative
way to quantize a classical theory is by finding out position and
momentum variables that satisfy the following Poisson brackets,
\begin{equation}
\{x_i,p_j\}=\delta_{ij}.
\end{equation}
Afterwards, these classical canonical variables are promoted to
operators that satisfy the commutation relation,
\begin{equation}\label{qzd:qp}
[x_i,p_j]=i\hbar\delta_{ij},
\end{equation}
That is, all we need to do is to substitute the Poisson bracket
$\{x_i,p_j\}$ by the commutator $[x_i,p_j]/i\hbar$. Based on the
commutation relation, these variables can be written explicitly
using either the differential-operator representation or the
matrix representation. Once this is done, one can proceed to
obtain the eigenvalues and eigenstates of the Hamiltonian $H({\bf
x},{\bf p})$.

Even though one can always have canonical pairs in a Hamiltonian
system, as guaranteed by the Darboux theorem \cite{arnold1989}, in
practice, however, finding them may not be a trivial task. For
example, the center-of-mass variables ${\bf r}_c$ and ${\bf k}_c$
in the semiclassical dynamics in Eq.~\eqref{em:EOM} are not
canonical variables since their Poisson brackets are not of the
canonical form \cite{xiao2005,duval2006},
\begin{eqnarray}
\{r_i,r_j\}&=&\epsilon_{ijk}\Omega_k/\kappa,\label{qzd:xiao1}\\
\{k_i,k_j\}&=&-\epsilon_{ijk}eB_k/\kappa,\label{qzd:xiao2}\\
\{r_i,k_j\}&=&\left(\delta_{ij}+eB_i\Omega_j\right)/\kappa,\label{qzd:xiao3}
\end{eqnarray}
where $\kappa({\bf k})\equiv 1+e{\bf B}\cdot{\bf \Omega}({\bf k
})$. In order to carry out the canonical quantization, canonical
variables of position and momentum must be found.

Let us start with two special cases. The first is a solid with
zero Berry curvature that is under the influence of a magnetic
field (${\bf \Omega}=0, {\bf B}\neq 0$). In this case, the factor
$\kappa$ in Eq.~\eqref{qzd:xiao2} reduces to one and the position
variables commute with each other. Obviously, if one assumes
$\hbar{\bf k}_c={\bf p}+e{\bf A}({\bf x})$ and requires ${\bf x}$
and ${\bf p}$ to be canonical conjugate variables, then the
quantized version of Eq.~\eqref{qzd:xiao2} (with $i\hbar$
inserted) can easily be satisfied. This is the familiar Peierls
substitution \cite{peierls1933}.

In the second case, consider a system with Berry curvature but not
in a magnetic field (${\bf \Omega}\neq 0, {\bf B}= 0$). In this
case, again we have $\kappa=1$. Now the roles of ${\bf r}_c$ and
${\bf k}_c$ in the commutators are reversed. The momentum
variables commute with each other, but not the coordinates. One
can apply a Peierls-like substitution to the coordinate variables
and write ${\bf r}_c={\bf x}+\bm{\mathcal A}({\bf q})$. It is not
difficult to see that the commutation relations arising from
Eq.~\eqref{qzd:xiao1} can indeed be satisfied. After the canonical
quantization, ${\bf x}$ becomes $i\partial/\partial {\bf q}$ in
the quasi-momentum representation. In \citet{blount1962c}, the
position operator ${\bf r}$ in the one-band approximation acquires
a correction, which is our Berry connection ${\bm{\mathcal A}}$.
Therefore, ${\bf r}_c$ can be identified with the projected
position operator $P{\bf r}P$, where $P$ projects to the energy
band of interest.

When both ${\bf B}$ and ${\bf \Omega}$ are nonzero, applying both
of the Peierls substitutions simultaneously is not enough to
produce the correct commutation relations, mainly because of the
non-trivial factor $\kappa$ there. In general, exact canonical
variables cannot be found easily. However, since the semiclassical
theory itself is valid to linear order of field, we only need to
find the canonical variables correct to the same order in
practice. The result is \cite{chang2008},
\begin{eqnarray}
{\bf r}_c&=&{\bf x}+\bm{\mathcal
A}({\mbox{\boldmath$\pi$}})+{\bf G}(\bm{\pi}),\nonumber\\
\hbar{\bf k}_c&=&{\bf p}+e{\bf A}({\bf x})+e{\bf
B}\times\bm{\mathcal A}({\mbox{\boldmath$\pi$}}),
\end{eqnarray}
where ${\mbox{\boldmath$\pi$}}={\bf p}+e{\bf A}({\bf x})$, and
$G_\alpha({\bf k}_c)\equiv(e/\hbar)(\bm{\mathcal A}\times{\bf
B})\cdot\partial\bm{\mathcal A}/\partial k_{c\alpha}$. This is the
generalized Peierls substitution for systems with Berry connection
$\bm{\mathcal A}$ and vector potential ${\bf A}$. With these
equations, one can verify
Eqs.~\eqref{qzd:xiao1},\eqref{qzd:xiao2}, and \eqref{qzd:xiao3} to
linear orders of ${\bf B}$ and $\bm{\Omega}$.

A few comments are in order: First, if a physical observable is a
product of several canonical variables, the order of the product
may become a problem after the quantization since the variables
may not commute with each other. To preserve the hermitian
property of the physical observable, the operator product needs to
be symmetrized. Second, the Bloch energy, Berry curvature, and
orbital moment of the semiclassical theory contains sufficient
information for building a quantum theory that accounts for all
physical effects to first order in external fields. We will come
back to this in Sec.~\ref{sec:nab}, where the non-Abelian
generalization of the canonical quantization method is addressed.

\section{\label{sec:mbb}Magnetic Bloch band}

The semiclassical dynamics in previous sections is valid when the
external field is weak, such that the latter can be treated as a
perturbation to the Bloch states. Such a premise is no longer
valid if the external field is so strong that the structure of the
Bloch bands is significantly altered. This happens, for example,
in quantum Hall systems where the magnetic field is of the order
of Tesla and a Bloch band would break into many subbands. The
translational symmetry and the topological property of the subband
are very different from those of the usual Bloch band. To
distinguish between the two, the former is called the magnetic
Bloch band (MBB).

The MBB usually carries non-zero quantum Hall conductance and has
a nontrivial topology. Compared to the usual Bloch band, the MBB
is a more interesting playground for many physics phenomena. In
fact, the Berry curvature of the Bloch electron is first revealed
in the MBB. In this section, we review some basic fact of the MBB,
as well as the semiclassical dynamics of the magnetic Bloch
electron when it is subject to {\it further} electromagnetic
perturbation. Such an elegant formulation provides a clear picture
of the hierarchical subbands split by the strong magnetic field
(the Hofstadter spectrum) \cite{hofstadter1976}.

\subsection{\label{sec:mbba}Magnetic translational symmetry}

In the presence of a strong magnetic field, one needs to treat the
magnetic field and the lattice potential on equal footing and
solve the following Schrodinger equation,
\begin{equation}
\left\{\frac{1}{2m}\left[{\bf p}+e{\bf A}({\bf
r})\right]^2+V_L({\bf r})\right\}\psi({\bf r})=E\psi({\bf r}),
\end{equation}
where $V_L$ is the periodic lattice potential. For convenience of
discussion, we assume the magnetic field is uniform along the
$z$-axis and the electron is confined to the $x$-$y$ plane.
Because of the vector potential, the Hamiltonian $H$ above no
longer has the lattice translation symmetry.

However, since the lattice symmetry of the charge density is not
broken by an uniform magnetic field, one should be able to define
translation operators that differ from the usual ones only by
phase factors \cite{lifshitz1980}. First, consider a system
translated by a lattice vector ${\bf a}$,
\begin{equation}
\left\{\frac{1}{2m}\left[{\bf p}+e{\bf A}({\bf r}+{\bf a
})\right]^2+V_L({\bf r})\right\}\psi({\bf r}+{\bf a })=E\psi({\bf
r}+{\bf a}),
\end{equation}
where $V_L({\bf r}+{\bf a})=V_L({\bf r})$ has been used. One can
write
\begin{equation}
{\bf A}({\bf r}+{\bf a})={\bf A}({\bf r})+\nabla f({\bf r}),
\end{equation}
where $\nabla f({\bf r})={\bf A}({\bf r}+{\bf a})-{\bf A}({\bf
r})\equiv \Delta {\bf A}({\bf a})$. For an uniform magnetic field,
the vector potential must be a linear function of ${\bf r}$.
Therefore, $\Delta {\bf A}$ is independent of ${\bf r}$ and
$f=\Delta{\bf A}\cdot{\bf r}$. The extra vector potential $\nabla
f$ can be removed by a gauge transformation,
\begin{equation}
\left\{\frac{1}{2m}\left[{\bf p}+e{\bf A}({\bf
r})\right]^2+V_L({\bf r})\right\}e^{i\frac{e}{\hbar}f}\psi({\bf
r}+{\bf a })=Ee^{i\frac{e}{\hbar}f}\psi({\bf r}+{\bf a}).
\end{equation}
We now identify the state above as the magnetic translated state
$T_{\bf a}\psi({\bf r})$,
\begin{equation}
T_{\bf a}\psi({\bf r})=e^{i\frac{e}{\hbar}\Delta{\bf A}\cdot{\bf
r}}\psi({\bf r}+{\bf a}).
\end{equation}
The operator $T_{\bf a}$ being defined this way has the desired
property that $[H,T_{\bf a}]$=0.

\begin{figure}
\center
\includegraphics[width=2.5 in,angle=90]{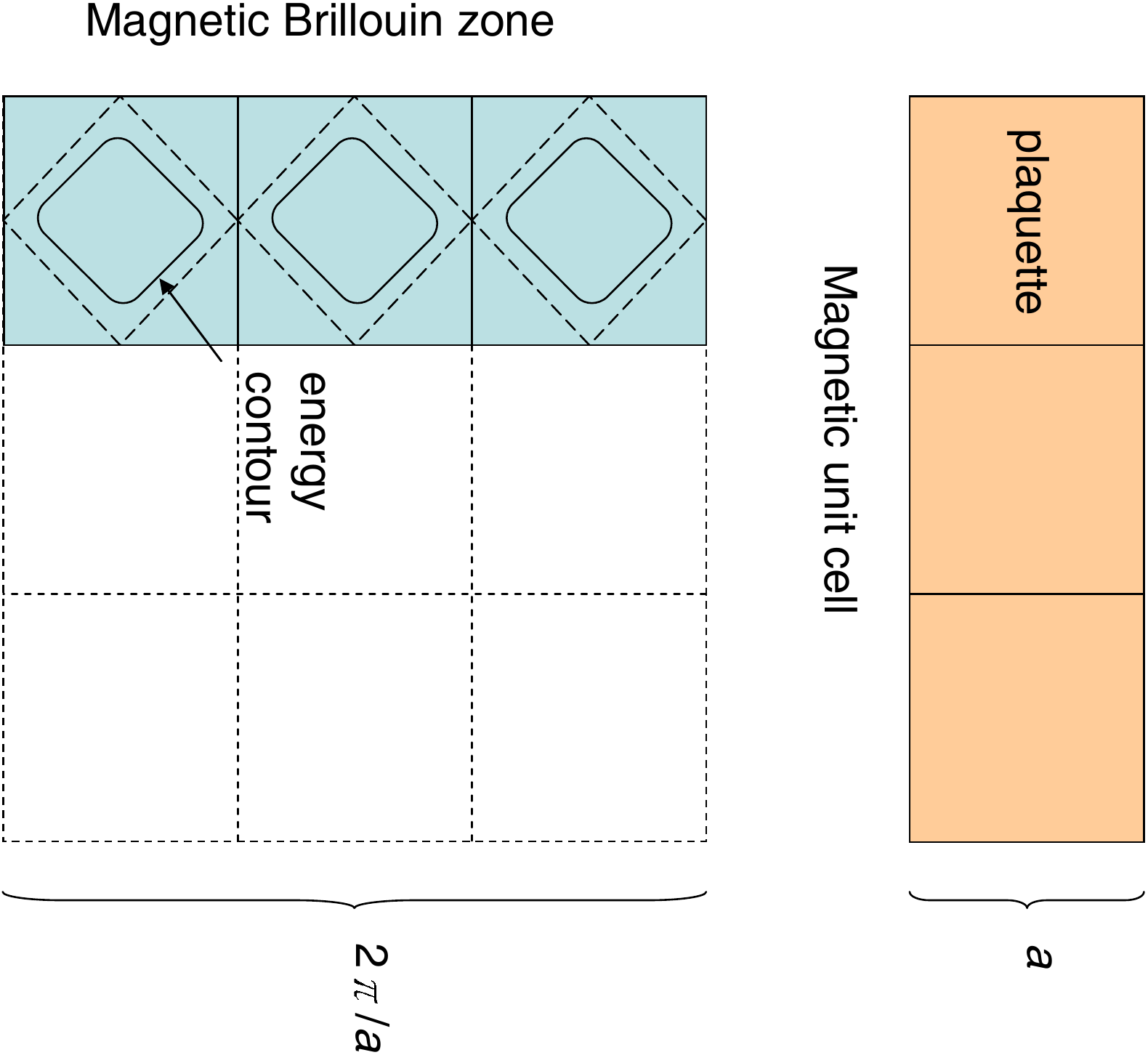}
\caption{(color online) When the magnetic flux per plaquette is
$\phi_0/3$, the magnetic unit cell is composed of three
plaquettes. The magnetic Brillouin zone is three times smaller
than the usual Brillouin zone. Furthermore, the magnetic Bloch
states are three-fold degenerate.} \label{figc1}
\end{figure}

Unlike usual translation operators, magnetic translations along
different directions usually do not commute. For example, let
${\bf a}_1$ and ${\bf a}_2$ be lattice vectors, then
\begin{equation}\label{mbba:commutator}
T_{{\bf a}_2}T_{{\bf a}_1}=T_{{\bf a}_1}T_{{\bf
a}_2}e^{i\frac{e}{\hbar}\oint{\bf A}\cdot d{\bf r}},
\end{equation}
where $\oint{\bf A}\cdot d{\bf r}$ is the magnetic flux going
through the unit cell defined by ${\bf a}_1$ and ${\bf a}_2$. That
is, the non-commutativity is a result of the Aharonov-Bohm phase.
$T_{{\bf a}_1}$ commutes with $T_{{\bf a}_2}$ only if the flux
$\phi$ is an integer multiple of the flux quantum $\phi_0=e/h$.

If the magnetic flux $\phi$ enclosed by a plaquette is
$(p/q)\phi_0$, where $p$ and $q$ are co-prime integers, then
$T_{q{\bf a}_1}$ would commute with $T_{{\bf a}_2}$ (see
Fig.~\ref{figc1}). The simultaneous eigenstate of $H$, $T_{q{\bf
a}_1}$, and $T_{{\bf a}_2}$ is called a magnetic Bloch state, and
its energy the magnetic Bloch energy,
\begin{eqnarray}
H\psi_{n{\bf k}}&=&E_{n{\bf k}}\psi_{n{\bf k}},\\
T_{q{\bf a}_1}\psi_{n{\bf k}}&=&e^{i{\bf k}\cdot q{\bf a}_1}
({\bf k})\psi_{n{\bf k}},\label{mbba:t2}\\
T_{{\bf a}_2}\psi_{n{\bf k}}&=&e^{i{\bf k}\cdot{\bf a}_2}({\bf
k})\psi_{n{\bf k}}.
\end{eqnarray}

Since the magnetic unit cell is $q$ times larger than the usual
unit cell, the magnetic Brillouin zone (MBZ) has to be $q$ times
smaller. If ${\bf b}_1$ and ${\bf b}_2$ are defined as the lattice
vectors reciprocal to ${\bf a}_1$ and ${\bf a}_2$. Then, in this
example, the MBZ is folded back $q$ times along the ${\bf b}_1$
direction.

In addition, with the help of Eqs.~ \eqref{mbba:commutator} and
\eqref{mbba:t2}, one can show that the eigenvalues of the $T_{{\bf
a }_2}$ operator for the following translated states,
\begin{equation} T_{{\bf
a}_1}\psi_{n{\bf k}}, T_{2{\bf a}_1}\psi_{n{\bf
k}},\cdots,T_{(q-1){\bf a}_1}\psi_{n{\bf k}}
\end{equation}
are
\begin{equation}
e^{i\left({\bf k}+{\bf b}_2p/q \right)\cdot a_2}, e^{i\left({\bf
k}+2{\bf b}_2p/q \right)\cdot a_2},\cdots,e^{i\left({\bf
k}+(q-1){\bf b}_2p/q \right)\cdot a_2}
\end{equation}
respectively. These states are not equivalent, but have the same
energy as $\psi_{n{\bf k}}$ since $[H,T_{{\bf a}_1}]=0$.
Therefore, the MBZ has a $q$-fold degeneracy along the ${\bf b}_2$
direction. Each repetition unit in the MBZ is sometimes called a
reduced magnetic Brillouin zone. More discussions on the magnetic
translation group can be found in
\citet{zak1964a,zak1964b,zak1964c}.

\subsection{\label{sec:mbbb}Basics of magnetic Bloch band}

In this subsection, we review some basic properties of the
magnetic Bloch band. This includes the pattern of band splitting
due to a quantizing magnetic field, the phase of the magnetic
Bloch state in ${\bf k}$-space and its connection with the Hall
conductance.

The rules of band splitting are simple in two opposite limits,
which are characterized by the relative strength between the
lattice potential and the magnetic field. When the lattice
potential is much stronger than the magnetic field, it is more
appropriate to start with the zero-field Bloch band as a
reference. It was found that, if each plaquette encloses a
magnetic flux $(p/q)\phi_0$, then each Bloch band would split to
$q$ subbands \cite{kohmoto1989,hatsugai1990}. We know that if $N$
is the total number of lattice sites on the two dimensional plane,
then the number of allowed ${\bf k}$-states in the Brillouin zone
(and in one Bloch band) is $N$. Since the area of the MBZ (and the
number of states within) is smaller by a factor of $q$, each MBB
has $N/q$ states, sharing the number of states of the original
Bloch band equally.

On the other hand, if the magnetic field is much stronger than the
lattice potential, then one should start from the Landau level as
a reference. In this case, if each plaquette has a magnetic flux
$\phi=(p/q)\phi_0$, then after turning on the lattice potential,
each LL will split to $p$ subbands. The state counting is quite
different from the previous case: The degeneracy of the original
LL is $\Phi/\phi_0=Np/q$, where $\Phi=N\phi$ is the total magnetic
flux through the two dimensional sample. Therefore, after
splitting, each MBB again has only $N/q$ states, the number of
states in a MBZ.

Between the two limits, when the magnetic field is neither very
strong nor very weak, the band splitting does not follow a simple
pattern. When the field is tuned from weak to strong, the subbands
will split, merge, and interact with each other in a complicated
manner, such that in the end there are only $p$ subbands in the
strong-field limit.

\begin{figure}
\center
\includegraphics[width=3.5 in,angle=90]{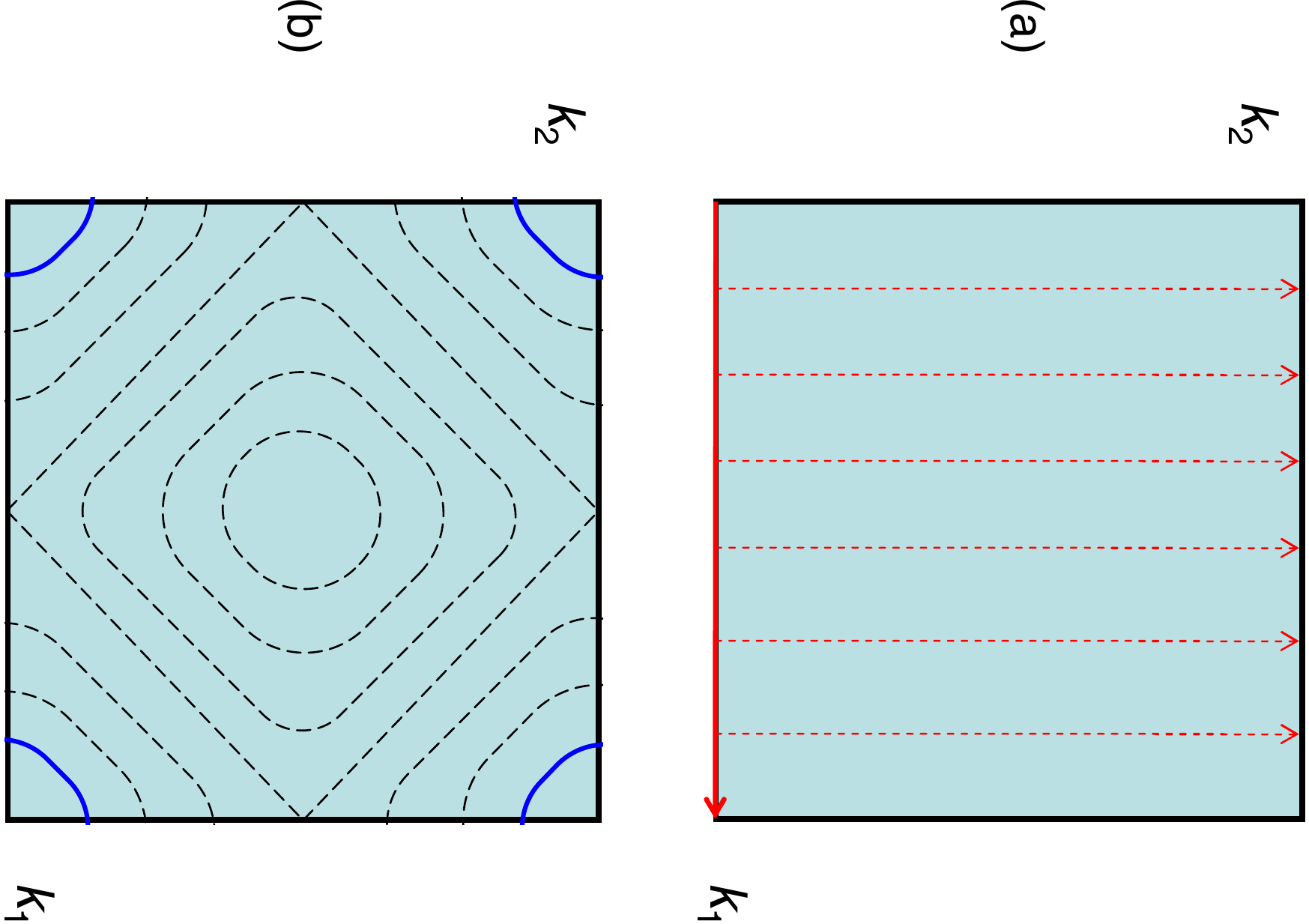}
\caption{(color online) (a) The phases of the MBS in the reduced
MBZ can be assigned using the parallel transport conditions, first
along the $k_1$-axis, then along the paths paralllel to the
$k_2$-axis. (b) Hyperorbits in a reduced MBZ. Their sizes are
quantized following the Bohr-Sommerfeld quantization condition.
The orbit enclosing the largest area is indicated by solid lines.}
\label{figc2}
\end{figure}

According to Laughlin's gauge-invariance argument
\cite{laughlin1981}, each of the isolated magnetic Bloch band
carries a quantized Hall conductivity (see Sec.~\ref{sec:quant}
and Sec.~\ref{sec:hallc}). This is closely related to the
non-trivial topological property of the magnetic Bloch state
\cite{kohmoto1985,morandi1988}. Furthermore, the distribution of
Hall conductivities among the split subbands follows a very simple
rule first discovered by \citet{thouless1982} {\it et al}. This
rule can be derived with the help of the magnetic translation
symmetry \cite{dana1985}. We show the derivation below following
Dana {\it et al}'s analysis since it reveals the important role
played by the Berry phase in the magnetic Bloch state.

In general, the phases of Bloch states at different ${\bf k}$'s
are unrelated and can be defined independently. However, the same
does not apply to a MBZ. For one thing, the phase has to be
non-integrable in order to account for the Hall conductivity. One
way to assign the phase of the MBS $u_{\bf k}({\bf r})$ is by
imposing the parallel-transport condition (see Thouless's article
in \citet{prange1987}),
\begin{eqnarray}
\left\langle u_{k_10}\left|\frac{\partial}{\partial{k_1}}\right|u_{k_10}\right\rangle&=&0;\label{mbbb:connection1}\\
\left\langle
u_{k_1k_2}\left|\frac{\partial}{\partial{k_2}}\right|u_{k_1k_2}\right\rangle&=&0.\label{mbbb:connection2}
\end{eqnarray}
The first equation defines the phase of the states on the
$k_1$-axis; the second equation defines the phase along a line
with fixed $k_1$ (see Fig.~\ref{figc2}(a)). As a result, the
phases of any two states in the MBZ have a definite relation.

The states on opposite sides of the MBZ boundaries represent the
same physical state. Therefore, they can only differ by a phase
factor. Following Eqs.~\eqref{mbbb:connection1} and
\eqref{mbbb:connection2}, we have
\begin{eqnarray}
u_{k_1+b_1/q,k_2}&=&u_{k_1,k_2};\\
u_{k_1,k_2+b_2}&=&e^{i\delta(k_1)}u_{k_1,k_2},
\end{eqnarray}
where $b_1$ and $b_2$ are the lengths of the primitive vectors
reciprocal to ${\bf a}_1,{\bf a}_2$. That is, the states on the
opposite sides of the $k_1$ boundaries have the same phase. The
same cannot also be true for the $k_2$ boundaries, otherwise the
topology will be too trivial to accommodate the quantum Hall
conductivity.

Periodicity of the MBZ requires that
\begin{equation}\label{mbbb:integer}
\delta(k_1+b_1/q)=\delta(k_1)+2\pi\times{\rm integer}.
\end{equation}
In order for the integral $(1/2\pi)\oint_{\partial{\rm MBZ}}d{\bf
k}\cdot\bm{\mathcal A}({\bf k})$ (which is nonzero only along the
upper horizontal boundary) to be the Hall conductivity $\sigma_H$
(in units of $h/e^2$), the integer in Eq.~\eqref{mbbb:integer}
obviously has to be equal to $\sigma_H$.

Following the periodicity condition in Eq.~\eqref{mbbb:integer},
it is possible to assign the phase in the form,
\begin{equation}
\delta(k_1)=\tilde\delta(k_1)+\sigma_H k_1qa_1,
\end{equation}
where $\tilde\delta(k_1+b_1/q)=\tilde\delta(k_1)$. On the other
hand, from the discussion at the end of the previous subsection,
we know that
\begin{equation}
T_{a_1}u_{k_1k_2}=e^{i\theta(k_1)}u_{k_1k_2+2\pi p/qa_2}.
\end{equation}
Again from the periodicity of the MBZ, one has
\begin{equation}
\theta(k_1+b_1/q)=\theta(k_1)+2\pi m; m\in Z,
\end{equation}
which gives
\begin{equation}
\theta(k_1)=\tilde\theta(k_1)+m k_1qa_1.
\end{equation}
Choose $\tilde\delta(k_1)$ and $\tilde\theta(k_1)$ to be zero, one
finally gets
\begin{eqnarray}
T_{qa_1}u_{k_1k_2}&=&e^{iqmk_1qa_1}u_{k_1k_2+2\pi p/a_2}\nonumber\\
&=&e^{iqmk_1qa_1}e^{ip\sigma_Hk_1qa_1}u_{k_1k_2}.
\end{eqnarray}
But this state should also be equal to $e^{iqk_1a_1}u_{k_1k_2}$.
Therefore, the Hall conductivity should satisfy
\begin{equation}
p\sigma_H+qm=1.
\end{equation}
This equation determines the Hall conductivity (mod $q$) of a MBB
\cite{dana1985}. In Sec.~\ref{sec:mbbd}, we will see that the
semiclassical analysis can also help us finding out the Hall
conductivity of a MBB.

\subsection{\label{sec:mbbc}Semiclassical picture: hyperorbits}

When a weak magnetic field is applied to a Bloch band, the
electron experiences a Lorentz force and executes a cyclotron
motion on the surface of the Fermi sea. In the case of the MBB,
the magnetic field $B_0$ changes the band structure itself. On the
other hand, the {\it magnetic} quasi-momentum $\hbar{\bf k}$ is a
good quantum number with $\hbar\dot{\bf k}=0$. Therefore, there is
{\it no} cyclotron motion of ${\bf k}$ (even though there is a
magnetic field $B_0$ ). Similar to the case of the Bloch band, one
can construct a wave packet out of the magnetic Bloch states, and
study its motion in both the ${\bf r}$ and the ${\bf k}$ space
when it is subject to an additional weak electromagnetic field
${\bf E}$ and $\delta{\bf B}$. The semiclassical equations of
motion that is valid under the one-band approximation have exactly
the same form as Eq.~\eqref{em:EOM}. One simply needs to
reinterpret $\hbar{\bf k}$, $E_0({\bf k})$, and ${\bf B}$ in
Eq.~\eqref{em:EOM} as the magnetic momentum, the magnetic band
energy, and the extra magnetic field $\delta{\bf B}$ respectively
\cite{chang1995,chang1996}. As a result, when $\delta{\bf B}$ is
not zero, there exists similar circulating motion in the MBB. This
type of orbit will be called ``hyperorbit".

Let us first consider the case without the electric field (the
case with both ${\bf E}$ and $\delta{\bf B}$ will be considered in
the next subsection). By combining the following two equations of
motion (Cf. Eq.~\eqref{em:EOM}),
\begin{eqnarray}
\hbar\dot{\bf k}&=&-e\dot{\bf r}\times \delta{\bf B};\label{mbbc:eqk}\\
\hbar\dot{\bf r}&=&\frac{\partial E}{\partial {\bf
k}}-\hbar\dot{\bf k}\times \Omega,\label{mbbc:eqr}
\end{eqnarray}
one has,
\begin{equation}
\hbar\dot{\bf k}=-\frac{1}{\kappa}\frac{\partial E}{\partial{\bf k
}}\times\delta{\bf B}\frac{e}{\hbar},
\end{equation}
where $\kappa({\bf k})=1+\Omega({\bf k})\delta Be/\hbar$. This
determines the ${\bf k}$-orbit moving along a path with constant
$E({\bf k})=E_0({\bf k })-{\bf M}({\bf k})\cdot\delta{\bf B}$,
which is the magnetic Bloch band energy shifted by the
magnetization energy. Similar to the Bloch band case, it is not
difficult to see from Eq.~\eqref{mbbc:eqk} that the ${\bf
r}$-orbit is simply the ${\bf k}$-orbit rotated by $\pi/2$ and
(linearly) scaled by the factor $\hbar/e\delta B$. These orbits in
the MBB and their real-space counterparts are the hyperorbits
mentioned earlier \cite{chambers1965}.

The size of a real-space hyperorbit may be very large (if phase
coherence can be maintained during the circulation) since it is
proportional to the inverse of the residual magnetic field $\delta
B$. Furthermore, since the split magnetic subband is narrower and
flatter than the original Bloch band, the electron group velocity
is small. As a result, the frequency of the hyperorbit motion can
be very low. Nevertheless, it is possible to detect the hyperorbit
using, for example, resonant absorption of ultrasonic wave or the
conductance oscillation in an electron focusing device.

Similar to the cyclotron orbit, the hyperorbit motion can also be
quantized using the Bohr-Sommerfeld quantization rule (see
Eq.~\eqref{qza:bs}). One only needs to bear in mind that ${\bf k}$
is confined to the smaller MBZ and the magnetic field in
Eq.~\eqref{qza:bs} should be $\delta {\bf B}$. After the
quantization, there can only be a finite number of hyperorbits in
the MBZ. The area of the largest hyperorbit should be equal to or
slightly smaller (assuming $\delta B \ll B_0$ so that the number
of hyperorbits is large) than the area of the MBZ $(2\pi/a)^2/q$
(see Fig.~\ref{figc2}(b)). For such an orbit, the Berry phase
correction $\Gamma/2\pi$ in Eq.~\eqref{qza:bs} is very close to
the integer Hall conductivity $\sigma_H$ of the MBB. Therefore, it
is not difficult to see that the number of hyperorbits should be
$|1/(q\delta \phi)+\sigma_H|$, where $\delta \phi\equiv \delta B
a^2/\phi_0$ is the residual flux per plaquette.

Because the MBZ is $q$-fold degenerate (see Sec.~\ref{sec:mbba}),
the number of energy levels produced by these hyperorbits are
\cite{chang1995}
\begin{equation}\label{mbbc:D}
D=\frac{|1/(q\delta\phi)+\sigma_H|}{q}.
\end{equation}
If one further takes the tunnelling between degenerate hyperorbits
into account \cite{wilkinson1984b}, then each energy level will be
broadened into an energy band. These are the magnetic energy
subbands at a finer energy scale compared to the original MBB.

\subsection{\label{sec:mbbd}Hall conductivity of hyperorbit}

According to Laughlin's argument, each of the isolated subband
should have its own integer Hall conductivity. That is, as a
result of band splitting, the integer Hall conductivity $\sigma_H$
of the parent band is split to a distribution of integers
{$\sigma_r$} (there are $q$ of them). The sum of these integers
should be equal to the original Hall conductivity:
$\sigma_H=\sum_r\sigma_r$. There is a surprisingly simple way to
determine this distribution using the semiclassical formulation:
one only needs to study the response of the hyperorbit to an
electric field.

After adding a term $-e{\bf E}$ to Eq.~\eqref{mbbc:eqk}, one
obtains,
\begin{equation}\label{mbbc:drift}
\dot{\bf r}=\frac{\hbar}{e\delta B}\dot{\bf
k}\times\hat{z}+\frac{{\bf E}\times\hat{z}}{\delta B}.
\end{equation}
For a closed orbit, this is just a cyclotron motion superimposed
with a drift along the ${\bf E}\times \delta{\bf B}$ direction.
After time average, the former does not contribute to a net
transport. Therefore the Hall current density for a filled
magnetic band in a clean sample is,
\begin{equation}
{\bf J}_H=-e\int\frac{d^2{\bf k}}{(2\pi)^2}\dot{\bf
r}=-e\rho\frac{{\bf E}\times\hat{z}}{\delta B},
\end{equation}
where $\rho$ is the number of states in the MBZ divided by the
sample area. Therefore, the Hall conductivity is $\sigma^{\rm
close}_r=e\rho/\delta B$. If the areal electron density of a
sample is $\rho_0$, then after applying a flux $\phi=p/q$ per
plaquette, the MBZ shrinks by $q$ times and $\rho=\rho_0/q$.

How can one be sure that both the degeneracy in Eq.~\eqref{mbbc:D}
and the Hall conductivity $\sigma^{\rm close}_r$ are integers?
This is closely related to the following question: How does one
divide an uniform magnetic field $B$ into the quantizing part
$B_0$ and the perturbation $\delta B$? The proper way to separate
them was first proposed by \citet{azbel1964}. Since then, such a
recipe has been used widely in the analysis of the Hofstadter
spectrum \cite{hofstadter1976}.

One first expands the flux $\phi=p/q (<1)$ as a continued
fraction,
\begin{eqnarray}
\frac{p}{q}&=&\frac{1}{\displaystyle f_1+\frac{1}{\displaystyle
f_2+\frac{1}{\displaystyle f_3+\frac{1}{\cdots}}}}\nonumber\\
&\equiv&[f_1,f_2,f_3,\cdots],
\end{eqnarray}
then the continued fraction is truncated to obtain various orders
of approximate magnetic flux. For example, $\phi_1=[f_1]\equiv
p_1/q_1, \phi_2=[f_1,f_2]\equiv p_2/q_2,\phi_3=[f_1,f_2,f_3]\equiv
p_3/q_3 ,...$ etc. What is special about these truncations is that
$p_r/q_r$ is the closet approximation to $p/q$ among all fractions
with $q\leq q_r$ \cite{khinchin1964}.

As a reference, we show two identities that will be used below:
\begin{eqnarray}
q_{r+1}&=&f_{r+1}q_r+q_{r-1},\label{mbbd:pq1}\\
p_{r+1}q_r-p_rq_{r+1}&=&(-1)^r.\label{mbbd:pq2}
\end{eqnarray}

According to desired accuracy, one chooses a particular $\phi_r$
to be the quantizing flux, and takes
$\delta\phi_r\equiv\phi_{r+1}-\phi_r$ as a perturbation. With the
help of Eq.~\eqref{mbbd:pq2} one has
\begin{equation}
\delta \phi_r=\frac{(-1)^r}{q_rq_{r+1}}.
\end{equation}
As a result, the Hall conductivity for a closed hyperorbit
produced by $\delta B_{r-1}\equiv \delta\phi_{r-1}/a^2$ is (recall
that $\rho_r=\rho_0/q_r$),
\begin{equation}
\sigma^{\rm close}_r=\frac{e\rho_r}{\delta B_{r-1}}=(-1)^{r-1}
q_{r-1}.
\end{equation}
Substitute this value back to Eq.~\eqref{mbbc:D} for $D^{\rm
close}_r$ (the number of subbands split by  $\delta \phi_r$), and
use Eq.~\eqref{mbbd:pq1}, one has
\begin{equation}
D^{\rm close}_r=\frac{|1/(q_r\delta\phi_r)+\sigma_{r}^{\rm close
}|}{q_r}=f_{r+1}.
\end{equation}
This is the number of subbands split from a parent band that is
originated from a closed hyperorbit. One can see that the Hall
conductivity and the number of splitting subbands are indeed
integers.

\begin{figure}
\includegraphics[width=5 cm,angle=90]{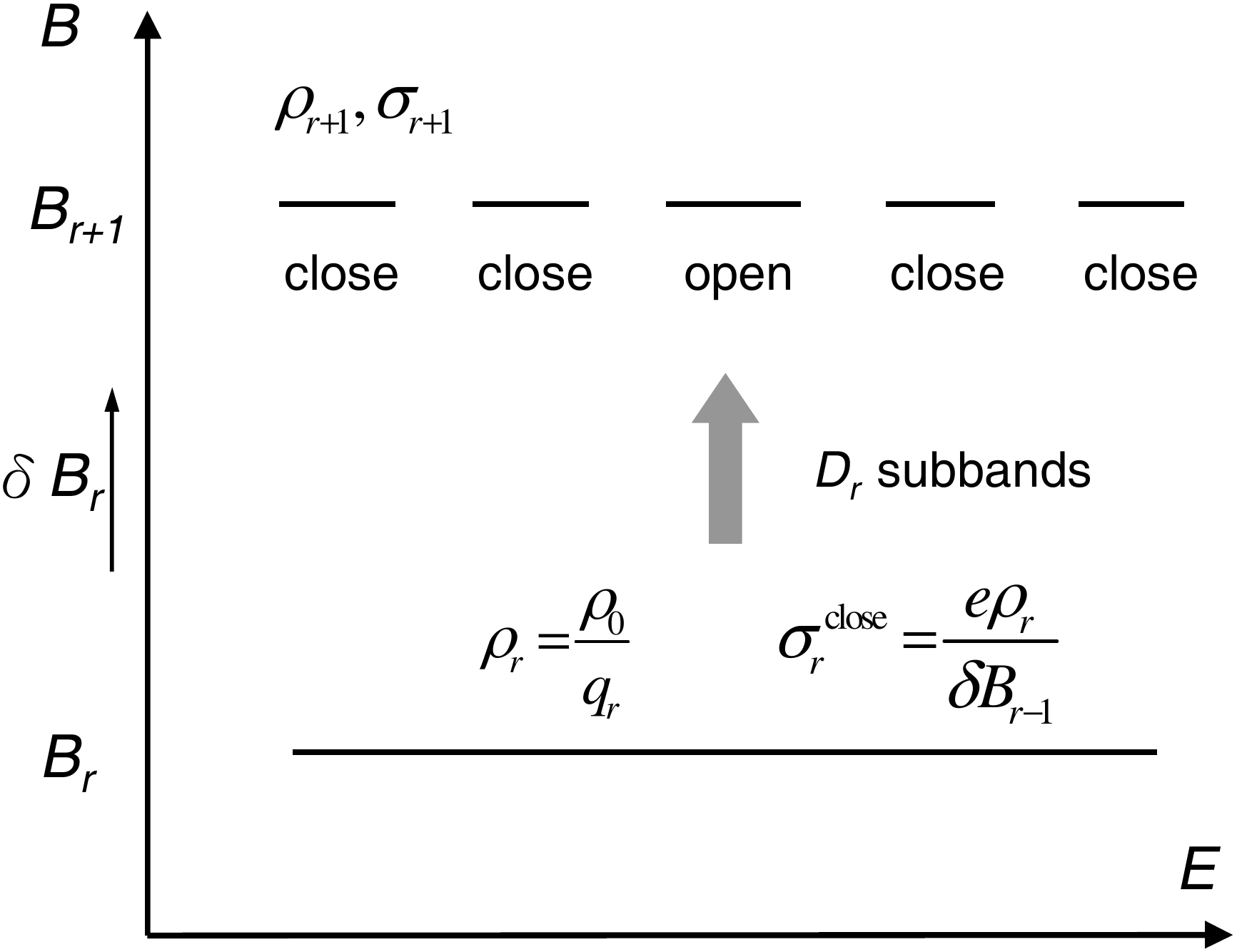}
\caption{A parent magnetic Bloch band at magnetic field $B_r$
splits to $D_r$ subbands ($D_r=5$ here) due to a perturbation
$\delta B_{r+1}$. The subbands near the band edges of the parent
band are usually originated from closed hyperorbits. The subband
in the middle is from an open hyperorbit.} \label{figc3}
\end{figure}

For lattices with square or triangular symmetry, there is one, and
only one, nesting (open) hyperorbit in the MBZ (for example, see
the diamond-shaped energy contour in Fig.~\ref{figc2}(b)). Because
of its open trajectory, the above analysis fails for the nesting
orbit since the first term in Eq.~\eqref{mbbc:drift} also
contributes to the Hall conductivity. However, since the total
number of hyperorbits in the parent band can be determined by the
quantization rule, we can easily pin down the value of
$\sigma^{\rm open}_r$ with the help of the sum rule:
$\sigma_H^{\rm parent}=\sum_r\sigma_r$. Furthermore, $D^{\rm
open}_r$ can be calculated from Eq.~\eqref{mbbc:D} once
$\sigma^{\rm open}_r$ is known. Therefore, both the distribution
of the $\sigma_r$'s and the pattern of splitting can be determined
entirely within the semiclassical formulation. The computation in
principle can be carried out to all orders of $r$. Interested
readers may consult \citet{chang1996} and \citet{gphase2003} (Chap
13) for more details.

\section{\label{sec:nab}Non-Abelian formulation}

In previous sections, we have considered the semiclassical
electron dynamics with an Abelian Berry curvature. Such a
formalism can be extended to the cases where the energy bands are
degenerate or nearly degenerate (eg., due to spin)
\cite{culcer2005,shindou2005}. Because the degenerate Bloch states
have multiple components, the Berry curvature becomes a matrix
with non-Abelian gauge structure. We will report recent progress
on re-quantizing the semiclassical theory that helps turning the
wavepacket energy into an effective quantum Hamiltonian
\cite{chang2008}. After citing the dynamics of the Dirac electron
as an example, this approach is applied to semiconductor electrons
with spin degrees of freedom. Finally, we point out that the
effective Hamiltonian is only part of an effective theory, and
that the variables in the effective Hamiltonian are often
gauge-dependent and therefore cannot be physical varaibles. In
order to have a complete effective theory, one also needs to
identify the correct physical variables relevant to experiments.

\subsection{\label{sec:naba}Non-Abelian electron wavepacket}

The wavepacket in an energy band with $D$-fold degeneracy is a
superposition of the Bloch states $\psi_{n{\bf q}}$ (Cf.
Sec.~\ref{sec:wave}),
\begin{equation}
|W\rangle=\sum_{n=1}^D\int d^3 q a({\bf q},t)\eta_n({\bf
q},t)|\psi_{n{\bf q}}\rangle,
\end{equation}
where $\sum_n|\eta_n({\bf q},t)|^2=1$ at each ${\bf q}$, and
$a({\bf q},t)$ is a normalized distribution that centers at ${\bf
q}_c(t)$. Furthermore, the wavepacket is built to be localized at
${\bf r}_c$ in the ${\bf r}$-space. One can first obtain an
effective Lagrangian for the wavepacket variables ${\bf r}_c$,
${\bf q}_c$, and $\eta_n$, then derive their dynamical equations
of motion. Without going into details, we only review primary
results of such an investigation \cite{culcer2005}.

Similar to the non-degenerate case, there are three essential
quantities in such a formulation. In addition to the Bloch energy
$E_0({\bf q})$, there are the Berry curvature and the magnetic
moment of the wavepacket (see Sec.~\ref{sec:wave}). However,
because of the spinor degree of freedom, the latter two become
vector-valued matrices, instead of the usual vectors. The Berry
connection becomes,
\begin{equation}
{\bf R}_{mn}({\bf q})=i\left\langle u_{m{\bf q}}|\frac{\partial
u_{{n{\bf q}}}}{\partial {\bf
q}}\right\rangle.\label{naba:Rmatrix}
\end{equation}
In  the rest of this section, boldfaced calligraphic fonts are
reserved for vector-valued matrices. Therefore, the Berry
connection in Eq.~\eqref{naba:Rmatrix} can simply be written as
$\bm{\mathcal R }$.

The Berry curvature is defined as,
\begin{equation}
\bm{\mathcal F}({\bf q})=\nabla_{\bf q}\times\bm{\mathcal
R}-i\bm{\mathcal R}\times\bm{\mathcal R}.\label{naba:Fmatrix}
\end{equation}
Recall that the Berry connection and Berry curvature in the
Abelian case have the same mathematical structures as the vector
potential and the magnetic field in electromagnetism. Here,
$\bm{\mathcal R}$ and $\bm{\mathcal F}$ also have the same
structure as the gauge potential and gauge field in the
non-Abelian $SU(2)$ gauge theory \cite{wilczek1984}. Redefining
the spinor basis $\{\psi_{n{\bf q}}\}$ amounts to a gauge
transformation. Assuming that the new basis is obtained from the
old basis by a gauge transformation $U$, then $\bm{\mathcal R}$
and $\bm{\mathcal F}$ would change in the following way:
\begin{eqnarray}
\bm{\mathcal R}' &=& U\bm{\mathcal R}U^\dagger+i\frac{\partial U}{\partial{\bm{\lambda}}}U^\dagger, \nonumber\\
\bm{\mathcal F}' &=& U \bm{\mathcal F}U^\dagger,
\end{eqnarray}
where ${\bm\lambda}$ is the parameter of adiabatic change.

The magnetic moment of the wavepacket can be found in
Eq.~\eqref{wave:m}. If the wavepacket is narrowly distributed
around ${\bf q}_c$, then it is possible to write it as the spinor
average of the following quantity \cite{culcer2005},
\begin{equation}
{\bf M}_{nl}({\bf
q}_c)=-i\frac{e}{2\hbar}\left\langle\frac{\partial u_n}{\partial
{\bf q}_c}\left|\times\left[{\tilde H}_0-E_0({\bf q
}_c)\right]\right|\frac{\partial u_l}{\partial {\bf
q}_c}\right\rangle,\label{naba:Lmatrix}
\end{equation}
where $\tilde{H}_0\equiv e^{-i{\bf q}\cdot{\bf r}}H_0e^{i{\bf
q}\cdot{\bf r}}$. That is, ${\bf M}=\langle \bm{\mathcal
M}\rangle=\bm{\eta}^\dagger \bm{\mathcal
M}\bm{\eta}=\sum_{nl}\eta^*_n{\bf M}_{nl}\eta_l$. Except for the
extension to multiple components, the form of the magnetic moment
remains the same as its Abelian counterpart (see
Eq.~\eqref{wave:m}).

As a reference, we write down the equations of motion for the
non-Abelian wavepacket \cite{culcer2005}:
\begin{eqnarray}
\hbar\dot{\bf k}_c&=&-e{\bf E}-e\dot{\bf r}_c\times{\bf
B},\label{naba:kdot}\\
\hbar\dot{\bf r}_c&=&\left\langle\left[\frac{\cal D}{\cal D{\bf
k}_c},{\cal H}\right]\right\rangle-\hbar\dot{\bf k}_c\times{\bf F},\label{naba:rdot}\\
i\hbar\dot{\bm{\eta}}&=&\left(-\bm{\mathcal M}\cdot{\bf
B}-\hbar{\dot{\bf k}_c}\cdot\bm{\mathcal
R}\right)\bm{\eta},\label{naba:etadot}
\end{eqnarray}
where ${\bf F}=\langle\bm{\mathcal F}\rangle$, and the covariant
derivative ${\cal D}/{\cal D}{\bf k}_c\equiv\partial/\partial{\bf
k}_c-i\bm{\mathcal R}$. The semiclassical Hamiltonian inside the
commutator in Eq.~\eqref{naba:rdot} is
\begin{equation}
{\cal H}({\bf r}_c,{\bf k}_c)=E_0({\bf k}_c)-e\phi({\bf
r}_c)-\bf\bm{\mathcal M}({\bf k}_c)\cdot{\bf B},\label{naba:H}
\end{equation}
where ${\bf k}_c={\bf q}_c+(e/\hbar){\bf A}({\bf r}_c)$.

Equation \eqref{naba:etadot} governs the dynamics of the spinor,
from which we can derive the equation for the spin vector
$\dot{\bf J}$, where ${\bf J}=\langle\bm{\mathcal J}\rangle$, and
$\bm{\mathcal J}$ is the spin matrix,
\begin{equation}
i\hbar\dot{\bf J}=\bm{\eta}^\dagger\left[\bm{\mathcal J},{\cal
H}-\hbar\dot{\bf k}_c\cdot\bm{\mathcal
R}\right]\bm{\eta}.\label{naba:Jdot}
\end{equation}
The spin dynamics in Eq.~\eqref{naba:Jdot} is influenced by the
Zeeman energy in ${\cal H}$, as it should be. However, the
significance of the other term that is proportional to the Berry
connection is less obvious here. Later we will see that it is in
fact the spin-orbit coupling.

Compared to the Abelian case in Eq.~\eqref{em:EOM}, the $\dot{\bf
k}_c$-equation also has the electric force and the Lorentz force.
The $\dot{\bf r}_c$-equation is slightly more complicated: The
derivative in the group velocity $\partial E/\partial{\bf k}_c$ is
replaced by the covariant derivative and the variables are now
matrices in general. The spinor-averaged Hamiltonian matrix is
nothing but the wavepacket energy, $E=\langle{\cal H}\rangle$.
Same as the Abelian case, it has three terms : the Bloch energy,
the electrostatic energy, and the magnetization energy. Also, the
anomalous velocity in Eq.~\eqref{naba:rdot} remains essentially
the same. One only needs to replace the Abelian Berry curvature
with the non-Abelian one.

\subsection{\label{sec:naba1}Spin Hall effect}

The anomalous velocity in Eq.~\eqref{naba:rdot} that is
proportional to the Berry curvature ${\bf F}$ is of great physical
significance. We have seen earlier that it is the transverse
current in the quantum Hall effect and the anomalous Hall effect
(Sec.~\ref{sec:hall}). The latter requires spinful electrons with
spin-orbit coupling and therefore the carrier dynamics is suitably
described by Eqs.~\eqref{naba:kdot}, \eqref{naba:rdot} and
\eqref{naba:Jdot}.

For the non-Abelian case, the Berry curvature ${\bf F}$ is often
proportional to the spin ${\bf S }$ (see Secs.~\ref{sec:mbbc} and
\ref{sec:nabd}). If this is true, then in the presence of an
electric field, the anomalous velocity is proportional to ${\bf E
}\times {\bf S}$. That is, the trajectories of spin-up and
spin-down electrons are parted toward opposite directions
transverse to the electric field. There can be a net transverse
current if the populations of spin-up and spin-down electrons are
different, such as the case in a ferromagnet. This then leads to
the anomalous Hall effect.

If the populations of different spins are equal, then the net
electric Hall current is zero. However, the spin Hall current can
still be nonzero. This is the source of the intrinsic spin Hall
effect (SHE) in semiconductors predicted by \citet{murakami2003}.
In the original proposal, a four-band Luttinger model is used to
describe the heavy-hole (HH) bands and light-hole (LH) bands. The
Berry curvature emerges when one restricts the whole Hilbert space
to a particular (HH or LH) subspace. As we shall see in
Sec.~\ref{sec:nabd}, such a projection of the Hilbert space almost
always generates a Berry curvature. Therefore, the SHE should be
common in semiconductors or other materials. Indeed, intrinsic SHE
has also been theoretically predicted in metals \cite{guo2008}.
The analysis of the SHE from the semiclassical point of view can
also be found in \citet{culcer2005}.

In addition to the Berry curvature, impurity scattering is another
source of the (extrinsic) SHE. This is first predicted by
\citet{dyakonov1971a,dyakonov1971b} (also see
\cite{chazalviel1975}) and the same idea is later revived by
\citet{hirsch1999}. Because of the spin-orbit coupling between the
electron and the (spinless) impurity, the scattering amplitude is
not symmetric with respect to the transverse direction. This is
the same skew scattering (or Mott scattering) in AHE (see
Sec.~\ref{sec:halld1}).

To date, most of the experimental evidences for the SHE belong to
the extrinsic case. They are first observed in semiconductors
\cite{kato2004,wunderlich2005,sih2005}, and later in metals
\cite{valenzuela2006,kimura2007,seki2008}. The spin Hall
conductivity in metals can be detected at room temperature and can
be several orders of magnitude larger than that in semiconductors.
Such a large effect could be due to the resonant Kondo scattering
from the Fe impurities \cite{guo2009}. This fascinating subject is
still in rapid progress. Complete understanding of the intrinsic
or extrinsic SHE is crucial to future designs that would generate
a significant amount of spin current.

\subsection{\label{sec:nabb}Quantization of electron dynamics}

In Sec.~\ref{sec:qz}, we have introduced the Bohr-Sommerfeld
quantization, which helps predicting quantized energy levels. Such
a procedure applies to the Abelian case and is limited to closed
orbits in phase space. In this subsection, we report on the method
of canonical quantization that applies to more general situations.
With both the semiclassical theory and the method of
re-quantization at hand, one can start from a quantum theory of
general validity (such as the Dirac theory of electrons) and
descend to an effective quantum theory with a smaller range of
validity. Such a procedure can be applied iteratively to generate
a hierarchy of effective quantum theories.

As we have mentioned in Sec.~\ref{sec:qzd}, even though a
Hamiltonian system always admits canonical variables, it is not
always easy to find them. In the wavepacket theory, the variables
${\bf r}_c$ and ${\bf k}_c$ have very clear physical meaning, but
they are not canonical variables. The canonical variables ${\bf
r}$ and ${\bf p}$ accurate to linear order of the fields are
related to the center-of-mass variables as follows
\cite{chang2008},
\begin{eqnarray}
{\bf r}_c&=&{\bf r}+\bm{\mathcal R}({\mbox{\boldmath$\pi$}})+\bm{\mathcal G}(\bm{\pi}),\nonumber\\
\hbar{\bf k}_c&=&{\bf p}+e{\bf A}({\bf r})+e{\bf
B}\times\bm{\mathcal
R}({\mbox{\boldmath$\pi$}}),\label{nabb:peierls}
\end{eqnarray}
where ${\mbox{\boldmath$\pi$}}={\bf p}+e{\bf A}({\bf r})$, and
${\mathcal G}_\alpha(\bm{\pi})\equiv(e/\hbar)(\bm{\mathcal
R}\times{\bf B})\cdot\partial\bm{\mathcal R}/\partial \pi_\alpha$.
This is a generalization of the Peierls substitution to the
non-Abelian case. The last terms in both equations can be
neglected in some occasions. For example, they will not change the
force and the velocity in Eqs.~\eqref{naba:kdot} and
\eqref{naba:rdot}.

When expressed in the new variables, the semiclassical Hamiltonian
in Eq.~\eqref{naba:H} can be written as,
\begin{eqnarray}
{\cal H}({\bf r},{\bf p})&=&E_0({\bm{\pi}})-e\phi({\bf r})+e{\bf
E}\cdot\bm{\mathcal R}({\bm{\pi}})\nonumber\\
&-&{\bf B}\cdot\left[\bm{\mathcal M}({\bm{\pi}})-e\bm{\mathcal
R}\times \frac{\partial E_0}{\partial
{\bm{\pi}}}\right],\label{nabb:Heff}
\end{eqnarray}
where we have used the Taylor expansion and neglected terms
nonlinear in fields. Finally, one promotes the canonical variables
to quantum conjugate variables to convert ${\cal H}$ to an
effective quantum Hamiltonian.

Compared to the semiclassical Hamiltonian in Eq.~\eqref{naba:H},
the quantum Hamiltonian has two additional terms from the Taylor
expansion. The dipole-energy term $e{\bf E}\cdot \bm{\mathcal R}$
is originated from the shift between the charge center ${\bf r}_c$
and the center of the canonical variable ${\bf r}$. Although the
exact form of the Berry connection $\bm{\mathcal R}$ depends on
the physical model, we will show that for both the Dirac electron
(\ref{sec:nabc}) and the semiconductor electron (\ref{sec:nabd}),
the dipole term is closely related to the spin-orbit coupling. The
correction to the Zeeman energy is sometimes called the Yafet
term, which vanishes near a band edge \cite{yafet1963}.

Three remarks are in order.  First, the form of the Hamiltonian,
especially the spin-orbit term and Yafet term, is clearly gauge
dependent because of the gauge-dependent Berry connection. Such
gauge dependence has prevented one from assigning a clear physical
significance to the Yafet term. For that matter, it is also
doubtful that the electric dipole, or the spin-orbit energy can be
measured independently. Second, in a neighborhood of a $k$-point,
one can always choose to work within a particular gauge.  However,
if the first Chern number (or its non-Abelian generalization) is
not zero, one cannot choose a global gauge in which $\bm{\mathcal
R}$ is smooth everywhere in the Brillouin zone. In such a
non-trivial topological situation, one has to work with patches of
the Brillouin zone for a single canonical quantum theory.  Third,
the semiclassical theory based on the variables $\bm{\mathcal F}$
and $\bm{\mathcal M}$, on the other hand, are gauge independent.
Therefore, the effective quantum theory can be smooth globally.

\subsection{\label{sec:nabc}Dirac electron}

To illustrate the application of the non-Abelian wavepacket theory
and its re-quantization, we use the Dirac electron as an example.
The starting quantum Hamiltonian is
\begin{eqnarray}\label{nabc:dirac}
H&=&c{\bm{\alpha}}\cdot\left({\bf p}+e{\bf A}\right)+\beta
mc^2-e\phi({\bf r})\nonumber\\
&=&H_0+ce{\bm{\alpha}}\cdot{\bf A}-e\phi({\bf r}),
\end{eqnarray}
where $\bm{\alpha}$ and $\beta$ are the Dirac matrices
\cite{strange1998} and $H_0$ is the free-particle Hamiltonian. The
energy spectrum of $H_0$ has positive-energy branch and
negative-energy branch, each with two-fold degeneracy due to the
spin. This two branches are separated by a huge energy gap $mc^2$.
One can construct a wavepacket out of the positive-energy
eigenstates and study its dynamics under the influence of an
external field. The size of the wavepacket can be as small as the
Compton wavelength $\lambda_c=\hbar/mc$ (but not smaller), which
is two orders of magnitude smaller than the Bohr radius.
Therefore, the adiabatic condition on the external electromagnetic
field can be easily satisfied: the spatial variation of the
potential only needs to be much smoother than $\lambda_c$. In this
case, even the lattice potential in a solid can be considered as a
semiclassical perturbation. Furthermore, because of the huge gap
between branches, interbranch tunnelling happens (and the
semiclassical theory fails) only if the field is so strong that
electron-positron pair-production can no longer be ignored.

Since the wavepacket is living on a branch with two-fold
degeneracy, the Berry connection and curvature are $2\times 2$
matrices \cite{chang2008},
\begin{eqnarray}
\bm{\mathcal R}({\bf q
})&=&\frac{\lambda_c^2}{2\gamma(\gamma+1)}{\bf
q}\times\bm{\sigma},\\
\bm{\mathcal F}({\bf q
})&=&-\frac{\lambda_c^2}{2\gamma^3}\left(\bm{\sigma}+\lambda_c^2\frac{{\bf
q}\cdot\bm{\sigma}}{\gamma+1}{\bf q}\right),
\end{eqnarray}
where $\gamma({q})\equiv\sqrt{1+(\hbar q/mc)^2}$ is the
relativistic dilation factor. To calculate these quantities, we
only need the free particle eigenstates of $H_0$ (see
Eqs.~\eqref{naba:Rmatrix} and \eqref{naba:Fmatrix}). That is, the
non-trivial gauge structure exists in the free particle already.

It may come as a big surprise that the free wavepacket also
possess an intrinsic magnetic moment. Straightforward application
of Eq.~\eqref{naba:Lmatrix} gives \cite{chuu},
\begin{equation}
\bm{\mathcal M}({\bf q
})=-\frac{e\hbar}{2m\gamma^2}\left(\bm{\sigma}+\lambda_c^2\frac{{\bf
q}\cdot\bm{\sigma}}{\gamma+1}{\bf q}\right).\label{nabc:Ldirac}
\end{equation}
This result agrees with the one calculated from the abstract spin
operator $\hat{\bf S}$ in the Dirac theory \cite{chuu},
\begin{equation}
\bm{\mathcal M}= -g\frac{e\hbar}{2m\gamma({\bf q})}\langle
W|\hat{\bf S}|W\rangle,
\end{equation}
in which the $g$-factor is two. The Zeeman coupling in the
wavepacket energy is $-\bm{\mathcal M}\cdot{\bf B}$. Therefore,
this magnetic moment gives the correct magnitude of the Zeeman
energy with the correct $g$-factor. Recall that
Eq.~\eqref{naba:Lmatrix} is originated from Eq.~\eqref{wave:m},
which is the magnetic moment due to circulation charge current.
Therefore, the magnetic moment here indeed is a result of the
spinning wavepacket.

The present approach is a revival of Uhlenbeck and Goudsmit's
rotating sphere model of the electron spin but without its
problem. The size of the wavepacket $\lambda_c$ constructed from
the positive-energy states is two orders of magnitude larger than
the classical electron radius $e^2/mc^2$. Therefore, the
wavepacket does not have to rotate faster than the speed of light
to have the correct magnitude of spin. This semiclassical model
for spin is certainly very pleasing since it gives a clear and
heuristic picture of the electron spin. Also, one does not have to
resort to the more complicated Foldy-Wouthuysen approach to
extract the spin from the Dirac Hamiltonian \cite{foldy1950}.

From the equation of motion in Eq.~\eqref{naba:Jdot}, one obtains
\begin{equation}
\langle\dot{\bm{\sigma}}\rangle=\frac{e}{\gamma m}\left[{\bf B }
+{\bf E}\times\frac{\hbar{\bf k}_c}{(\gamma+1)mc^2
}\right]\times\langle{\bm{\sigma}}\rangle.\label{bmt}
\end{equation}
This is the Bargmann-Michel-Telegdi (BMT) equation for a
relativistic electron \cite{bargmann1959}. More discussions on the
equations of motion for ${\bf r}_c$ and ${\bf k}_c$ can be found
in \citet{chang2008}.

Finally, substituting the Berry connection and the magnetic moment
into Eq.~\eqref{nabb:Heff} and using
$E_0(\bm{\pi})=\sqrt{c^2\bm{\pi}^2+m^2c^4}$, one can obtain the
effective quantum Hamiltonian,
\begin{eqnarray}
{\cal H}({\bf r},{\bf p})&=&\gamma(\bm{\pi}) mc^2-e\phi({\bf r })
+\frac{\mu_B}{\gamma
(\gamma+1)}\frac{\bm{\pi}}{mc^2}\times\bm{\sigma}\cdot{\bf
E}\nonumber\\
&+&\frac{\mu_B}{\gamma}\bm{\sigma}\cdot{\bf B},\label{nabc:pauli}
\end{eqnarray}
in which all the $\gamma$'s are functions of $\bm{\pi}$ and
$\mu_B=e\hbar/2m$. This is the relativistic Pauli Hamiltonian. At
low velocity, $\gamma\simeq 1$, and it reduces to the more
familiar form. Notice that the spin-orbit coupling comes from the
dipole energy term with the Berry connection, as we have mentioned
earlier (also see \citet{mathur1991,shankar1994}).

\subsection{\label{sec:nabd}Semiconductor electron}

\begin{figure}
\center
\includegraphics[width=2.5 in,angle=90]{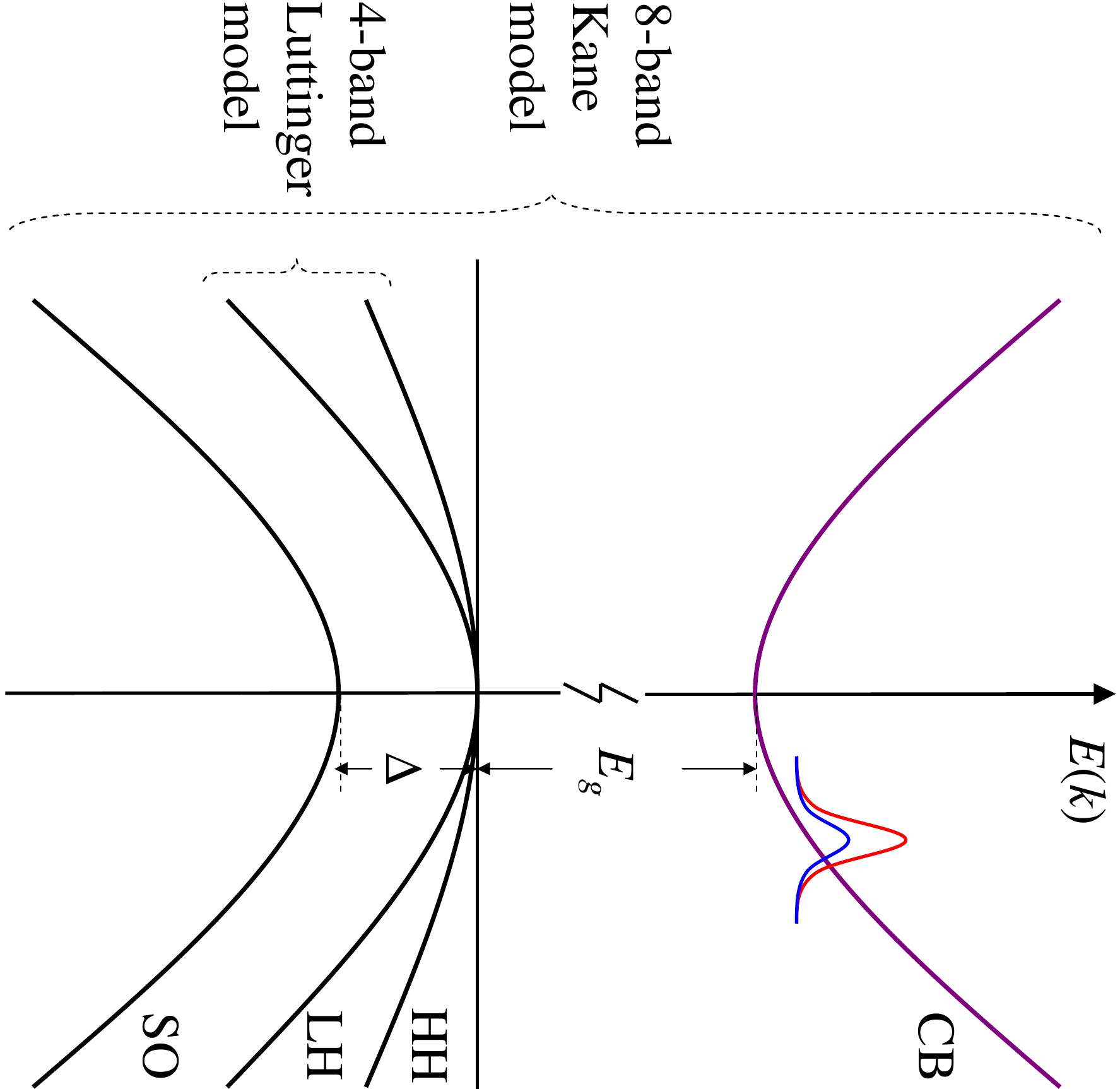}
\caption{(color online) Schematic plot of the semiconductor band
structure near the fundamental gap. The wavepacket in the
conduction band is formed from a two-component spinor.}
\label{figc4}
\end{figure}

When studying the transport properties of semiconductors, one
often only focus on the carriers near the fundamental gap at the
$\Gamma$-point. In this case, the band structure far away from
this region is not essential. It is then a good approximation to
use the ${\bf k}\cdot{\bf p}$ expansion and obtain the 4-band
Luttinger model or the 8-band Kane model
\cite{luttinger1951,kane1957,winkler2003} to replace the more
detailed band structure (see Fig.~\ref{figc4}). In this
subsection, we will start from the 8-band Kane model and study the
wavepacket dynamics in one of its subspace: the conduction band.
It is also possible to investigate the wavepacket dynamics in
other subspaces: the HH-LH complex or the spin-orbit split-off
band. The result of the latter is not reported in this review.
Interested readers can consult \citet{chang2008} for more details,
including the explicit form of the Kane Hamiltonian that the
calculations are based upon.

To calculate the Berry connection in Eq.~\eqref{naba:Rmatrix}, one
needs to obtain the eigenstates of the Kane model, which have
eight components. Similar to the positive-energy branch of the
Dirac electron, the conduction band is two-fold degenerate.
Detailed calculation shows that, the Berry connection is a
$2\times 2$ matrix of the form,
\begin{equation}
{\bm{\mathcal
R}}=\frac{V^2}{3}\left[\frac{1}{E_g^2}-\frac{1}{(E_g+\Delta)^2}\right]
{\bm{\sigma}}\times{\bf k}\label{nabd:R},
\end{equation}
where $E_g$ is the fundamental gap, $\Delta$ is the spin-orbit
spit-off gap, and $V=\frac{\hbar}{m_0}\langle
S|\hat{p}_x|X\rangle$ is a matrix element of the momentum
operator.

As a result, the dipole term $e{\bf E}\cdot\bm{\mathcal R}$
becomes,
\begin{equation}
{\cal H}_{so}=e{\bf E}\cdot{\bm{\mathcal R}}=\alpha{\bf
E}\cdot{{\bm{\sigma}}\times{\bf k}},
\end{equation}
where $\alpha\equiv(eV^2/3)[1/E_g^2-1/(E_g+\Delta)^2]$. The
coefficient $\alpha$ and the form of the spin-orbit coupling are
the same as the Rashba coupling \cite{rashba1960,bychkov1984}.
However, unlike the usual Rashba coupling that requires structural
inversion asymmetry to generate an internal field, this term
exists in a bulk semiconductor with inversion symmetry but
requires an external field ${\bf E}$.

From the Berry connection, we can calculate the Berry curvature in
Eq.~\eqref{naba:Fmatrix} to the leading order of ${\bf k}$ as,
\begin{equation}
\bm{\mathcal
F}=\frac{2V^2}{3}\left[\frac{1}{E_g^2}-\frac{1}{(E_g+\Delta)^2}\right]
{\mbox{\boldmath$\sigma$}}.
\end{equation}
In the presence of an electric field, this would generate the
transverse velocity in Eq.~\eqref{naba:rdot},
\begin{equation}
{\bf v}_T=2e\alpha{\bf E}\times\langle\bm{\sigma}\rangle.
\end{equation}
As a result, spin-up and spin-down electrons move toward opposite
directions, which results in a spin-Hall effect (see
Sec.~\ref{sec:naba1} for related discussion).

The wavepacket in the conduction band also spontaneously rotates
with respect to its own center of mass. To the lowest order of
${\bf k}$, it has the magnetic moment,
\begin{equation}
\bm{\mathcal
M}=\frac{eV^2}{3\hbar}\left(\frac{1}{E_g}-\frac{1}{E_g+\Delta}\right)
{\mbox{\boldmath$\sigma$}}.
\end{equation}

With these three basic quantities, $\bm{\mathcal R}$,
$\bm{\mathcal F}$, and $\bm{\mathcal M}$, the re-quantized
Hamiltonian in Eq.~\eqref{nabb:Heff} can be established as
\begin{equation}
{\cal H}({\bf r},{\bf p})=E_0({\bm{\pi}})-e\phi({\bf
r})+\alpha{\bf E}\cdot{\bm{\sigma}}\times{\bm{\pi}}+\delta
g\mu_B{\bf B}\cdot\frac{\hbar\bm{\sigma}}{2},
\end{equation}
where $E_0$ includes the Zeeman energy from the bare spin and
\begin{equation}
\delta
g=-\frac{4}{3}\frac{mV^2}{\hbar^2}\left(\frac{1}{E_g}-\frac{1}{E_g
+\Delta}\right).\label{g}
\end{equation}
In most textbooks on solid state physics, one can find this
correction of the $g$-factor. However, a clear identification with
electron's angular momentum is often lacking. In the wavepacket
formulation, we see that $\delta g$ is indeed originated from the
electron's rotating motion.

\subsection{\label{sec:nabe}Incompleteness of effective Hamiltonian}

Once the effective Hamiltonian ${\cal H}({\bf r},{\bf p})$ is
obtained, one can go on to study its spectra and states, without
referring back to the original Hamiltonian. Based on the spectra
and states, any physics observables of interest can be calculated.
These physics variables may be position, momentum, or other
related quantities. Nevertheless, we would like to emphasize that,
the canonical variables in the effective Hamiltonian may not be
physical observables. They may differ, for example, by a Berry
connection in the case of the position variable. The effective
Hamiltonian itself is not enough for correct prediction, if the
physical variables have not been identified properly.

This is best illustrated using the Dirac electron as an example.
At low velocity, the effective Pauli Hamiltonian is (see
Eq.~\eqref{nabc:pauli}),
\begin{eqnarray}
{\cal H}({\bf r},{\bf p})&=&\frac{\pi^2}{2m}-e\phi({\bf r })
+\frac{\mu_B}{2}\frac{\bm{\pi}}{mc^2}\times{\bm{\sigma}}\cdot{\bf
E}\nonumber\\&+&\mu_B{\bm{\sigma}}\cdot{\bf B},
\end{eqnarray}
which is a starting point of many solid-state calculations. It is
considered accurate for most of the low-energy applications in
solid state. When one applies an electric field, then according to
the Heisenberg equation of motion, the velocity of the electron is
\begin{equation}
\dot{\bf r}=
\frac{\bm{\pi}}{m}+\frac{e\lambda_c^2}{4\hbar}\bm{\sigma}\times\bf{E},
\end{equation}
where $\lambda_c$ is the Compton wavelength.

However, if one calculates the velocity of a Dirac electron
according to Eq.~\eqref{naba:rdot}, then the result is,
\begin{equation}
\dot{\bf r}_c= \frac{\hbar{\bf k
}}{m}+\frac{e\lambda_c^2}{2\hbar}\langle\bm{\sigma}\rangle\times\bf{E}.
\end{equation}
That is, the transverse velocity is larger by a factor of two. The
source of this discrepancy can be traced back to the difference
between the two position variables: ${\bf r}_c$ and ${\bf r}$ (see
Eq.~\eqref{nabb:peierls}). One should regard the equation for
$\dot{\bf r}_c$ as the correct one since it is based on the Dirac
theory (also see \citet{bliokh2005}).

Such a discrepancy between the same physical variable in different
theories can also be understood from the perspective of the
Foldy-Wouthuysen transformation. The Pauli Hamiltonian can also be
obtained from block-diagonalizing the Dirac Hamiltonian using an
unitary transformation. Since the basis of states has been
rotated, the explicit representations of all of the observables
should be changed as well. For example, the ${\bf r}_c$ in
Eq.~\eqref{nabb:peierls} can be obtained by a FW rotation,
followed by a projection to the positive-energy subspace
\cite{foldy1950}.

\begin{figure}
\center
\includegraphics[width=2.2 in,angle=90]{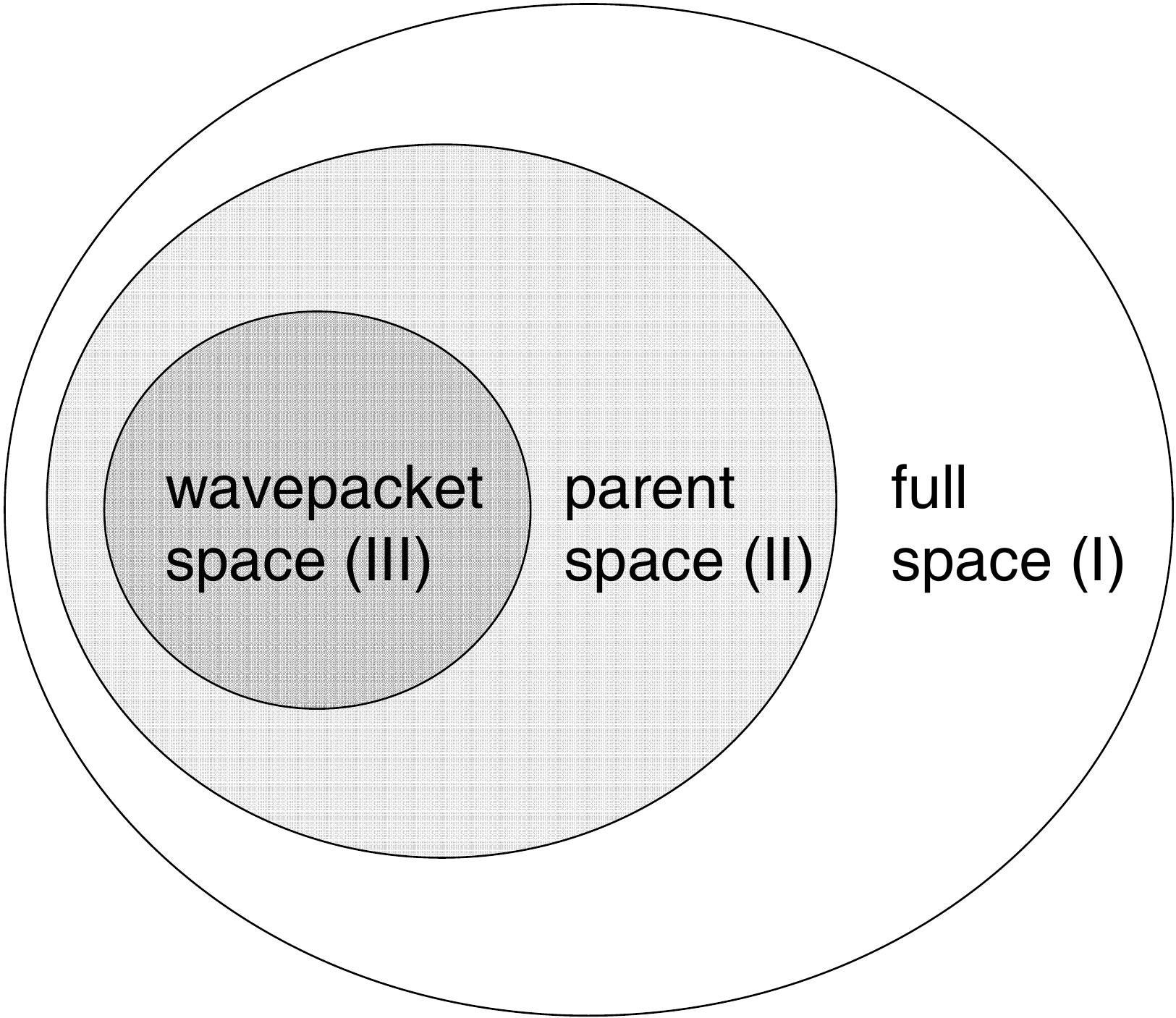}
\caption{The extent of wavepacket space, parent space, and full
space.} \label{figc5}
\end{figure}

\subsection{\label{sec:nabf}Hierarchy structure of effective theories}

Finally, we report on the hierarchical relations for the basic
quantities, the Berry curvature $\bm{\mathcal F}$ and the magnetic
moment $\bm{\mathcal M}$. Let us consider theories on three
different levels of hierarchy -- I, II, and III -- with
progressively smaller and smaller Hilbert spaces. These spaces
will be called the full space, the parent space, and the
wavepacket space respectively (see Fig.~\ref{figc5}).

Alternative to Eqs.~\eqref{naba:Fmatrix} and \eqref{naba:Lmatrix},
the Berry curvature and the magnetic moment can be written in the
following forms \cite{chang2008},
\begin{eqnarray}
{\bf F}_{mn}&=&i\sum_{l\in {out}}{\bf R}_{ml}\times{\bf
R}_{ln},\label{nabe:FF}\\
{\bf M}_{mn}&=&\frac{ie}{2\hbar}\sum_{l\in
{out}}(E_{0,m}-E_{0,l}){\bf R}_{ml}\times{\bf
R}_{ln}.\label{nabe:LL}
\end{eqnarray}
where ${\bf R}_{ml}$ is the Berry connection, and $l$ sums over
the states {\it outside} of the space of interest. From
Eqs.~\eqref{nabe:FF} and \eqref{nabe:LL}, one sees that the Berry
curvature and the magnetic moment for theory I are zero since
there is no state outside the full space. With the help of the
states in the full space, one can calculate the Berry curvatures
and the magnetic moment in theory II and theory III. They are
designated as $(\bm{\mathcal F}_p,\bm{\mathcal M}_p)$ and
$(\bm{\mathcal F},\bm{\mathcal M})$ respectively. These two sets
of matrices have different ranks since the parent space and the
wavepacket space have different dimensions.

On the other hand, if one starts from the parent space, then the
Berry curvature and the magnetic moment for theory II is zero
(instead of $\bm{\mathcal F}_p$ and $\bm{\mathcal M}_p$). The
Berry curvature and the magnetic moment for theory III are now
designated as $\bm{\mathcal F}'$ and $\bm{\mathcal M}'$. They are
different from $\bm{\mathcal F}$ and $\bm{\mathcal M}$ since the
former are obtained from the summations with more outside states
from the full space. It is straightforward to see from
Eqs.~\eqref{nabe:FF} and \eqref{nabe:LL} that
\begin{eqnarray}
\bm{\mathcal F}&=&\bm{\mathcal
F}'+P\bm{\mathcal F}_pP,\nonumber\\
\bm{\mathcal M}&=&\bm{\mathcal M}'+P\bm{\mathcal
M}_pP,\label{nabe:correction}
\end{eqnarray}
where $P$ is a dimension-reduction projection from the parent
space to the wavepacket subspace. This means that starting from
theory II, instead of theory I, as the parent theory, one would
have the errors $P\bm{\mathcal F}_pP$ and $P\bm{\mathcal M}_pP$.
On the other hand, however, whenever the scope of the parent
theory needs to be extended, e.g, from II to I, instead of
starting all of the calculations anew, one only needs additional
input from $\bm{\mathcal F}_p$ and $\bm{\mathcal M}_p$ and the
accuracy can be improved easily.

For example, in \citet{murakami2003,murakami2004}'s original
proposal of the spin Hall effect of holes, the parent space is the
HH-LH complex. The heavy hole (or the light hole) acquires a
non-zero Berry curvature as a result of the projection from this
parent space to the HH band (or the LH band). This Berry curvature
corresponds to the $\bm{\mathcal F}'$ above. It gives rise to a
spin-dependent transverse velocity $e{\bf E}\times{\bm{\mathcal
F}}'$ that is crucial to the spin Hall effect.

Instead of the HH-LH complex, if one chooses the eight bands in
Fig.~\ref{figc4} as the full space, then the Berry curvatures of
the heavy hole and the light hole will get new contributions from
$P{\bm{\mathcal F}}_pP$. The projection from the full space with
eight bands to the HH-LH complex of four bands generates a Berry
curvature ${\bm{\mathcal F}}_p=-(2V^2/3E_g^2){\bm{\mathcal J}}$
\cite{chang2008}, where ${\bm{\mathcal J}}$ is the spin-3/2
matrix. Therefore, after further projections, we would get
additional anomalous velocities $(eV^2/E_g^2){\bf E}\times
{\bm\sigma}$ and $(eV^2/3E_g^2){\bf E}\times {\bm\sigma}$ for HH
and LH respectively.

\section{Outlook}

In most of the researches mentioned this review, the Berry phase
and semiclassical theory are explored in the single-particle
context. The fact that they are so useful and that in some of the
materials the manybody effect is crucial naturally motivates one
to extend this approach to manybody regime. There has also been
effort to include the Berry phase in the density functional theory
with spin degree of freedom \cite{niu1998,niu1999}. Recently,
Haldane studied the Berry phase and relevant quantities in the
context of Fermi-liquid theory \cite{haldane2004}. Also, the Berry
curvature on the Fermi surface, if strong enough, is predicted to
modify a repulsive interaction between electrons to an attractive
interaction and causes pairing instability \cite{shi2009}. In
addition to the artificial magnetic field generated by the
monopole of Berry curvature, a slightly different Berry curvature
involving the time-component is predicted to generate an
artificial electric field, which would affect the normalization
factor and the transverse conductivity \cite{shindou2006}. This
latter work has henceforth been generalized to multiple-band Fermi
liquid with non-Abelian Berry phase \cite{shindou2008}. Researches
along such a path is exciting and still at its early stage.

There has been a growing number of researches on the Berry phase
effect in light-matter interaction. The Berry curvature is
responsible for a transverse shift (side jump) of the light beam
reflecting off an interface
\cite{onoda2004a,sawada2005,onoda2006a}. The shift is of the order
of the wavelength and is a result of the conservation of angular
momentum. The direction of the shift depends on the circular
polarization of the incident beam. This ``optical Hall effect" can
be seen as a rediscovery of the Imbert-Federov effect
\cite{federov1955,imbert1972}. More detailed study of the optical
transport involving spin has also been carried out by Bliokh and
other researchers \cite{bliokh2006a,bliokh2006b,duval2006a}.
The similarity between the side jump of a light beam and analogous
``jump" of an electron scattering off an impurity has been noticed
quite early in Berger and Bergmann's review \cite{hallbook}. In
fact, the side jump of the electromagnetic wave and the electron
can be unified using similar dynamical equations. This shows that
the equation of motion approach being focused in this review has
very general validity. Indeed, similar approach has also been
extended to the quasiparticle dynamics in Bose-Einstein condensate
\cite{zhang2006}.

Even though the Berry curvature plays a crucial role in the
electronic structure and electron dynamics of crystals, direct
measurement of such a quantity is still lacking. There does exist
sporadic and indirect evidences of the effect of the Berry phase
or the Berry curvature through the measurement of, for example,
the quantum Hall conductance, the anomalous Hall effect, or the
Hall plateau in graphene. However, this is just a beginning. In
this review, one can see clearly that in many circumstances, the
Berry curvature should be as important as the Bloch energy.
Condensed matter physicists over the years have compiled a huge
database on the band structures and Fermi surfaces of all kinds of
materials. It is about time to add theoretical and experimental
results of the Berry curvature that will deepen our understanding
of material properties. There is still plenty of room in the
quasi-momentum space!

\begin{acknowledgments}
QN acknowledges the support from NSF (DMR-0404252/0606485), DOE
(DE-FG03-02ER45958), and the Welch Foundation.  MCC was supported by
Taiwan's NSC (NSC 96-2112-M-003-010) This research was supported in
part by an appointment to the ORNL Postdoctoral Research Associates
Program which is sponsored by Oak Ridge National Laboratory and
administered jointly by Oak Ridge National Laboratory and by the Oak
Ridge Institute for Science and Education under contract numbers
DE-AC05-00OR22725 and DE-AC05-00OR22750, respectively.
\end{acknowledgments}

\bibliography{wp.bbl}

\end{document}